\documentclass[10pt,a4paper]{article} 
\usepackage{arxiv}          
\usepackage{tikz}           
\usepackage[utf8]{inputenc} 
\usepackage[T1]{fontenc}    
\usepackage{hyperref}       
\usepackage{url}            
\usepackage{booktabs}       
\usepackage{amsfonts}       
\usepackage{nicefrac}       
\usepackage{microtype}      
\usepackage{lipsum}         
\usepackage{miller}         
\usepackage{array}          
\usepackage{tabularx}
\usepackage{longtable}
\usetikzlibrary{matrix, positioning}
\usetikzlibrary{calc}
\usepackage{amssymb}
\usepackage{lineno}         
\usepackage{amsmath}
\usepackage{mathtools}
\usepackage{optidef}
\usepackage{bm}
\usepackage{bbm}
\usepackage{subcaption}
\usepackage{graphicx}       
\usepackage{svg}            
\usepackage{rotating}       
\usepackage{lscape}         
\usepackage{array,multirow} 
\usepackage[title]{appendix}
\usepackage{float}
\usepackage{stmaryrd}       
\usepackage{textcomp}       
\usepackage{algorithm}      
\usepackage{algpseudocode}  
\usepackage{amsthm}         
\usepackage{listings}       
\usepackage{xcolor}         
\usepackage{siunitx}        
\usepackage{blindtext}       
\usepackage{pifont}         
\usepackage{import}         
\usepackage{physics, xparse}
\usepackage{comment}        
\usepackage{ragged2e, blindtext}
\usepackage{paralist}
\usepackage[font=small,labelfont=bf, justification=justified, format=plain]{caption}
\DeclareFontFamily{OT1}{pzc}{} 
\DeclareFontShape{OT1}{pzc}{m}{it}{<-> s * [1.10] pzcmi7t}{}
\DeclareMathAlphabet{\mathpzc}{OT1}{pzc}{m}{it}
\usepackage{natbib}         
\newcommand\Algphase[1]{%
\vspace*{-.7\baselineskip}\Statex\hspace*{\dimexpr-\algorithmicindent-2pt\relax}\rule{\textwidth}{0.4pt}%
\Statex\hspace*{-\algorithmicindent}\textit{#1}%
\vspace*{-.7\baselineskip}\Statex\hspace*{\dimexpr-\algorithmicindent-2pt\relax}\rule{\textwidth}{0.4pt}%
}

\title{Physics-informed Data-driven Discovery of Constitutive Models with Application to Strain-Rate-sensitive Soft Materials}

\author{{Kshitiz Upadhyay} \\
	Department of Mechanical and Industrial Engineering\\
	Louisiana State University\\
	Baton Rouge, LA 70803, USA\\
    \texttt{kshitizu@lsu.edu}
	\AND
	{Jan N. Fuhg} \\
	Sibley School of Mechanical and Aerospace Engineering\\
	Cornell University\\
	Ithaca, NY 14850, USA\\
    \texttt{jf853@cornell.edu}
    \AND
	{Nikolaos Bouklas} \\
	Sibley School of Mechanical and Aerospace Engineering\\
	Center for Applied Mathematics\\
	Cornell University\\
	Ithaca, NY 14850, USA\\
    \texttt{nb589@cornell.edu}
	\AND
	{K.T. Ramesh} \\
	Department of Mechanical Engineering\\
	Hopkins Extreme Materials Institute\\
	Johns Hopkins University\\
	Baltimore, MD 21210, USA\\
    \texttt{ramesh@jhu.edu}
}

\DeclareMathOperator*{\argmax}{arg\,max}

\begin{document}
\maketitle

\begin{abstract}
A novel data-driven constitutive modeling approach is proposed, which combines the physics-informed nature of modeling based on continuum thermodynamics with the benefits of machine learning. This approach is demonstrated on strain-rate-sensitive soft materials. This model is based on the viscous dissipation-based visco-hyperelasticity framework where the total stress is decomposed into volumetric, isochoric hyperelastic, and isochoric viscous overstress contributions. It is shown that each of these stress components can be written as linear combinations of the components of an irreducible integrity basis. Three Gaussian process regression-based surrogate models are trained (one per stress component) between principal invariants of strain and strain rate tensors and the corresponding coefficients of the integrity basis components. It is demonstrated that this type of model construction enforces key physics-based constraints on the predicted responses: the second law of thermodynamics, the principles of local action and determinism, objectivity, the balance of angular momentum, an assumed reference state, isotropy, and limited memory. The three surrogate models that constitute our constitutive model are evaluated by training them on small-size numerically generated data sets corresponding to a single deformation mode and then analyzing their predictions over a much wider testing regime comprising multiple deformation modes. Our physics-informed data-driven constitutive model predictions are compared with the corresponding predictions of classical continuum thermodynamics-based and purely data-driven models. It is shown that our surrogate models can reasonably capture the stress–strain–strain rate responses in both training and testing regimes, and provide improvements in terms of prediction accuracy, generalizability to multiple deformation modes, and compatibility with limited data.
 
\end{abstract}
%
\keywords{Data-driven constitutive models, Visco-hyperelasticity, Large deformations, Physics-informed machine learning, Gaussian process regression, Hyperelasticity}
\section{Introduction} \label{Section_1}

\begin{figure}[t]
    \centering
    \includegraphics[width=13 cm]{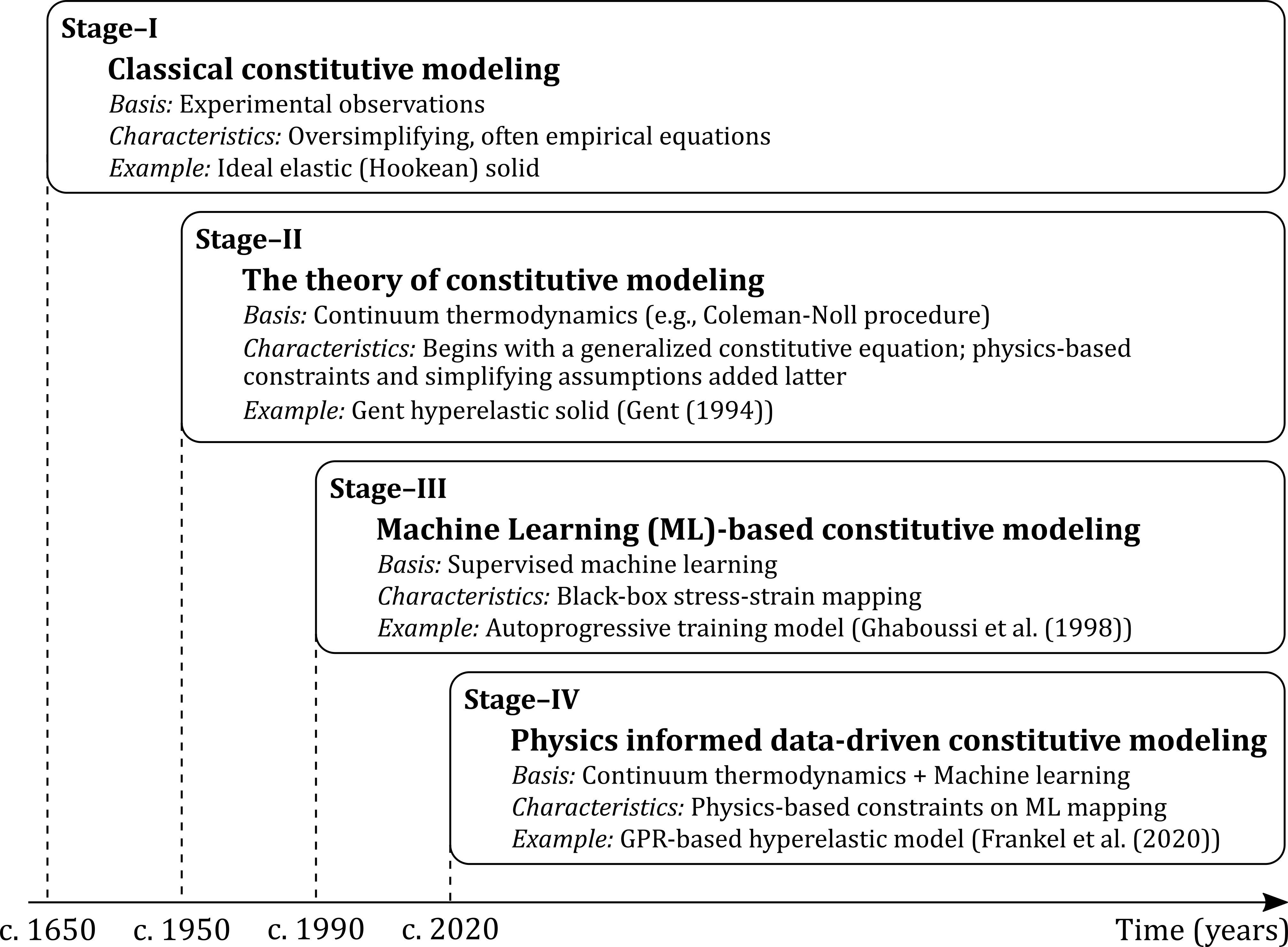}
    \caption{Schematic diagram showing the different stages in the evolution of constitutive modeling.}
    \label{fig:Evolution}
\end{figure}

Constitutive models are equations (or sets of equations) that describe the response of a material to imposed loads, deformations, or temperature changes. The philosophy behind formulating constitutive models has evolved considerably over the years, developing through four stages which are illustrated in Fig. \ref{fig:Evolution}. The oldest stage can be traced back to Robert Hooke in the Latin anagram ``ut tensio, sic vis'' (as the extension, so the force) (\cite{Hooke:1678}), and involved writing simplified and often empirical equations to describe observations of a material's response within a restricted range of loading conditions (e.g., the ideal Hookean elastic solid and the ideal Newtonian viscous fluid). The emergence of the modern continuum theory of constitutive modeling in the 1950s introduced a different philosophy: it begins with a very general functional constitutive equation that conforms to a set of constraints imposed by certain physical laws and thermodynamic considerations, and simplifying assumptions to obtain the finalized model (based on the observed material response) are imposed as late and as little as possible (\cite{Coleman_Noll:1963,malvern:1969,truesdell_noll:2004}). This approach especially benefited the formulation of constitutive models of materials that exhibit complex deformation features (e.g., large deformations) by mitigating the requirement of an extensive experimental exploration to propose constitutive equations. One such class of materials is \emph{soft materials}, whose mechanical response involves large deformation, nonlinear stress-strain behavior, and strain rate (or time)-dependence. The present study focuses on the constitutive modeling of soft materials (e.g., elastomers, hydrogels, and soft tissues) with a rate-sensitive response.

The third stage in the evolution of constitutive modeling emerged with the popularity of machine learning (ML) tools in computer sciences. These "black-box" models create simple mappings between stress and deformation, and have been extensively applied in the past 30 years on many material classes including soft materials (\cite{Wu_etal:1990,Ghaboussi_etal:1998,lefik_schrefler:2003,Jung_Ghaboussi:2006,Huang_etal:2020,Fung_etal:2021,Logarzo_etal:2021}). ML offers numerous advantages in regard to describing mechanical responses: (i) it can capture complex trends when given sufficient training data, (ii) it can directly utilize experimental data with no requirement of physical laws or expert knowledge of response trends, and (iii) it can offer potentially lower computational cost and high-throughput. Despite these advantages, the accuracy of these black-box models is limited due to their requirement of a large amount of training data, which is often not available in real experimental settings. In addition, recent studies have shown that because these models are not restricted by physical laws, they show very poor prediction accuracy in the deformation regimes (say, $>$ 10\% strain) that are not considered during model training (say, $\leqslant$ 10\% strain) (\cite{Fuhg_Bouklas:2022}).

The latest stage in the constitutive modeling evolution attempts to combine the physics-informed nature of the continuum theory of constitutive modeling (i.e., the second stage) with the flexibility and efficiency of classical ML (i.e., the third stage) to formulate "physics-informed data-driven constitutive models". For example, \cite{Liu_etal:2020} recently proposed a physics-informed neural network material model (NNMat) for isotropic hyperelastic soft tissues. This model is based on a mapping between the invariants of the right Cauchy--Green deformation tensor and the derivatives of the strain energy density with respect to the invariants, and imposes convexity constraints in the NN loss function to ensure physically reasonable predictions. Unlike this study, \cite{Frankel_etal:2020} employed Gaussian process regression (GPR) to create a mapping between the invariants of the right Cauchy--Green deformation tensor and the coefficients of the irreducible integrity basis of the stress tensor (for isotropic hyperelastic solids). \cite{Fuhg_Bouklas:2022} showed that this type of generalized functional constitutive equation (i.e., stress written as a linear function of the irreducible integrity basis) automatically ensures a number of physical constraints on material behavior: material frame-indifference, material symmetry, the balance of angular momentum, and the second law of thermodynamics (as the Clausius--Planck inequality). Further, owing to the ability of GPR to exactly predict a training data point (called the exact inference property), the physical constraint of a stress-free reference state was achieved by simply including the zero stress--zero strain data point in the training data.

Although important and useful, these GPR-based models are limited in applicability to hyperelastic solids that assume no strain-rate-dependence in material behavior, and were trained using a large amount of artificially generated data. In practice, soft materials often show strain-rate dependence in material response. In addition, sparse data at only a few strain rate levels and in a limited range of strain values is usually available from mechanical experiments (\cite{Upadhyay_etal:2019,Upadhyay_etal:2021a,Luo_etal:2019}). Thus, the goal of the present study is to develop a physics-informed data-driven constitutive model that can capture the strain-rate-sensitive mechanical response of visco-hyperelastic soft materials from limited training data, consistent with current experimental practice. To this end, this work leverages the recent contributions by \cite{Frankel_etal:2020} and \cite{Fuhg_Bouklas:2022} for hyperelastic soft materials. Note that strain rate-dependence in visco-hyperelasticity can be classified into short-time (e.g., high strain rate deformation) and long-time (e.g., creep and relaxation) responses (\cite{Upadhyay_etal:2020}). This work focuses only on the short-time response, which is of special interest in the field of injury biomechanics (e.g., simulations of crashes, blast, and ballistic impact) and in the design of protective equipment (\cite{Bracq_etal:2018,Harrigan_etal:2010,Upadhyay_etal:2021b,Payne_etal:2015}).

The paper is organized as follows: Section \ref{Section_2} formulates a generalized functional constitutive equation for visco-hyperelastic soft materials, which forms the basis for the data-driven model proposed in this study. In Section \ref{Section_3}, a data-driven mapping is defined for the generalized constitutive equation. The physics-based constraints on this data-driven mapping, which stem from both the generalized model construction and the mapping approach, are also summarized. Gaussian process regression is the primary supervised learning method utilized in this work, which allows the imposition of several additional physical constraints. These are described in Section \ref{Section_4}. Section \ref{Section_5} demonstrates the fitting and prediction performance of our GPR-based physics-informed data-driven constitutive model on several numerical tests that consider multiple deformation modes and a wide range of loading rates. The performance of our model is also compared against both conventional visco-hyperelastic constitutive models (i.e, from the second stage of the constitutive modeling evolution) and the classical ML mapping models (i.e., from the third stage). Finally, Section \ref{Section_6} presents a summary of this work.

\section{Generalized Visco-hyperelastic Constitutive Framework} \label{Section_2}

In general, every admissible thermomechanical process (and so the choice of a constitutive model) must satisfy the Clausius--Duhem inequality,
\begin{equation} \label{eq:1}
    \rho_0\dot{\eta}_0 + \nabla_0\cdot \left(\frac{\mathbf{Q}}{\theta}\right) - \rho_0 \frac{R}{\theta} \geq 0
\end{equation}
where $\rho_0$ is the mass density, $\eta_0$ is the specific entropy, $\mathbf{Q}$ is the heat flux, $R$ is the rate of internal heating per unit mass, and $\theta$ is the absolute temperature. Subscript $0$ denotes the reference system configuration, and the operator $\left(\nabla_0\cdot\right)$ represents divergence. Further, considering the reference configuration local form of the conservation of energy,
\begin{equation} \label{eq:2}
    \rho_0\dot{u}_0 = \mathbf{T}^0 \cdot\cdot \dot{\mathbf{F}} - \nabla_0 \cdot \mathbf{Q} + \rho_0 R
\end{equation}
where $u_0$ is the specific internal energy, $\mathbf{T}^0$ is the nominal stress tensor, and $\mathbf{F}$ is the deformation gradient tensor. The symbol $\cdot\cdot$ represents the tensor scalar product, such that $\mathbf{T}^0 \cdot\cdot \dot{\mathbf{F}} = T^0_{ij}\dot{F}_{ji}$ in rectangular Cartesians. By substituting $R$ from Eq. (\ref{eq:2}) in Eq. (\ref{eq:1}) and introducing the Helmholtz free energy function, $\psi = u_0 - \theta\eta_0$,
\begin{equation} \label{eq:3}
    \mathbf{T}^0 \cdot\cdot \dot{\mathbf{F}} - \rho_0\left(\dot{\psi} + \dot{\theta}\eta_0\right) - \frac{\mathbf{Q}}{\theta}\left(\nabla_0 \cdot \theta\right) \geq 0 
\end{equation}
Restricting the focus to isothermal deformations only, we particularize the second law of thermodynamics through the Clausius–Planck inequality as
\begin{equation} \label{eq:4}
    \Xi_{int} = \mathbf{T}^0 \cdot\cdot \dot{\mathbf{F}} - \rho_0 \dot{\psi} \geq 0
\end{equation}
where $\Xi_{int}$ is the internal dissipation or local entropy production (\cite{Limbert_Middleton:2004}). Equation \ref{eq:4} represents the primary physical constraint on the possible mathematical forms of our generalized constitutive equation. 

In addition to the Clausius--Planck inequality, another common physical law that constrains the forms of constitutive equations is the principle of local action, which states that material response at a given point depends only on conditions in the close vicinity of that point (e.g., $\psi=\psi(\theta,\mathbf{F},\nabla_0\mathbf{F},\dots)$). Assuming that the Helmholtz free energy function depends only on $\theta$ and $\mathbf{F}$ (i.e., neglecting higher-order gradients) so that $\psi=\psi(\theta,\mathbf{F})$, Eq. (\ref{eq:4}) becomes
\begin{equation} \label{eq:5}
    \Xi_{int} = \left(\mathbf{T}^{0^\mathrm{T}} - \rho_0 \frac{\partial\psi}{\partial\mathbf{F}}\right) : \dot{\mathbf{F}} \geq 0
\end{equation}

In terms of the right Cauchy--Green deformation tensor, $\mathbf{C} = \mathbf{F}^\mathrm{T}\mathbf{F}$, and the hyperelastic strain energy density function, $W_h$,
\begin{equation} \label{eq:6}
    \Xi_{int} = \left(\mathbf{T}^{0^\mathrm{T}} - 2\mathbf{F} \frac{\partial W_h(\mathbf{C})}{\partial\mathbf{C}}\right) : \dot{\mathbf{F}} \geq 0
\end{equation}

For an ideal hyperelastic material, the inequality in Eq. (\ref{eq:6}) reduces to equality under the assumption that the material undergoes zero dissipation during deformation. This leads to a rate-independent form of the constitutive model. While this assumption holds reasonably well in the case of quasi-static deformations, the response of soft materials under high strain rate deformation is associated with irreversible thermodynamic processes resulting from viscous dissipation effects stemming from the microscale level (\cite{Vogel_etal:2017,Pioletti:2006}). Note that "high strain rate" here is synonymous to the short-time response, when the deformation time scale is very small compared to the time it takes for the internal material microstructure to rearrange or relax under load. Under these conditions, the existence of a viscous dissipation potential (also called pseudo-potential) $W_v$ for accounting energy dissipation is usually assumed (\cite{Germain:1998,Pioletti_etal:1998,Pioletti_Rakotomanana:2000,Zhurov_etal:2007,Upadhyay_etal:2020}),
\begin{equation} \label{eq:7}
    \Xi_{int} = \left(2\mathbf{F}\frac{\partial W_v(\mathbf{C},\mathbf{\dot{C}})}{\partial\dot{\mathbf{C}}}\right) : \dot{\mathbf{F}}= \left(\mathbf{T}^{0^\mathrm{T}} - 2\mathbf{F} \frac{\partial W_h(\mathbf{C})}{\partial\mathbf{C}}\right) : \dot{\mathbf{F}} \geq 0
\end{equation}

Rearranging Eq.(\ref{eq:7}) and introducing the second Piola-Kirchhoff stress $\mathbf{S}$ ($\mathbf{S}= \mathbf{F}^\mathrm{-1} \mathbf{T}^{0^\mathrm{T}}$),
\begin{equation} \label{eq:8}
    \mathbf{S} = \mathbf{S}_h + \mathbf{S}_v = 2\frac{\partial W_h(\mathbf{C})}{\partial\mathbf{C}} + 2\frac{\partial W_v(\mathbf{C},\mathbf{\dot{C}})}{\partial\dot{\mathbf{C}}}
\end{equation}

From Eq. (\ref{eq:8}, the total stress in the material is additively decomposed into a hyperelastic stress component ($\mathbf{S}_h$) and a viscous overstress component ($\mathbf{S}_v$). Note that because both the hyperelastic strain energy density and the viscous dissipation potential are functions of symmetric tensors (i.e., $\mathbf{C}$ and $\dot{\mathbf{C}}$), the stress tensor $\mathbf{S}$ is also symmetric. This ensures the physical constraint of the balance of angular momentum. Further, it can be readily confirmed that Eq. (\ref{eq:8}) also satisfies the principle of determinism or causality (i.e., material response at a given instant is only a function of past and present events)

Next, imposing another physical constraint of isotropic material symmetry, Eq. (\ref{eq:8}) can be written in terms of certain scalar invariants of the tensors $\mathbf{C}$ and $\dot{\mathbf{C}}$ (\cite{malvern:1969,Upadhyay_etal:2020}),
\begin{equation} \label{eq:9}
    \mathbf{S} = \mathbf{S}_h + \mathbf{S}_v = 2\frac{\partial W_h(I_1,I_2,I_3)}{\partial\mathbf{C}} + 2\frac{\partial W_v(I_1,I_2,I_3,J_1,J_2,J_3,J_4,J_5,J_6,J_7)}{\partial\dot{\mathbf{C}}}
\end{equation}
where
\begin{subequations} \label{eq:10}
\begin{align}
    I_1 = \mathrm{tr}\mathbf{C},\quad I_2 = \frac{1}{2}[(\mathrm{tr}\mathbf{C})^2 - \mathrm{tr}(\mathbf{C}^2)], \quad I_3 = \mathrm{det}\mathbf{C}
\end{align}
\begin{align}
    J_1 = \mathrm{tr}\dot{\mathbf{C}}, \quad J_2 = \mathrm{tr}({\dot{\mathbf{C}}}^2), \quad J_3 = \mathrm{det}(\dot{\mathbf{C}})
\end{align}
\begin{align}
    J_4 = \mathrm{tr}(\mathbf{C}\dot{\mathbf{C}}), \quad J_5 = \mathrm{tr}(\mathbf{C}{\dot{\mathbf{C}}}^2), \quad J_6 = \mathrm{tr}(\mathbf{C}^2\dot{\mathbf{C}}), \quad J_7 = \mathrm{tr}(\mathbf{C}^2{\dot{\mathbf{C}}}^2)
\end{align}
\end{subequations}

The generalized constitutive equation in Eq. (\ref{eq:9}) will now be converted into an equivalent form, without loss of generality, to be compatible with our data-driven mapping approach (described later). To this end, following the standard approach of modeling compressible materials (\cite{Holzapfel:2000}), the tensors $\mathbf{F}$ and $\mathbf{C}$ are multiplicatively decomposed into dilatational and isochoric components,
\begin{equation} \label{eq:11}
    \mathbf{F} = J^{1/3}\Bar{\mathbf{F}}, \quad \mathbf{C} = J^{2/3}\Bar{\mathbf{C}}
\end{equation}
where $J$ is an invariant that is given as
\begin{equation} \label{eq:12}
    J = \sqrt{I_3}=\sqrt{\mathrm{det} \mathbf{C}}=\mathrm{det}\mathbf{F}
\end{equation}
and the modified deformation gradient tensor $\Bar{\mathbf{F}}$ along with the modified right Cauchy--Green deformation tensor $\Bar{\mathbf{C}}$ are introduced. These are associated with the isochoric (volume-preserving) part of the deformation, such that,
\begin{equation} \label{eq:13}
    \mathrm{det}\Bar{\mathbf{F}} = 1, \quad \mathrm{det}\Bar{\mathbf{C}} = 1
\end{equation}

The decomposition of the deformation gradient in Eq. (\ref{eq:11}) allows an equivalent decomposition of the stress tensor in Eq. (\ref{eq:9}) into dilatational and isochoric stress components (see \cite{Zhurov_etal:2007}),
\begin{equation} \label{eq:14}
    \mathbf{S} = \mathbf{S}_\mathrm{vol} + \mathbf{S}_{h,\mathrm{iso}} + \mathbf{S}_{v,\mathrm{iso}} = 2\frac{\partial U(J)}{\partial \mathbf{C}} + 2\frac{\partial \Bar{W}_h(\Bar{I}_1,\Bar{I}_2)}{\partial\mathbf{C}} + 2\frac{\partial \Bar{W}_v(\Bar{I}_1,\Bar{I}_2,\Bar{J}_1,\Bar{J}_2,\Bar{J}_3,\Bar{J}_4,\Bar{J}_5,\Bar{J}_6,\Bar{J}_7)}{\partial\dot{\mathbf{C}}}
\end{equation}
where
\begin{subequations} \label{eq:15}
\begin{align}
    \Bar{I}_1 = \mathrm{tr}\Bar{\mathbf{C}},\quad \Bar{I}_2 = \frac{1}{2}[(\mathrm{tr}\Bar{\mathbf{C}})^2 - \mathrm{tr}(\Bar{\mathbf{C}}^2)]
\end{align}
\begin{align}
    \Bar{J}_1 = \mathrm{tr}\dot{\Bar{\mathbf{C}}}, \quad \Bar{J}_2 = \mathrm{tr}({\dot{\Bar{\mathbf{C}}}}^2), \quad \Bar{J}_3 = \mathrm{det}(\dot{\Bar{\mathbf{C}}})
\end{align}
\begin{align}
    \Bar{J}_4 = \mathrm{tr}(\Bar{\mathbf{C}}\dot{\Bar{\mathbf{C}}}), \quad \Bar{J}_5 = \mathrm{tr}(\Bar{\mathbf{C}}{\dot{\Bar{\mathbf{C}}}}^2), \quad \Bar{J}_6 = \mathrm{tr}({\Bar{\mathbf{C}}}^2\dot{\Bar{\mathbf{C}}}), \quad \Bar{J}_7 = \mathrm{tr}({\Bar{\mathbf{C}}}^2{\dot{\Bar{\mathbf{C}}}}^2)
\end{align}
\end{subequations}
are the invariants. In Eq. (\ref{eq:14}), $\mathbf{S}_\mathrm{vol}$ is the volumetric stress component, $\mathbf{S}_{h,\mathrm{vol}}$ is the isochoric hyperelastic stress component, and $\mathbf{S}_{v,\mathrm{vol}}$ is the isochoric viscous overstress component. These three stress components are captured by the volumetric energy density function $U(J)$, the modified hyperelastic strain energy density $\Bar{W}_h$, and the modified viscous dissipation potential $\Bar{W}_v$, respectively. For convenience, $\Bar{W}_h$ and $\Bar{W}_v$ will henceforth be simply referred to as the strain energy density and the viscous dissipation potential, respectively. 

Equation (\ref{eq:14}) is the generalized functional form of constitutive equation considered in this work. This generalized framework (referring to both Eq. (\ref{eq:9}) and Eq. (\ref{eq:14})) is called external state variable-driven viscous dissipation-based visco-hyperelasticity (\cite{Pioletti_Rakotomanana:2000,Upadhyay_etal:2020}). Here, $\mathbf{C}$ and $\dot{\mathbf{C}}$ are the external thermodynamic state variables that are employed to relate the strain rate stiffening or softening of material response to viscous dissipation. This is in contrast to the internal state variable-driven framework in which the non-equilibrium part of the Helmholtz free energy function is described via a set of internal variables that a priori lack any physical meaning (e.g., see \cite{Simo:1987,Reese_Govindjee:1998,Holzapfel_Gasser:2001,Garcia-Gonzalez_etal:2018}). Unlike this alternative class of visco-hyperelastic models, many of which result in a hereditary-integral-based equation for stress, the external state variable-driven viscous dissipation-based visco-hyperelastic constitutive framework of this work is bound by the constraint of limited memory (a subset of the principle of fading memory). Here, limited memory means that the viscous material behavior is dependent only on the instantaneous deformation rate (i.e., very recent history), and dependence on the entire previous loading history (often described via a hereditary-integral) is neglected. First proposed by \cite{Pioletti_etal:1998}, constitutive models of this framework have been employed to successfully capture the rate-dependent response of numerous soft materials under rapid loading: human ligaments and tendons (\cite{Jiang_etal:2015,Zhurov_etal:2007,Ahsanizadeh_Li:2015}), skeletal muscles (\cite{Lu_etal:2010}), hydrogels and elastomers (\cite{Upadhyay_etal:2020,Upadhyay_etal:2021b}), tongue tissue (\cite{Yousefi_etal:2018}), and the brain and pericardium (\cite{Kulkarni_etal:2016,Upadhyay_etal:2021a,Upadhyay_etal:2022}), among others. These studies assume specific mathematical forms for $U(J)$, $\Bar{W}_h(\Bar{I}_1,\Bar{I}_2)$ and $\Bar{W}_v(\Bar{I}_1,\Bar{I}_2,\Bar{J}_1,\Bar{J}_2,\Bar{J}_3,\Bar{J}_4, \Bar{J}_5,\Bar{J}_6,\Bar{J}_7)$ based on expert knowledge and experience. The physics-informed data-driven mapping approach of the present study (Section \ref{Section_3}) aims to eliminate this need for expert intervention by employing ML to flexibly discover the mapping between stress- and strain-like variables directly from limited experimental data.

The partial derivatives representing the three stress components in Eq. (\ref{eq:14}) can be expanded via the chain rule, leading to the following generalized equations for the three stress components (see derivation in the \ref{Appendix_A}):
\begin{equation} \label{eq:16}
    \mathbf{S}_\mathrm{vol} = \zeta_1(J)\mathbf{C}^{-1}
\end{equation}
\begin{equation} \label{eq:17}
    \mathbf{S}_{h,\mathrm{iso}} = J^{-2/3} \left[ \Gamma_1(\Bar{I}_1,\Bar{I}_2) \mathrm{Dev}(\mathbf{I}) + \Gamma_2(\Bar{I}_1,\Bar{I}_2) \mathrm{Dev}(\Bar{\mathbf{C}}) \right]
\end{equation}
\vspace{-12 pt}
\begin{equation} \label{eq:18}
\begin{split}
    \mathbf{S}_{v,\mathrm{iso}} = {}& J^{-2/3} \bigl[ \Phi_1(\Bar{I}_1,\Bar{I}_2, \Bar{J}_1, \dots, \Bar{J}_7) \mathrm{Dev}(\mathbf{I}) + \Phi_2(\Bar{I}_1,\Bar{I}_2, \Bar{J}_1, \dots, \Bar{J}_7) \mathrm{Dev}(\Bar{\mathbf{C}}) + {}\\& \Phi_3(\Bar{I}_1,\Bar{I}_2, \Bar{J}_1, \dots, \Bar{J}_7) \mathrm{Dev}({\Bar{\mathbf{C}}}^{-1}) + \Phi_4(\Bar{I}_1,\Bar{I}_2, \Bar{J}_1, \dots, \Bar{J}_7) \mathrm{Dev} (\dot{\Bar{\mathbf{C}}}) + {} \\&\Phi_5(\Bar{I}_1,\Bar{I}_2, \Bar{J}_1, \dots, \Bar{J}_7) \mathrm{Dev} ({\dot{\Bar{\mathbf{C}}}}^{-1}) +\Phi_6(\Bar{I}_1,\Bar{I}_2, \Bar{J}_1, \dots, \Bar{J}_7) \mathrm{Dev} (\Bar{\mathbf{C}}\dot{\Bar{\mathbf{C}}} + \dot{\Bar{\mathbf{C}}}\Bar{\mathbf{C}}) + {} \\&\Phi_7(\Bar{I}_1,\Bar{I}_2, \Bar{J}_1, \dots, \Bar{J}_7) \mathrm{Dev} (\Bar{\mathbf{C}}^{2}\dot{\Bar{\mathbf{C}}} + \dot{\Bar{\mathbf{C}}}\Bar{\mathbf{C}}^{2})\bigr]
\end{split}
\end{equation}
where $\mathrm{Dev}(\cdot) = (\cdot) - \frac{1}{3}((\cdot):\textbf{C})\textbf{C}^{-1}$ is the deviatoric operator in the Lagrangian description (\cite{Holzapfel:2000}). Notice that individual stress components in these equations are linear combinations of the components of an integrity basis $\mathbbm{G}$ of the total stress $\mathbf{S}$, defined as
\begin{equation} \label{eq:19}
       \mathbb{G} = \{\mathbb{G}_1, \mathbb{G}_2, \mathbb{G}_3, \mathbb{G}_4, \mathbb{G}_5, \mathbb{G}_6, \mathbb{G}_7, \mathbb{G}_8\}
\end{equation}
where
\begin{subequations} \label{eq:20}
\begin{align}
    \mathbb{G}_1 = \mathbf{C}^{-1},\quad \mathbb{G}_2 = \mathrm{Dev}(\mathbf{I}), \quad \mathbb{G}_3 = \mathrm{Dev}(\Bar{\mathbf{C}})
\end{align}
\begin{align}
    \mathbb{G}_4 = \mathrm{Dev}({\Bar{\mathbf{C}}}^{-1}),\quad \mathbb{G}_5 = \mathrm{Dev} (\dot{\Bar{\mathbf{C}}}), \quad \mathbb{G}_6 = \mathrm{Dev} ({\dot{\Bar{\mathbf{C}}}}^{-1})
\end{align}
\begin{align}
    \mathbb{G}_7 = \mathrm{Dev} (\Bar{\mathbf{C}}\dot{\Bar{\mathbf{C}}} + \dot{\Bar{\mathbf{C}}}\Bar{\mathbf{C}}),\quad \mathbb{G}_8 = \mathrm{Dev} (\Bar{\mathbf{C}}^{2}\dot{\Bar{\mathbf{C}}} + \dot{\Bar{\mathbf{C}}}\Bar{\mathbf{C}}^{2})
\end{align}
\end{subequations}
Further, $\zeta_1, \Gamma_1, \Gamma_2, \Phi_1, \dots, \Phi_7$ are coefficients of the integrity basis components in the stress equations (Eqs. (\ref{eq:16}--\ref{eq:18})), and are functions of the invariants of the tensors $\Bar{\mathbf{C}}$ and $\dot{\Bar{\mathbf{C}}}$. 
For several commonly used constitutive models in the literature (i.e., for a given choice of $U$, $\Bar{W}_h$, and $\Bar{W}_v$), the explicit mathematical forms of these coefficients are provided in Tables \ref{table:A1}--\ref{table:A3} of the \ref{Appendix_A}. 
The physics-informed data-driven mapping approach of the present study discovers a mapping between invariants and coefficients of the integrity basis components directly from stress--strain--strain rate data.

\section{Proposed Physics-Informed Data-Driven Mapping Approach} \label{Section_3}
Consider a dataset $\mathcal{D}$ consisting of strain and strain rate as input and stress components as output,
\begin{equation} \label{eq:21}
    \mathcal{D} = \mathcal{D}_\mathrm{vol} \cup \mathcal{D}_{h,\mathrm{iso}} \cup \mathcal{D}_{v,\mathrm{iso}} = \{ \mathbf{C}^i, \mathbf{S}_\mathrm{vol}^i \}_{i=1}^{N_\mathrm{vol}} \cup \{ \mathbf{C}^j, \mathbf{S}_{h,\mathrm{iso}}^j \}_{j=1}^{N_{h,\mathrm{iso}}} \cup \{ \mathbf{C}^k, \dot{\mathbf{C}}^k, \mathbf{S}_{v,\mathrm{iso}}^k\}_{k=1}^{N_{v,\mathrm{iso}}}
\end{equation}
Here, constituent datasets $\mathcal{D}_\mathrm{vol}$, $\mathcal{D}_{h,\mathrm{iso}}$ and $\mathcal{D}_{v,\mathrm{iso}}$ correspond to the material response under hydrostatic, isochoric quasi-static, and isochoric dynamic conditions, respectively. Superscripts $i$,$j$, and $k$ denote a particular data point in the datasets $\mathcal{D}_\mathrm{vol}$, $\mathcal{D}_{h,\mathrm{iso}}$, and $\mathcal{D}_{v,\mathrm{iso}}$, respectively. A dataset like $\mathcal{D}$ for a soft material can be obtained in practice by compiling its hydrostatic stress--strain data as $\mathcal{D}_\mathrm{vol}$ (\cite{Nie_etal:2013,Saraf_etal:2007}), quasi-static stress--strain data under uniaxial and/or shear deformations as $\mathcal{D}_{h,\mathrm{iso}}$ (\cite{Yang_etal:2000,Upadhyay_etal:2020a}), and viscous overstress (total stress minus stress under quasi-static loading)--strain data from high strain rate testing at multiple strain rate levels under uniaxial and/or shear deformations as $\mathcal{D}_{v,\mathrm{iso}}$ (\cite{Yang_etal:2000,Saraf_etal:2007,Bracq_etal:2017,Upadhyay_etal:2021b,Upadhyay_etal:2019}).
For every data point in the dataset $\mathcal{D}$, the inputs $\mathbf{C}$ and $\dot{\mathbf{C}}$ can be used to compute the set of invariants (using Eqs. (\ref{eq:12}) and (\ref{eq:15})) as well as the components of the integrity basis $\mathbbm{G}$ (using Eq. (\ref{eq:20})). The coefficients of the integrity basis in Eqs. (\ref{eq:16}--\ref{eq:18}) can then be obtained by solving the following systems of equations, respectively:
\begin{align}
\begin{split} \label{eq:22}
    \Bigl[\mathrm{vec}(\mathbf{S}_\mathrm{vol})\Bigr] ={}& \Bigl[ \mathrm{vec}(\mathbb{G}_1)\Bigr] \Bigl[ \zeta_1 \Bigr]
\end{split} \\
\begin{split} \label{eq:23}
    \left[\mathrm{vec}\left(\frac{\mathbf{S}_{h,\mathrm{iso}}}{J^{-2/3}}\right)\right] ={}& \Bigl[ \mathrm{vec}(\mathbb{G}_2) \quad \mathrm{vec}(\mathbb{G}_3))\Bigr] \begin{bmatrix}
            \vspace{4 pt}
            \Gamma_{1} \\
            \Gamma_{2}
         \end{bmatrix}
\end{split} \\
\begin{split} \label{eq:24}
    \left[\mathrm{vec}\left(\frac{\mathbf{S}_{v,\mathrm{iso}}}{J^{-2/3}}\right)\right] ={}& \Bigl[ \mathrm{vec}(\mathbb{G}_2) \quad \mathrm{vec}(\mathbb{G}_3) \quad \mathrm{vec}(\mathbb{G}_4) \quad \mathrm{vec}(\mathbb{G}_5) \quad \mathrm{vec}(\mathbb{G}_6) \quad \mathrm{vec}(\mathbb{G}_7) \quad \mathrm{vec}(\mathbb{G}_8)\Bigr] \begin{bmatrix}
               \vspace{4 pt}
               \Phi_{1} \\
               \vspace{4 pt}
               \Phi_{2} \\
               \vspace{4 pt}
               \Phi_{3} \\
               \vspace{4 pt}
               \Phi_{4} \\
               \vspace{4 pt}
               \Phi_{5} \\
               \vspace{4 pt}
               \Phi_{6} \\
               \Phi_{7}
         \end{bmatrix}
\end{split}
\end{align}
where $\mathrm{vec}(\cdot)$ denotes the Voigt form of a matrix (for an arbitrary second-order symmetric tensor $\mathbf{Z}$, $\mathrm{vec}(\mathbf{Z}) = \bigl[\mathbf{Z}_{11}, \mathbf{Z}_{22}, \mathbf{Z}_{33}, \mathbf{Z}_{23}, \mathbf{Z}_{13}, \mathbf{Z}_{12}\bigr]^\mathrm{T}$). Notice that the above systems of equations are identical to Eqs. (\ref{eq:16}--\ref{eq:18}), and that each of these systems is written in the form of $\bigl[\boldsymbol{b}\bigr] = \bigl[\boldsymbol{A}\bigr] \bigl[\boldsymbol{x}\bigr]$. The vectors $\bigl[\boldsymbol{x}\bigr]$ containing the coefficients of the integrity basis can be obtained by solving the corresponding optimization problems---$\min_{x} \left\lVert \bigl[\boldsymbol{A}\bigr] \bigl[\boldsymbol{x}\bigr] - \bigl[\boldsymbol{b}\bigr]\right\lVert_2^2$---via, for example, QR-decomposition.

The above procedure of obtaining invariants and the coefficients of the integrity basis is carried out for every data point in $\mathcal{D}$. Using this information, an alternative dataset is generated that comprises invariants and integrity basis coefficients corresponding to each of the stress component types,
\begin{equation} \label{eq:25}
\begin{split}
     \mathcal{D}^* = \mathcal{D}_\mathrm{vol}^* \cup \mathcal{D}_{h,\mathrm{iso}}^* \cup \mathcal{D}_{v,\mathrm{iso}}^* = \{ J^i, \zeta_1^i \}_{i=1}^{N_\mathrm{vol}} \cup \{[ \Bar{I}_1^j, \Bar{I}_2^j ], [\Gamma_1^j, \Gamma_2^j] \}_{j=1}^{N_{h,\mathrm{iso}}} \cup \\ \{[ \Bar{I}_1^k, \Bar{I}_2^k, \Bar{J}_1^k, \Bar{J}_4^k, \Bar{J}_6^k], [\Phi_1^k, \dots,\Phi_7^k] \}_{k=1}^{N_{v,\mathrm{iso}}}
\end{split}
\end{equation}
The constituent datasets in $\mathcal{D}^*$ are used in the present study to train \emph{the three surrogate models that comprise our physics-informed data-driven constitutive model}:
\begin{align}
    \begin{split} \label{eq:26}
    \widetilde{\mathcal{M}}_{\mathrm{vol}}: {}& \mathcal{J}_\mathrm{vol} \in \mathbb{R}^1 \rightarrow \mathcal{\zeta} \in \mathbb{R}^1
    \end{split} \\
    \begin{split} \label{eq:27}
    \widetilde{\mathcal{M}}_{h,\mathrm{iso}}: {}& \mathcal{I} \in \mathbb{R}^2 \rightarrow \mathit{\Gamma} \in \mathbb{R}^2
    \end{split} \\
    \begin{split} \label{eq:28}
    \widetilde{\mathcal{M}}_{v,\mathrm{iso}}: {}& \mathcal{J} \in \mathbb{R}^5 \rightarrow \mathit{\Phi} \in \mathbb{R}^7
    \end{split}
\end{align}
Specifically, $\widetilde{\mathcal{M}}_{\mathrm{vol}}$ captures the volumetric stress component and is a mapping between the random vector of the invariant $J$ (i.e., $\mathcal{J}_\mathrm{vol}$) and the random vector of the coefficient $\zeta_1$ (i.e., $\mathcal{\zeta}$). $\widetilde{\mathcal{M}}_{h,\mathrm{iso}}$ captures the isochoric hyperelastic stress component and is a mapping between the random vector of the invariants $[\Bar{I}_1, \Bar{I}_2]$ (i.e., $\mathcal{I}$) and the random vector of the corresponding coefficients $[\Gamma_1, \Gamma_2]$ (i.e., $\mathit{\Gamma}$). Finally, $\widetilde{\mathcal{M}}_{v,\mathrm{iso}}$ captures the isochoric viscous overstress and is a mapping between the random vector of the invariants $[\Bar{I}_1, \Bar{I}_2, \Bar{J}_1, \Bar{J}_4, \Bar{J}_6]$ (i.e., $\mathcal{J}$) and the random vector of the corresponding coefficients $[\Phi_1, \dots, \Phi_7]$ (i.e., $\mathit{\Phi}$). In a more explicit form, we have
\begin{align}
    \begin{split} \label{eq:29}
    \widetilde{\mathcal{M}}_{\mathrm{vol}}: {}& \Bigl[ J \Bigr] \rightarrow \Bigl[ \zeta_1 \Bigr]
    \end{split} \\
    \begin{split} \label{eq:30}
    \widetilde{\mathcal{M}}_{h,\mathrm{iso}}: {}& \begin{bmatrix}
            \vspace{4 pt}
            \Bar{I}_1 \\
            \Bar{I}_2
         \end{bmatrix} \rightarrow \begin{bmatrix}
            \vspace{4 pt}
            \Gamma_{1} \\
            \Gamma_{2}
         \end{bmatrix}
    \end{split} \\
    \begin{split} \label{eq:31}
    \widetilde{\mathcal{M}}_{v,\mathrm{iso}}: {}& \begin{bmatrix}
            \vspace{4 pt}
            \Bar{I}_1 \\
            \vspace{4 pt}
            \Bar{I}_2 \\
            \vspace{4 pt}
            \Bar{J}_1 \\
            \vspace{4 pt}
            \Bar{J}_4 \\
            \Bar{J}_6 
         \end{bmatrix} \rightarrow \begin{bmatrix}
            \vspace{4 pt}
            \Phi_{1} \\
            \vspace{4 pt}
            \Phi_{2} \\
            \vspace{4 pt}
            \Phi_{3} \\
            \vspace{4 pt}
            \Phi_{4} \\
            \vspace{4 pt}
            \Phi_{5} \\
            \vspace{4 pt}
            \Phi_{6} \\
            \Phi_{7}
         \end{bmatrix}
    \end{split}
\end{align}

Note, the rationale behind not considering invariants $\Bar{J}_2$, $\Bar{J}_3$, $\Bar{J}_5$ and $\Bar{J}_7$ in the surrogate model $\widetilde{\mathcal{M}}_{v,\mathrm{iso}}$ (Eq. (\ref{eq:28}/\ref{eq:31})) is described in the next section. Briefly, this assumption of a negligible effect of certain invariants on the viscous overstress allows imposition of the physical constraint of a stress-free reference state; i.e., to ensure $\mathbf{S}_{v,\mathrm{iso}}(\mathbf{C} = \mathbf{I}) = \mathbf{0}$ regardless of the applied rate of the right Cauchy--Green deformation tensor ($\dot{\mathbf{C}}$).
\begin{algorithm}[t]
    \caption{Development of a physics-informed data-driven constitutive model for strain-rate-sensitive soft materials}\label{euclid}
    \hspace*{-2 pt} \textbf{Data:} Training datasets $\mathcal{D}_\mathrm{vol} = \{ \mathbf{C}^i, \mathbf{S}_\mathrm{vol}^i \}_{i=1}^{N_\mathrm{vol}}$, $\mathcal{D}_{h,\mathrm{iso}} = \{ \mathbf{C}^j, \mathbf{S}_{h,\mathrm{iso}}^j \}_{j=1}^{N_{h,\mathrm{iso}}}$, and $\mathcal{D}_{v,\mathrm{iso}} =  \{ \mathbf{C}^k, \dot{\mathbf{C}}^k, \mathbf{S}_{v,\mathrm{iso}}^k\}_{k=1}^{N_{v,\mathrm{iso}}}$\\
    \hspace*{-2 pt} \textbf{Result:} Corresponding surrogate models $\widetilde{\mathcal{M}}_{\mathrm{vol}}$, $\widetilde{\mathcal{M}}_{h,\mathrm{iso}}$, and $\widetilde{\mathcal{M}}_{v,\mathrm{iso}}$
    \begin{algorithmic}[1]
        \Algphase{1 --- Volumetric stress part}
        \Require $\{ \mathbf{C}^1, ..., \mathbf{C}^{N_\mathrm{vol}}\}$ and $\{ \mathbf{S}_{\mathrm{vol}}^1, ..., \mathbf{S}_{\mathrm{vol}}^{N_\mathrm{vol}}\}$, where $\mathbf{C}^i,\mathbf{S}_{\mathrm{vol}}^i \in \mathbb{R}^{3 \times 3}$
        \For{{$i \leftarrow 1, N_\mathrm{vol}$}}
        \State Obtain the invariant $J^i$ from $\mathbf{C}^i$ using Eq. (\ref{eq:12})
        \State Obtain the integity basis component $\mathbb{G}^i_\mathrm{1}$ using Eq. (\ref{eq:20}).
        \State Obtain the Voigt form of $\mathbb{G}^i_\mathrm{1}$: $\mathrm{vec}({\mathbb{G}^i_\mathrm{1}}) \in \mathbb{R}^{6 \times 1}$
        \State Obtain the Voigt form of $\mathbf{S}^i_\mathrm{vol}$: $\mathrm{vec}({\mathbf{S}^i_\mathrm{vol}}) \in \mathbb{R}^{6}$
        \State Solve the system of equations in Eq. (\ref{eq:22}) to obtain $\zeta_1^i$ ($\mathbb{R}^1$)
        \EndFor
        \State Construct $\mathcal{D}^*_{\mathrm{vol}} = \{ J^i, \zeta_1^i \}_{i=1}^{N_\mathrm{vol}}$ comprising $\{ J^1, ..., J^{N_\mathrm{vol}}\} \in \mathbb{R}^{N_{\mathrm{vol}} \times 1}$ and $\{ \zeta^1_1, ..., \zeta^{N_\mathrm{vol}}_1\} \in \mathbb{R}^{N_{\mathrm{vol}} \times 1}$
        \State Using $\mathcal{D}^*_{\mathrm{vol}}$, construct a standard GPR approximation $\widetilde{\zeta}_1 = \widetilde{M}_\mathrm{vol} (J)$
        
        \Algphase{2 --- Isochoric hyperelastic stress part}
        \Require $\{ \mathbf{C}^1, ..., \mathbf{C}^{N_{h,\mathrm{iso}}}\}$ and $\{ \mathbf{S}_{h,\mathrm{iso}}^1, ..., \mathbf{S}_{h,\mathrm{iso}}^{N_{h,\mathrm{iso}}}\}$ where $\mathbf{C}^j, \mathbf{S}_{h,\mathrm{iso}}^j \in \mathbb{R}^{3 \times 3}$
        \For{{$j \leftarrow 1, N_{h,\mathrm{iso}}$}}
        \State Obtain the invariants $J^j$, $\Bar{I}^j_1$, and $\Bar{I}^j_2$ using Eqs. (\ref{eq:12}) and (\ref{eq:15})
        \State Obtain the integrity basis components $\mathbb{G}^j_\mathrm{2}$ and $\mathbb{G}^j_\mathrm{3}$ using Eq. (\ref{eq:20})
        \State Obtain the Voigt forms and construct the matrix $[\mathrm{vec}({\mathbb{G}^j_\mathrm{2}}) \quad \mathrm{vec}({\mathbb{G}^j_\mathrm{3}})] \in \mathbb{R}^{6 \times 2}$
        \State Obtain the Voigt form of $\mathbf{S}^j_{h,\mathrm{iso}}$: $\mathrm{vec}({\mathbf{S}^j_{h,\mathrm{iso}}}) \in \mathbb{R}^{6}$
        \State Solve the system of equations in Eq. (\ref{eq:23}) to obtain $[\Gamma_1^j, \Gamma_2^j] \in \mathbb{R}^2$
        \EndFor
        \State Construct $\mathcal{D}^*_{h,\mathrm{iso}} = \{ [\Bar{I}_1^j, \Bar{I}_2^j], [\Gamma_1^j, \Gamma_2^j] \}_{j=1}^{N_{h,\mathrm{iso}}}$ comprising $\{ [\Bar{I}_1^1, \Bar{I}_2^1], ..., [\Bar{I}_1^{N_{h,\mathrm{iso}}}, \Bar{I}_2^{N_{h,\mathrm{iso}}}]\} \in \mathbb{R}^{N_{h,\mathrm{iso}} \times 2}$ and $\{ [\Gamma_1^1, \Gamma_2^1], $ $..., [\Gamma_1^{N_{h,\mathrm{iso}}}, \Gamma_2^{N_{h,\mathrm{iso}}}]\} \in \mathbb{R}^{N_{h,\mathrm{iso}} \times 2}$
        \State Using $\mathcal{D}^*_{h,\mathrm{iso}}$, construct a standard GPR approximation $[\widetilde{\Gamma}_1, \widetilde{\Gamma}_2] = \widetilde{M}_{h,\mathrm{iso}} (\Bar{I}_1, \Bar{I}_2)$

        \Algphase{3 --- Isochoric viscous overstress part}
        \Require $\{ \mathbf{C}^1, ..., \mathbf{C}^{N_{v,\mathrm{iso}}}\}$, $\{ \dot{\mathbf{C}}^1, ..., \dot{\mathbf{C}}^{N_{v,\mathrm{iso}}}\}$ and $\{ \mathbf{S}_{v,\mathrm{iso}}^1, ..., \mathbf{S}_{v,\mathrm{iso}}^{N_{v,\mathrm{iso}}}\}$, where $\mathbf{C}^k, \dot{\mathbf{C}}^k, \mathbf{S}_{v,\mathrm{iso}}^k \in \mathbb{R}^{3 \times 3}$
        \For{{$k \leftarrow 1, N_{v,\mathrm{iso}}$}}
        \State Obtain the invariants $J^k$, $\Bar{I}^k_1$, $\Bar{I}^k_2$, $\Bar{J}^k_1$, $\Bar{J}^k_4$, and $\Bar{J}^k_5$ using Eqs. (\ref{eq:12}) and (\ref{eq:15})
        \State Obtain the integrity basis components $\mathbb{G}^k_\mathrm{2}$, $\mathbb{G}^k_\mathrm{3}$, $\mathbb{G}^k_\mathrm{4}$, $\mathbb{G}^k_\mathrm{5}$, $\mathbb{G}^k_\mathrm{6}$, $\mathbb{G}^k_\mathrm{7}$, and $\mathbb{G}^k_\mathrm{8}$ using Eq. (\ref{eq:20})
        \State Obtain the Voigt forms and construct the matrix $[\mathrm{vec}({\mathbb{G}^k_\mathrm{2}}) \quad \dots \quad \mathrm{vec}({\mathbb{G}^k_\mathrm{8}})] \in \mathbb{R}^{6 \times 7}$
        \State Obtain the Voigt form of $\mathbf{S}^k_{v,\mathrm{iso}}$: $\mathrm{vec}({\mathbf{S}^k_{v,\mathrm{iso}}}) \in \mathbb{R}^{6}$
        \State Solve the system of equations in Eq. (\ref{eq:23}) to obtain $[\Phi_1^k, \Phi_2^k, \Phi_3^k, \Phi_4^k, \Phi_5^k, \Phi_6^k, \Phi_7^k] \in \mathbb{R}^7$
        \EndFor
        \State Construct $\mathcal{D}^*_{v,\mathrm{iso}} = \{ [ \Bar{I}_1^k, \Bar{I}_2^k, \Bar{J}_1^k, \Bar{J}_4^k, \Bar{J}_6^k], [\Phi_1^k, \dots, \Phi_7^k] \}_{k=1}^{N_{v,\mathrm{iso}}}$ comprising $\{ [ \Bar{I}_1^1, \Bar{I}_2^1, \Bar{J}_1^1, \Bar{J}_4^1, \Bar{J}_6^1],...,[ \Bar{I}_1^{N_{v,\mathrm{iso}}},$$ \Bar{I}_2^{N_{v,\mathrm{iso}}}, $ $ \Bar{J}_1^{N_{v,\mathrm{iso}}}, \Bar{J}_4^{N_{v,\mathrm{iso}}}, \Bar{J}_6^{N_{v,\mathrm{iso}}}]\} \in \mathbb{R}^{N_{v,\mathrm{iso}} \times 5}$ and $\{ [\Phi_1^1, \dots, \Phi_7^1], ..., [\Phi_1^{N_{v,\mathrm{iso}}}, \dots, \Phi_7^{N_{v,\mathrm{iso}}}]\} \in \mathbb{R}^{N_{v,\mathrm{iso}} \times 7}$
        \State Choose a set of constraint points: $\mathcal{N}_c = \{ \mathbf{C}^1, ..., \mathbf{C}^{N_c}\}$ and $\{ \dot{\mathbf{C}}^1, ..., \dot{\mathbf{C}}^{N_c}\}$
        \State Using $\mathcal{D}^*_{v,\mathrm{iso}}$ and $\mathcal{N}_c$, construct a constrained GPR approximation $[\widetilde{\Phi}_1, \dots, \widetilde{\Phi}_7] = \widetilde{M}_{v,\mathrm{iso}} (\Bar{I}_1, \Bar{I}_2, \Bar{J}_1, \Bar{J}_4, \Bar{J}_6)$
    \end{algorithmic}
\end{algorithm}

In this work, the Gaussian Process Regression (GPR) supervised learning method (\cite{Williams_etal:1995}) is chosen for surrogate modeling purposes (see details of GPR in Section \ref{Section_4}), i.e., to learn the mappings in Eqs. (\ref{eq:26}--\ref{eq:28}) from a given input--output training dataset $\mathcal{D}$ and then predict integrity basis coefficients for a new set of input tensors $\mathbf{C}$ and $\dot{\mathbf{C}}$. Given the linear relationship between individual stress components and their corresponding integrity basis coefficients (Eq. (\ref{eq:16}--\ref{eq:18})), the trained surrogate models allow prediction of the three stress components as
\begin{equation} \label{eq:32}
    \widetilde{\mathbf{S}}_\mathrm{vol}^i = \widetilde{\zeta}_1^i \mathbb{G}_1^i
\end{equation}
\begin{equation} \label{eq:33}
    \widetilde{\mathbf{S}}_{h,\mathrm{iso}}^j = J^{-2/3} \left(\widetilde{\Gamma}_1^j \mathbb{G}_2^j + \widetilde{\Gamma}_2^j \mathbb{G}_3^j\right)
\end{equation}
\begin{equation} \label{eq:34}
    \widetilde{\mathbf{S}}_{v,\mathrm{iso}}^k = J^{-2/3} \left(\widetilde{\Phi}_1^k \mathbb{G}_2^k + \widetilde{\Phi}_2^k \mathbb{G}_3^k + \widetilde{\Phi}_3^k \mathbb{G}_4^k + \widetilde{\Phi}_4^k \mathbb{G}_5^k +\widetilde{\Phi}_5^k \mathbb{G}_6^k + \widetilde{\Phi}_6^k \mathbb{G}_7^k + \widetilde{\Phi}_7^k \mathbb{G}_8^k\right)
\end{equation}
In Eqs. (\ref{eq:32}--\ref{eq:34}), the accent symbols $\widetilde{\cdot}$ over stress and coefficients of the integrity basis denote that these are approximated values predicted via surrogate models. Note, for any given input strain and strain rate, the integrity basis components ($\mathbb{G}^{i/j/k}$) can be obtained using Eq. (\ref{eq:20})). \cite{FUHG2022105022} have recently proposed a data-driven method that allows discovering which of the integrity basis components play an active role in the stress response, i.e. discovery of both the type and orientation of the anisotropy. The entire process of the development of our physics-informed data-driven constitutive model from a stress--strain--strain rate dataset (i.e., $\mathcal{D}$) is summarized in Algorithm \ref{euclid}.

As the proposed surrogate models in Eqs. (\ref{eq:26}--\ref{eq:28}) are based on the generalized functional forms of the visco-hyperelastic stress equations (i.e., Eq. (\ref{eq:16}--\ref{eq:18})), our data-driven constitutive model and its predictions (Eqs. (\ref{eq:32}--\ref{eq:34})) automatically satisfy a number of physics-based constraints:
\begin{itemize}
    \item \textit{Principle of local action}: Because the integrity basis only captures spatially local deformation (via $\mathbf{C}$ and $\dot{\mathbf{C}}$).
    \item \textit{Balance of angular momentum}: Because the integrity basis is necessarily symmetric.
    \item \textit{Principle of determinism (or causality)}: Because the predicted stress is only a function of past and present events.
    \item \textit{Principle of material frame-indifference (objectivity)}: Because $\mathbf{S}$, $\mathbf{C}$, and $\dot{\mathbf{C}}$ are all objective tensors associated with the reference configuration and thus remain unaffected under change of observer. 
    \item \textit{Isotropic material symmetry}: Because the surrogate model stress equations (Eqs. (\ref{eq:32}--\ref{eq:34})) remain unaffected by any transformation from the proper orthogonal material symmetry group $SO(3)$ (i.e., under rigid body rotation). This is a direct consequence of the generalized model formulation in Eq. (\ref{eq:14}) in the form of isotropic invariants of tensors $\Bar{\mathbf{C}}$ and $\dot{\Bar{\mathbf{C}}}$.
    \item \textit{Limited memory constraint (a subset of the principle of fading memory)}: Because the viscous overstress component (Eq. (\ref{eq:34})) depends only on the instantaneous deformation rate (i.e., very recent history).
\end{itemize}

Two additional physics-based constraints will be imposed via the employment of GPR as the chosen surrogate modeling technique: 
\begin{inparaitem}
\item the stress-free reference state, and
\item the second law of thermodynamics.
\end{inparaitem}
These additional physical constraints along with the GPR technique will now be described.

\section{Gaussian Process Regression and the Enforcement of Additional Physical Constraints } \label{Section_4}
Gaussian process regression has recently seen increased interest as a tool for building surrogate models for describing material behaviors (\cite{rocha2021fly,fuhg_etal:2022,chen2023hybrid}). This is in part because GPR is derived from a convenient statistical background, offers strict convergence guarantees (unlike, for example, artificial neural networks), and has empirically shown excellent performance for out-of-sample model predictions (\cite{park2014parametric,datta2016nearest}). For an excellent overview of GPR, the reader is referred to \cite{rasmussen2003gaussian}. In this section, the basic ideas behind GPR and its specific employment for regression in our surrogate models are discussed. Two distinct but related GPR formulations are introduced: the standard GPR and the constrained GPR. The latter allows enforcement of thermodynamic consistency in the viscous overstress component of our data-driven model.

\subsection{Standard Gaussian Process Regression (GPR)}
Consider a general dataset $\mathcal{D}_g$ of $N$ data points,
\begin{equation}\label{eq:DatasetTraining}
    \mathcal{D}_{g} = \lbrace \mathbf{x}_{i}, \mathbf{y}_{i} \rbrace_{i=1}^{N}
\end{equation}
where $\mathbf{x} \in \mathbb{R}^{n_{i}}$ and $\mathbf{y} \in \mathbb{R}^{n_{o}}$ represent input and output data points of dimensions $n_{i}$ and $n_{o}$, respectively. The sets of all input and output data points can be reformulated as matrices $\mathbf{X} \in \mathbb{R}^{N \times n_{i}}$ and $\mathbf{Y} \in \mathbb{R}^{N \times n_{o}}$, respectively. The objective of a GPR model that is trained using the dataset $\mathcal{D}_g$ is to predict an output $\widetilde{\mathbf{y}}$ for a new input data point $\mathbf{x}_{\star}$ that is not contained in the training dataset.

In GPR, the notion of the similarity between input data points is critical. The assumption is that two input data points that are closer together are more likely to have closer target values than the two input data points that are farther away from each other. Typically, the similarity between two input data points $\mathbf{x}$ and $\mathbf{x}'$ is modeled by a user-defined covariance function. The choice of this function is an important part of GPR. Based on the work of
\cite{laurent2019overview}, which compared different functional forms of the covariance function for computer experiments when no prior knowledge is available, we restrict ourselves to the Mat\'{e}rn 3/2 kernel \citep{matern1960spatial}, given as
\begin{equation}\label{eq:autocorr}
\begin{aligned}
k(\mathbf{ x}, \mathbf{ x}')  =   \sigma_{f}^{2} \left( 1 + \dfrac{\sqrt{3} \norm{\mathbf{ x} - \mathbf{ x}'}_2}{l} \right) \exp \left(-\dfrac{\sqrt{3} \norm{\mathbf{ x} - \mathbf{ x}'}_2 }{l}  \right) + \alpha \delta_{x, x'} 
\end{aligned}
\end{equation}
where $\lVert \left( \cdot \right) \rVert_2$ represents the L2-norm, $\sigma_{f}$ is a scaling factor, and $l$ is the characteristic length-scale of the covariance function. $\sigma_{f}$ and $l$ can be combined into a vector $\boldsymbol{\theta}=\left[ \sigma_{f}, l \right]^{T}$ that collects the trainable parameters. The term $\alpha$ denotes a small positive value that helps to overcome numerical instabilities \cite{peng2014choice}. $\delta_{x, x'}$ is the Kronecker delta function. Using the functional form of the covariance function in Eq. (\ref{eq:autocorr}), a covariance matrix $\mathbf{K}(\mathbf{X},\mathbf{X}')$ can be constructed to define similarity between sets of points $\mathbf{X}$ and $\mathbf{X}'$.

The optimal set of parameters ${\bm{\theta}}$ that best describes the training dataset (Eq. (\ref{eq:DatasetTraining})) is obtained using a maximum log-likelihood approach \citep{fuhg2021state},
\begin{equation} \label{eq:optim_ML}
\begin{aligned}
\widetilde{{\boldsymbol{\theta}}} =\argmax_{ {\boldsymbol{\theta}}^{\star}}  \text{log} \, p(\mathbf{Y} | \mathbf{X}, \boldsymbol{\theta}) =  \argmax_{ {\bm{\theta}}^{\star}} &\left[-\frac{1}{2} \mathbf{Y}^{T} \mathbf{K}(\mathbf{X},\mathbf{X}) ^{-1} \mathbf{Y} - \frac{1}{2} \text{log}( \det ( \mathbf{K}(\mathbf{X},\mathbf{X}) )) - \frac{N}{2} \log (2 \pi) \right]
\end{aligned}
\end{equation}
After finding the best parameters, the GPR regression model is fully defined. Given a new input data point $\mathbf{ x}_{\star}$, the predicted output value $\widetilde{\mathbf{y}}$ of the Gaussian process regressor reads
\begin{equation}\label{eq:mean}
\begin{aligned}
\widetilde{\mathbf{y}}({\mathbf{ x}}_{\star}) &=   \mathbf{K}(\mathbf{X},\mathbf{x}_{\star})^{\mathrm{T}} \, \mathbf{K}(\mathbf{X},\mathbf{X}) ^{-1} \, \mathbf{Y}
\end{aligned}
\end{equation}
As $\alpha \rightarrow 0$ in Eq. (\ref{eq:autocorr}), the GPR predictor becomes an exact interpolator (see \cite{marrel2008efficient}), i.e. $\widetilde{\mathbf{y}}(\mathbf{ x}_{i})= \mathbf{ y}_{i}$ for all points $i=1, \ldots, N$ of the training dataset of Eq. (\ref{eq:DatasetTraining}). This is termed as the exact inference property of GPR. Since the prediction is probabilistic, the variance $\widetilde{\boldsymbol{\sigma}}^{2}$ of the regressor fit can also be obtained,
\begin{equation}\label{eq:var}
\begin{aligned}
\widetilde{\boldsymbol{\sigma}}^{2} &= \mathbf{K}(\mathbf{x}_{\star},\mathbf{x}_{\star}) -  \mathbf{K}(\mathbf{X},\mathbf{x}_{\star})^{T} \, \mathbf{K}(\mathbf{X},\mathbf{X}) ^{-1}\,  \mathbf{K}(\mathbf{X},\mathbf{x}_{\star})
\end{aligned}
\end{equation}

In this study, standard GPR (as defined above) is employed to fit the surrogate models $\widetilde{\mathcal{M}}_{\mathrm{vol}}$ (Eq. (\ref{eq:26}) and $\widetilde{\mathcal{M}}_{h,\mathrm{iso}}$ (Eq. (\ref{eq:27}) using datasets of the form of $\mathcal{D}_{\mathrm{vol}}^*$ and $\mathcal{D}_{h,\mathrm{iso}}^*$ (Eq. (\ref{eq:25}), respectively; remember, the datasets $\mathcal{D}_{\mathrm{vol}}^*$ and $\mathcal{D}_{h,\mathrm{iso}}^*$ are first derived from the initial training datasets $\mathcal{D}_{\mathrm{vol}}$ and $\mathcal{D}_{h,\mathrm{iso}}$ (Eq. (\ref{eq:21}), respectively (see Algorithm \ref{euclid}). Further, the exact inference property of GPR is harnessed to enforce the physical constraint of a stress-free reference (i.e., undeformed) state, also called the normalization condition (\cite{Holzapfel:2000}). For the two stress components $\mathbf{S}_{v,\mathrm{iso}}$ and $\mathbf{S}_{h,\mathrm{iso}}$ that are modeled using standard GPR, the normalization condition reads
\begin{equation} \label{eq:40}
    \mathbf{S}_\mathrm{vol}(\mathbf{C} = \mathbf{I}) = \mathbf{0}, \quad \mathbf{S}_{h,\mathrm{iso}}(\mathbf{C} = \mathbf{I}) = \mathbf{0}
\end{equation}

To enforce the above constraint, the stress-free reference states---$\{\mathbf{C}=\mathbf{I}, \mathbf{S}_{\mathrm{vol}}=\mathbf{0}\}$ and $\{\mathbf{C}=\mathbf{I}, \mathbf{S}_{h,\mathrm{iso}}=\mathbf{0}\}$---are included in the training dataset. This guarantees that the normalization condition is achieved, provided the stabilization constant $\alpha$ of Eq. (\ref{eq:autocorr}) is kept small enough. Note, in terms of invariants, $\mathbf{C}=\mathbf{I}$ translates to $\Bar{I}_1 = \Bar{I}_2 = 3$.

Another physics-based constraint of interest in this work is the second law of thermodynamics. In light of the Clausius--Planck inequality, this constraint requires that the internal dissipation $\Xi_{int}$ in a visco-hyperelastic soft material as described in Eq. (\ref{eq:7}) is non-negative. Among the three stress components $\mathbf{S}_\mathrm{vol}$, $\mathbf{S}_{h,\mathrm{iso}}$ and $\mathbf{S}_{v,\mathrm{iso}}$, the first two are rate-independent and therefore do not contribute to dissipation. Thus, enforcing the second law of thermodynamics constraint on the surrogate models $\widetilde{\mathcal{M}}_{\mathrm{vol}}$ and $\widetilde{\mathcal{M}}_{h,\mathrm{iso}}$ that are based on standard GPR becomes trivial. In other words, the specific construction of the generalized constitutive equation in this work automatically ensures the second law constraint on the rate-independent portion of our data-driven constitutive model. 

\subsection{Constrained Gaussian Process Regression (C-GPR)}
Unlike the rate-independent stress components $\mathbf{S}_\mathrm{vol}$ and $\mathbf{S}_{h,\mathrm{iso}}$, the rate-dependent isochoric viscous overstress $\mathbf{S}_{v,\mathrm{iso}}$ results in viscous dissipation. In this regard, the Clausius--Planck inequality in Eq. (\ref{eq:7}) can be rewritten as (\cite{Pioletti_Rakotomanana:2000})
\begin{equation} \label{eq:41}
    \Xi_{int} = 2 \left( \mathbf{F} \mathbf{S}_{v,\mathrm{iso}} \right) : \dot{\mathbf{F}} = \mathbf{S}_{v,\mathrm{iso}} : \dot{\mathbf{C}} \geq 0
\end{equation}

For arbitrary input strain and strain rates, the isochoric viscous overstress prediction of the surrogate model $\widetilde{\mathcal{M}}_{v,\mathrm{iso}}$ (Eq. (\ref{eq:34})) must satisfy the above inequality. This type of inequality constraint on input-output mappings can be enforced by employing the constrained GPR (C-GPR) formulation recently proposed by \cite{pensoneault2020nonnegativity}. In standard GPR, the model output remains unconstrained.

The basic idea of C-GPR is to restrict the solution space in the hyperparameter optimization problem (i.e., Eq. (\ref{eq:optim_ML})) such that the predicted output at a set of user-defined "constraint points" follows the desired inequality constraint. Accordingly, in the present study, the following optimization problem is devised for training the GPR regressor of the surrogate model $\widetilde{\mathcal{M}}_{v,\mathrm{iso}}$ (cf. Eq. (\ref{eq:optim_ML})):
\begin{equation} \label{eq:42}
    \begin{aligned}
        &\widetilde{{\boldsymbol{\theta}}} =\argmax_{ {\boldsymbol{\theta}}^{\star}}  \text{log}\, p(\mathbf{Y} | \mathbf{X}, \boldsymbol{\theta}) \\
        & \text{s.t.} \quad \left(\widetilde{\mathbf{S}}_{v,\mathrm{iso}}(\mathbf{C}^{k},\dot{\mathbf{C}}^{k}) \right): \dot{\mathbf{C}}^{k} \geq 0, \quad \forall  \,  \, k=1, \ldots, N_{c}
    \end{aligned}
\end{equation}
Here, $\widetilde{\mathbf{S}}_{v,\mathrm{iso}}(\mathbf{C}^{k},\dot{\mathbf{C}}^{k})$ denotes the isochoric viscous overstress prediction of the surrogate at the $k^\text{th}$ constraint point (i.e., for the $\{\mathbf{C}^k, \dot{\mathbf{C}}^k\}$ input). $N_c$ is the total number of constraint points. The constrained optimization of Eq. (\ref{eq:42}) imposes the Clausius--Planck inequality (Eq. (\ref{eq:41})) on the rate-dependent part of our physics-informed data-driven constitutive model. Note that the inequality condition in Eq. (\ref{eq:42}) is not a functional constraint; rather, it is applied at a finite set of input data points. Therefore, C-GPR only applies a "weak" constraint on the model output.

Next, the normalization condition constraint for the isochoric viscous overstress is written as
\begin{equation} \label{eq:43}
    \mathbf{S}_{v,\mathrm{iso}}(\mathbf{C} = \mathbf{I}, \dot{\mathbf{C}}) = \mathbf{0}
\end{equation}
The above condition requires that the isochoric viscous overstress vanishes in the undeformed state (i.e., $\mathbf{C} = \mathbf{I}$) regardless of the applied rate of deformation $\dot{\mathbf{C}}$ in that state. Owing to the particular choice of the strain rate invariants $\Bar{J}_1$, $\Bar{J}_4$, and $\Bar{J}_6$ in the mapping of the $\widetilde{\mathcal{M}}_{v,\mathrm{iso}}$ surrogate (Eq. (\ref{eq:31})), this condition can be simply enforced by including the stress-free reference state---$\{\mathbf{C}=\mathbf{I}, \mathbf{S}_{h,\mathrm{iso}}=\mathbf{0}\}$---in the training data. This is possible because in the undeformed state, these three invariants hold a fixed value of zero regardless of the loading rate, i.e., 
\begin{equation} \label{eq:44}
    \Bar{J}_1(\mathbf{C} = \mathbf{I}, \dot{\mathbf{C}}) = 0, \quad \Bar{J}_2(\mathbf{C} = \mathbf{I}, \dot{\mathbf{C}}) = 0, \quad \Bar{J}_2(\mathbf{C} = \mathbf{I}, \dot{\mathbf{C}}) = 0
\end{equation}
For a derivation of the above result, see \ref{Appendix_B}. From Eq. (\ref{eq:44}), $\mathbf{C}=\mathbf{I}$ translates to a single set of input values in the mapping of $\widetilde{\mathcal{M}}_{v,\mathrm{iso}}$: $[ \Bar{I}_1, \Bar{I}_2, \Bar{J}_1, \Bar{J}_4, \Bar{J}_6] = [3,3,0,0,0]$, at which the exact inference property of GPR will always lead to a zero viscous overstress value.

Unlike the invariants $\Bar{J}_1$, $\Bar{J}_4$, and $\Bar{J}_6$, the invariants $\Bar{J}_2$, $\Bar{J}_3$, $\Bar{J}_5$, and $\Bar{J}_7$ in the undeformed state of $\mathbf{C}=\mathbf{I}$ are functions of the tensor $\dot{\mathbf{C}}$ (see \ref{Appendix_B}). Therefore, if these invariants were included in the mapping of $\widetilde{\mathcal{M}}_{v,\mathrm{iso}}$ (Eq. (\ref{eq:31})), there would be infinitely many possible combinations of invariant values (i.e., input data points) that would correspond to the stress-free reference state, making it impossible to utilize the GPR exact inference property for ensuring the normalization condition.



\section{Model Performance} \label{Section_5}
In this section, the performance of our physics-informed data-driven constitutive model is demonstrated in terms of (i) the accuracy of fitting training datasets from various deformation modes (e.g., hydrostatic compression, and unconfined uniaxial tension and compression), (ii) the accuracy and physical plausibility of out-of-sample predictions outside the training regime, and (iii) the training dataset size requirement. As the three surrogate models that make up our constitutive model---$\widetilde{\mathcal{M}}_{\mathrm{vol}}$, $\widetilde{\mathcal{M}}_{h,\mathrm{iso}}$, and $\widetilde{\mathcal{M}}_{v,\mathrm{iso}}$---are independent of each other in terms of the specific dataset type required for training them (i.e., $\mathcal{D}_\mathrm{vol}$, $\mathcal{D}_\mathrm{vol}$, and $\mathcal{D}_\mathrm{vol}$ in Eq. (\ref{eq:21}), respectively), the following subsections will evaluate these surrogate models individually using multiple deformation modes for training and testing purposes, for a deeper insight. In practice, given multiple datasets spanning volumetric, isochoric quasi-static, and isochoric dynamic deformation modes, the individually trained $\widetilde{\mathcal{M}}_{\mathrm{vol}}$, $\widetilde{\mathcal{M}}_{h,\mathrm{iso}}$, and $\widetilde{\mathcal{M}}_{v,\mathrm{iso}}$ surrogate models can be conveniently combined to form one constitutive model.

To assess the fitting accuracy of the surrogate models, the following percent relative error metric is employed:
\begin{equation} \label{eq:46}
    \mathrm{err} (\widetilde{\mathbf{S}}, \mathbf{S}) = 100 \times \frac{\left \lVert \mathrm{vec}(\widetilde{\mathbf{S}}) - \mathrm{vec}(\mathbf{S})\right \rVert_F}{\left \lVert \mathrm{vec}(\mathbf{S})\right \rVert_F}
\end{equation}
where $\lVert \left( \cdot \right) \rVert_F$ represents the Frobenius norm, $\widetilde{\mathbf{S}}$ is the stress tensor predicted by a surrogate model, and $\mathbf{S}$ is the true value of the stress tensor (i.e., the ground truth). The scalar error, $\mathrm{err}$, is obtained with respect to individual data points. For convenience, a percent mean error value will also be reported,
\begin{equation} \label{eq:47}
    \overline{\mathrm{err}} = \frac{\sum_{i=1}^N \mathrm{err}_i}{N}
\end{equation}
where $\mathrm{err}_i$ is the error value at the $i$th data point, and $N$ is the total size of the dataset (training or testing).

To assess the accuracy and physical plausibility of model predictions outside the training regime, two types of out-of-sample "testing regions" will be considered: (i) Testing region that corresponds to the same deformation mode that was considered during training---for example, training a model using uniaxial tension stress--strain data in the [0, 0.25] strain range, and then predicting tensile stresses for the wider [0, 0.5] strain range; (ii) Testing region that corresponds to a different deformation mode from those considered during model training---for example, training a model using uniaxial tension stress--strain data and then using the trained model to predict material responses under uniaxial compression and simple shear deformation modes. Further, the effect of training dataset size on both fitting errors and out-of-sample predictions will be studied.

Finally, for each surrogate model case, the performance of our model will be compared to (i) a conventional visco-hyperelastic constitutive model, and to (ii) a corresponding surrogate model that employs a classical ML black-box mapping between the Voigt forms of tensors $\mathbf{C}$ and $\dot{\mathbf{C}}$ (as strain and strain rate input) and $\mathbf{S}$ (as the second Piola-Kirchoff stress output) (\cite{Ghaboussi_etal:1998,lefik_schrefler:2003}), i.e.,
\begin{equation} \label{eq:48}
    \textrm{Classical mapping:} {}
    \begin{bmatrix}
            \vspace{4 pt}
            \mathrm{vec} (\mathbf{C})\\
            \mathrm{vec} (\dot{\mathbf{C}})            
         \end{bmatrix} \rightarrow \begin{bmatrix}
            \mathrm{vec} (\mathbf{S})\\
         \end{bmatrix}
\end{equation}
Standard GPR will be utilized to learn the classical mappings. Of course, tensor $\dot{\mathbf{C}}$ will not be considered in the case of quasi-static loading conditions.

All the calculations presented in this work are performed in Python (\cite{python3}). For machine learning tasks, the Scikit-learn python package (\cite{pedregosa_etal:2011}) is utilized. Note that constrained optimization capability was added by the authors in the GPR module of this package for the development of C-GPR-based surrogate models.\footnote{After acceptance of the paper, the codes of this manuscript will be released under \url{https://github.com/kshitizupadhyay/viscohyperelastic-modeling}.}

\subsection{The \texorpdfstring{$\widetilde{\mathcal{M}}_{\mathrm{vol}}$} {} Surrogate Model Under Hydrostatic (Confined Uniaxial) Loading} \label{Section_5.1}
Consider the following deformation gradient tensor,
\begin{equation} \label{eq:49}
    \mathbf{F}_{app} = \mathbf{I} + J \mathbf{e}_1 \otimes \mathbf{E}_1 \quad \textrm{with} \hspace{2 pt} J \in [-0.75,0]
\end{equation}
Deformations of the form of Eq. (\ref{eq:49}) are applied in practice via confined compression experiments to study the hydrostatic bulk material response (\cite{Sasson_etal:2012,Kim_etal:2019,Nie_etal:2013}). These experiments provide hydrostatic stress / pressure versus volumetric strain (i.e., $J$) response of the material, which in turn can be used to calibrate a particular form of the volumetric energy density function $U(J)$ (see Eq. (\ref{eq:14})). In this study, the commonly used Simo--Miehe volumetric energy density (\cite{Simo_Miehe:1992}) is employed to generate the hydrostatic stresses for training the $\widetilde{\mathcal{M}}_{\mathrm{vol}}$ surrogate model,
\begin{equation} \label{eq:50}
    U^{\mathrm{SM}} (J) = \frac{\kappa}{2} \left( \frac{J^2 - 1}{2} - \ln{J} \right)
\end{equation}
where $\kappa$ is the bulk modulus. We choose $\kappa$ = 10. The corresponding volumetric stress component $\mathbf{S}_\mathrm{vol}^\mathrm{SM}$ (see Eq. (\ref{eq:13})) is given by
\begin{equation} \label{eq:51}
    \mathbf{S}_\mathrm{vol}^\mathrm{SM} = 2 \frac{\partial U^{\mathrm{SM}}(J)}{\partial \mathbf{C}} = \frac{\kappa}{2} \left( J^2 - 1 \right) \mathbf{C}^{-1}
\end{equation}
where $\mathbf{C}$ = $\mathbf{C}_{app}$ = $\mathbf{F}_{app}^{\mathrm{T}} \mathbf{F}_{app}$.

$\mathbf{C}_{app}$ and $\mathbf{S}_\mathrm{vol}^\mathrm{SM}$ constitute the training dataset $\mathcal{D}_\mathrm{vol}$ of Eq. (\ref{eq:21}) (note, $\mathcal{D} = \mathcal{D}_\mathrm{vol}$ in this case) that is utilized to train the $\widetilde{\mathcal{M}}_{\mathrm{vol}}$ surrogate model (Eq. (\ref{eq:26})) by following the procedure detailed in Section \ref{Section_3}. Standard GPR is used for mapping. As the stress-free undeformed state (i.e., $\mathbf{F}_{app} = \mathbf{I}$) is included in the training data, the normalization condition is simply enforced by assigning a near-zero Gaussian noise of $\alpha = 10^{-4}$ (Eq. (\ref{eq:autocorr})) at all data points.

\begin{figure}[t]
    \centering
    \includegraphics[width=13 cm]{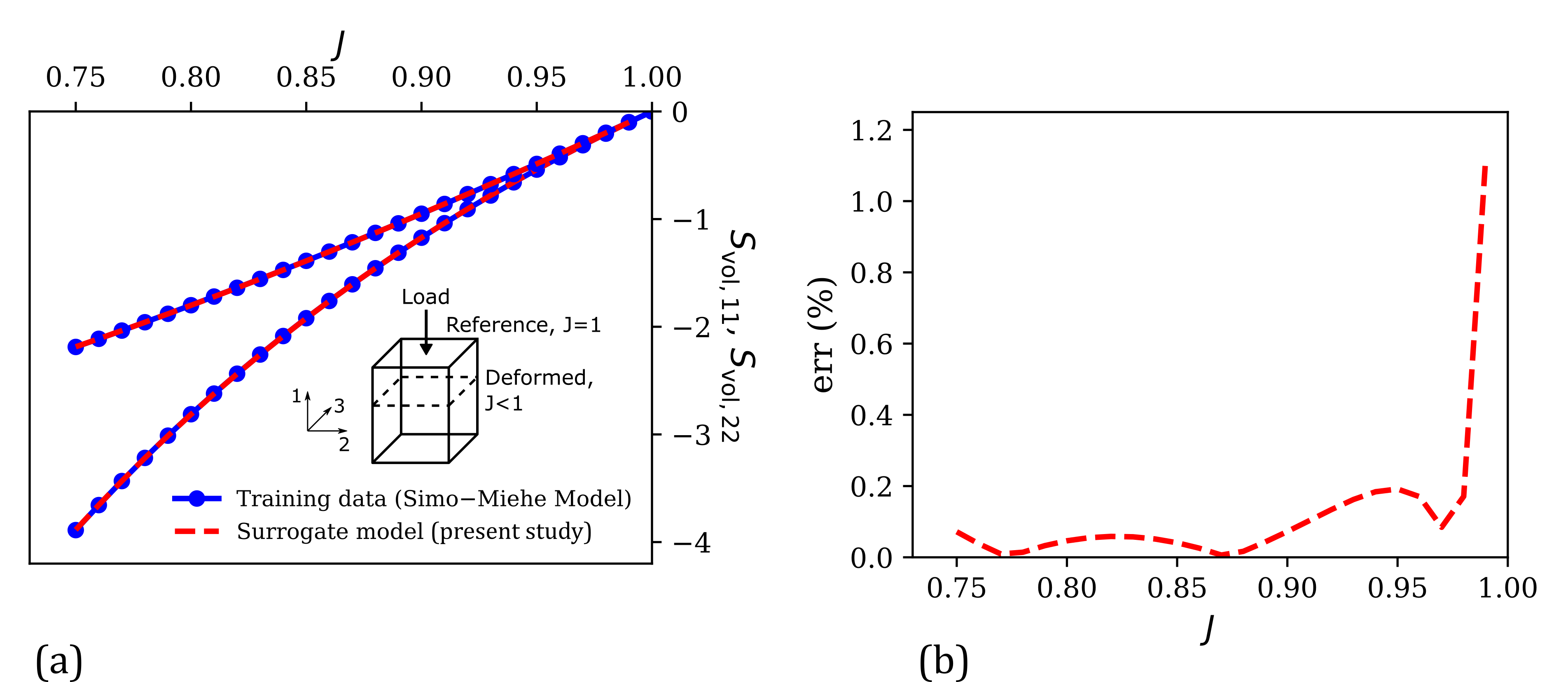}
    \caption{(a) Comparison of the numerically generated volumetric stress components ($\mathbf{S}_{\mathrm{vol,11}}$ and $\mathbf{S}_{\mathrm{vol,22}}$) versus volumetric strain ($J$) training data under confined compression with the corresponding surrogate model predictions. Inset shows a schematic illustration of the confined compression deformation mode, highlighting the reference and deformed states and the 11-loading direction. (b) Evolution of the percent relative error ($\mathrm{err}$) of surrogate model predictions versus $J$ in the training regime.}
    \label{fig:GPR_hydrostatic_1}
\end{figure}

The confined compression deformation mode in Eq. (\ref{eq:48}) generates three non-zero volumetric stress components: $\mathbf{S}_{\mathrm{vol,11}}$, $\mathbf{S}_{\mathrm{vol,22}}$, and $\mathbf{S}_{\mathrm{vol,33}}$. Given the 11-loading direction, $\mathbf{S}_{\mathrm{vol,22}} = \mathbf{S}_{\mathrm{vol,33}}$, resulting in only two independent stress components. Figure \ref{fig:GPR_hydrostatic_1}(a) shows the evolution of $\mathbf{S}_{\mathrm{vol,11}}$ and $\mathbf{S}_{\mathrm{vol,22}}$ versus $J$ from the training data and compares it with the corresponding $\mathbf{S}_{\mathrm{vol,11}}$--$J$ and $\mathbf{S}_{\mathrm{vol,22}}$--$J$ responses predicted by the trained surrogate model. 26 data points were considered for training in this case (i.e., a strain increment of 0.01). The corresponding percent relative error (Eq. (\ref{eq:46})) versus $J$ plot is shown in Fig. \ref{fig:GPR_hydrostatic_1}(b). From these plots, a very good agreement between training data and surrogate model predictions in the training regime is evident, which is expected from an ML-based model. Overall, the maximum and mean relative errors in the training regime are 1.12\% and 0.12\%, respectively, suggesting a very good fitting accuracy of our surrogate model.

Next, the performance of our surrogate model for a wider range of $J$ (i.e., the testing regime) is analyzed and compared with the corresponding predictions of (i) a conventional volumetric strain energy density function of the neo-Hookean model (\cite{DeRooij_Kuhl:2016}), and (ii) the classical ML-based mapping approach (Eq. (\ref{eq:48})). The neo-Hookean volumetric strain energy density and its corresponding volumetric stress tensor are given as
\begin{subequations} \label{eq:NH_volumetric_model}
\begin{align}
        U^{\mathrm{NH}} (J) = \frac{\kappa_\mathrm{NH}}{2} \left(J - 1\right)^2
\end{align}
\begin{align}
        \mathbf{S}_\mathrm{vol}^\mathrm{NH} = 2 \frac{\partial U^{\mathrm{NH}}(J)}{\partial \mathbf{C}} = \kappa_\mathrm{NH} J \left( J - 1 \right) \mathbf{C}^{-1}
\end{align}
\end{subequations}
The calibration of the volumetric neo-Hookean model against the training data yielded a $\kappa_\mathrm{NH}$ value of 11.56, which is close to the ground truth of $\kappa = $10 considered in the Simo--Miehe model for generating the training data. 

\begin{figure}[t]
    \centering
    \includegraphics[width=\textwidth]{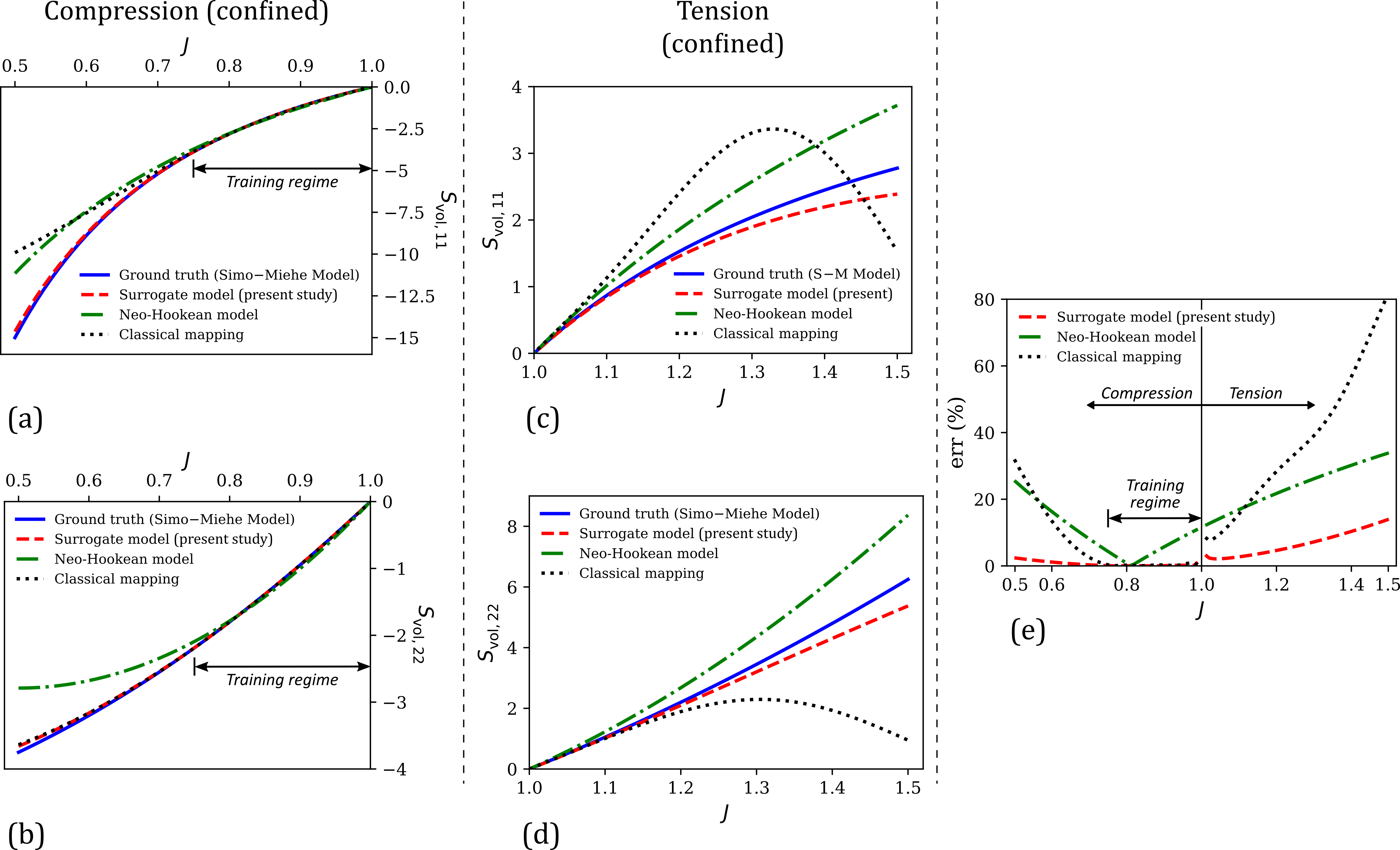}
    \caption{Comparison of the numerically generated volumetric stress--volumetric strain data from the Simo--Miehe model (i.e., ground truth) in the overall testing regime ($J \in [0.5, 1.5]$) with the corresponding predictions of our surrogate model, the volumetric neo-Hookean model, and the classical ML-based mapping model: (a) $\mathbf{S}_{\mathrm{vol,11}}$ versus $J$ under confined compression, (b) $\mathbf{S}_{\mathrm{vol,22}}$ versus $J$ under confined compression, (c) $\mathbf{S}_{\mathrm{vol,11}}$ versus $J$ under confined tension, and (d) $\mathbf{S}_{\mathrm{vol,22}}$ versus $J$ under confined tension. (d) Comparison of the percent relative error ($\mathrm{err}$) versus $J$ responses for the present surrogate model, the volumetric neo-Hookean model, and the classical mapping model.}
    \label{fig:GPR_hydrostatic_2}
\end{figure}

Figures \ref{fig:GPR_hydrostatic_2}(a-b) show the $\mathbf{S}_{\mathrm{vol,11}}$--$J$ and $\mathbf{S}_{\mathrm{vol,22}}$--$J$ responses as predicted by the surrogate model proposed in this work, the volumetric neo-Hookean model, and the classical mapping of Eq. (\ref{eq:48}), in a strain range ($J \in [0.5, 1]$) that outstretches the training regime in the compression deformation mode (i.e., the deformation mode considered in training). The ground truth (i.e., Simo--Miehe Model) in this extended strain regime is also shown in the same figure for comparison, with the corresponding percent relative error $\mathrm{err}$ (with respect to the ground truth) versus $J$ plots for the three models shown in the left half of Fig. \ref{fig:GPR_hydrostatic_2}(e). From these figures, it is clear that within the training regime (i.e., $J \in [0.75, 1]$), all three models exhibit reasonable accuracy in predicting the stress--strain data. Among the three models, the two ML-based data-driven models -- the surrogate model of this study and the classical mapping model -- resulted in much smaller percent mean relative errors ($\overline{\mathrm{err}}$) of 0.12\% and 0.27\% compared to the volumetric neo-Hookan model, which resulted in a $\overline{\mathrm{err}}$ of 4.90\%. This was expected because unlike ML-based mappings, conventional constitutive models have limited fitting accuracy due to their fixed mathematical forms. Outside the training regime when $J \in [0.5,0.75]$, our surrogate model clearly outperforms the other two models by predicting stress components that are in excellent agreement with the ground truth. In terms of the percent mean relative error, our surrogate model yields a $\overline{\mathrm{err}}$ of 1.07\% in this regime, which is an order-of-magnitude smaller than the corresponding errors of the volumetric neo-Hookean model (i.e., 14.86\%) and the classical mapping model (i.e., 12.72\%).

Figures \ref{fig:GPR_hydrostatic_2}(c-d) compare the predictions of the three models with the ground truth (Simo--Miehe model) in the tension (confined) deformation mode ($J \in [1, 1.5]$) that was not considered at all in the training data. Here, the classical mapping model results in physically implausible predictions in that the stresses at large volumetric strains start to decrease, violating thermodynamic stability for incremental deformations. In fact, for $J > 1.6$ (not shown in the plots), this model predicts compressive stresses for a confined tension loading, which is not possible. This type of erroneous model behavior is attributed to the purely black-box mapping in these conventional ML-based constitutive models, which do not respect any physical or mechanistic constraints on the response of the continuum. From Figs. \ref{fig:GPR_hydrostatic_2}(c-d), our physics-informed data-driven surrogate model does not suffer from this issue and makes physically-plausible and trustworthy predictions even in the deformation mode that was not considered in training. Not surprisingly, the predictions of the volumetric neo-Hookean model are also physically reasonable (as $\kappa_{\mathrm{NH}} > 0$). Further, our surrogate model outperforms the other two models in terms of agreement of the predicted stress components with the ground truth. The percent mean relative error of our model in the tensile regime is 6.66\%, which is significantly lower compared to the errors of the volumetric neo-Hookean model (i.e., 23.59 \%) and the classical mapping model (i.e., 37.23\%). 

\begin{figure}[t]
    \centering
    \includegraphics[width=13 cm]{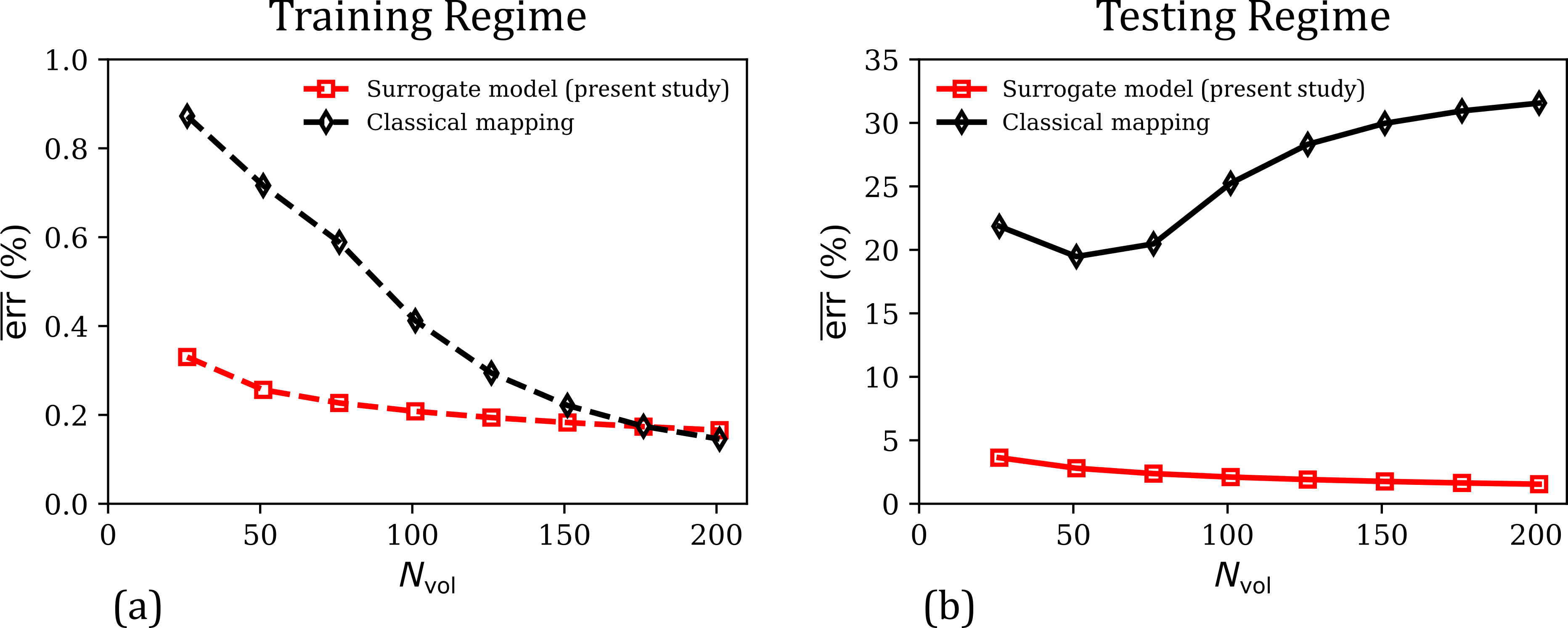}
    \caption{(a) Comparison of the evolution of mean percent relative error ($\overline{\mathrm{err}}$) in the predictions of the surrogate model and the classical mapping model in the training regime ($J \in [0.75,1]$), as a function of the training dataset size ($N_\mathrm{vol}$). (b) The corresponding $\overline{\mathrm{err}}$ versus $N_\mathrm{vol}$ responses of the present surrogate model and the classical mapping model, for their predicted responses in the overall testing regime ($J \in [0.5,1.5]$).}
    \label{fig:GPR_hydrostatic_3}
\end{figure}

Lastly, the effect of training dataset size on the performance of the two data-driven models is analyzed in Figs. \ref{fig:GPR_hydrostatic_3}(a-b), which plot the mean percent relative errors $\overline{\mathrm{err}}$ in the training regime ($J \in [0.75, 1]$) and the overall testing regime ($J \in [0.5, 1.5]$) as a function of the training dataset size $N_\mathrm{vol}$ (i.e., the number of data points in $\mathcal{D}_\mathrm{vol}$). In the training regime (Fig. (\ref{fig:GPR_hydrostatic_3}(a))), the errors of both the models decrease asymptotically with $N_\mathrm{vol}$. This is a typical behavior of ML-based models, which yield improved predictions in the training regime (i.e., interpolation) as larger volumes of data are used in model training. Overall, both the models show excellent fitting accuracy ($\overline{\mathrm{err}}$) even with a small training dataset of 26 stress--strain values. In the testing regime (Fig. (\ref{fig:GPR_hydrostatic_3}(b))), which includes both confined compression and tension deformation modes, the classical mapping model no longer shows improvement in the performance with an increase in training dataset size. In fact, for $N_\mathrm{vol}> 50$, $\overline{\mathrm{err}}$ versus $N_\mathrm{vol}$ increases monotonically, which coincides with highly physically implausible predictions in the confined tension regime (see an example of this in the supplementary material, Section SM1). Our surrogate model does not suffer from this limitation due to its physics-informed nature, and its predictions continue to improve with an increase in training dataset size in both the training and testing regimes. Further, as the inclusion of physics-based constraints confines the feature space of possible GPR hyperparameters, our surrogate model predictions are in very good agreement with the ground truth even at $N_\mathrm{vol} = 26$. Clearly, our physics-informed data-driven constitutive model combines the high fitting accuracy of ML-based models with the physical nature and small-data compatibility of conventional constitutive models.

\subsection{The \texorpdfstring{$\widetilde{\mathcal{M}}_{h,\mathrm{iso}}$} {} Surrogate Model Under Quasi-static Uniaxial and Shear Loading} \label{Section_5.2}
Now, consider the following isochoric (volume-preserving) uniaxial deformation mode,
\begin{equation} \label{eq:53}
    \mathbf{F}_{app} = \lambda \mathbf{e}_1 \otimes \mathbf{E}_1 + \frac{1}{\sqrt{\lambda}} \left(\mathbf{e}_2 \otimes \mathbf{E}_2 +  \mathbf{e}_3 \otimes \mathbf{E}_3\right) \quad \textrm{with} \hspace{2 pt} \lambda \in [1,1.25]
\end{equation}
where $\lambda$ is called the uniaxial stretch. Under the quasi-static assumption, the rate of stretching $\dot{\lambda}$ is approximately zero. Deformations like Eq. (\ref{eq:53}) are routinely applied in the mechanical characterization of soft materials during unconfined uniaxial tensile tests (\cite{Miller_Chinzei:2002,Gao_etal:2010,Brown:2006}). The uniaxial stress versus stretch responses obtained from these tests are used to calibrate hyperelastic constitutive models. In this study, we use a polynomial-type hyperelastic model known as the Mooney--Rivlin model (\cite{Mooney:1940,Rivlin:1948}) to generate training data for developing our $\widetilde{\mathcal{M}}_{h,\mathrm{iso}}$ surrogate model. The strain energy density $\Bar{W}_h$ and the corresponding isochoric hyperelastic stress $\mathbf{S}_{h,\mathrm{iso}}$ of the Mooney--Rivlin model are given by
\begin{subequations} \label{eq:MR_model}
\begin{align}
        \Bar{W}_h^\mathrm{MR} = A_{10}\left(\Bar{I}_1 - 3\right) + A_{01}\left(\Bar{I}_2 - 3\right)
\end{align}
\begin{align}
        \mathbf{S}_{h,\mathrm{iso}}^\mathrm{MR} = 2 \frac{\partial \Bar{W}_h^\mathrm{MR}}{\partial \mathbf{C}} = J^{-2/3}\left[ 2 \left( A_{10} + \Bar{I}_1 A_{01}\right) \mathrm{Dev}\mathbf{I} - 2 A_{01} \mathrm{Dev}\Bar{\mathbf{C}} \right]
\end{align}
\end{subequations}
where $A_{10}$ and $A_{01}$ are model parameters. We choose $A_{10} = 1$ and $A_{01} = 0.5$ for generating the training data. Also, $J = J_{app} = \mathrm{det}(\mathbf{F}_{app})$, $\mathbf{C} = \mathbf{C}_{app} = \mathbf{F}_{app}^\mathrm{T} \mathbf{F}_{app}$, and $\Bar{\mathbf{C}} = \Bar{\mathbf{C}}_{app} = J^{-2/3} \mathbf{C}_{app}$.

Using $\mathbf{C}_{app}$ and $\mathbf{S}_{h,\mathrm{iso}}^\mathrm{MR}$ as the training dataset $\mathcal{D}_{h,\mathrm{iso}}$ (in this case, $\mathcal{D} = \mathcal{D}_{h,\mathrm{iso}}$), a standard GPR-based mapping as in Eq. (\ref{eq:27}) was trained. Like the hydrostatic loading case, the normalization condition, in this case, is automatically enforced owing to the presence of the stress-free reference state in the training data. This trained surrogate model can predict the isochoric hyperelastic stress component $\widetilde{\mathbf{S}}_{h,\mathrm{iso}}$ for any arbitrary quasi-static deformation.
\begin{figure}[t]
    \centering
    \includegraphics[width=13 cm]{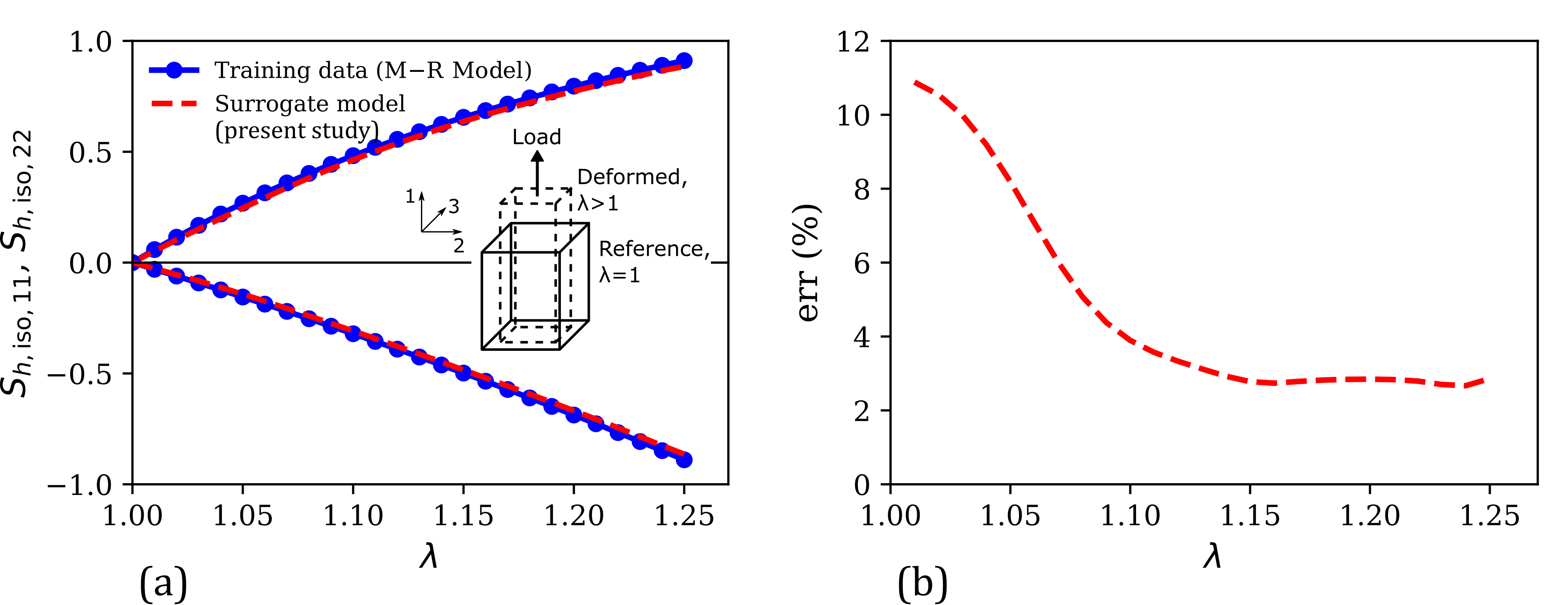}
    \caption{(a) Comparison of the numerically generated isochoric hyperelastic stress components ($\mathbf{S}_{h,\mathrm{iso,11}}$ and $\mathbf{S}_{h,\mathrm{iso,22}}$) versus uniaxial stretch ($\lambda$) training data under uniaxial tension with the corresponding surrogate model predictions. Inset shows a schematic illustration of the unconfined uniaxial tension deformation mode, highlighting the reference and deformed states and the 11-loading direction. (b) Evolution of the percent relative error ($\mathrm{err}$) of surrogate model predictions versus $\lambda$ in the training regime.}
    \label{fig:GPR_hyperelastic_1}
\end{figure}

Figure \ref{fig:GPR_hyperelastic_1} compares the stress--stretch response considered for training the surrogate model with the corresponding predictions of the trained surrogate model in this regime. Like the hydrostatic case (Section \ref{Section_5}.1), there are two independent stress components resulting from the uniaxial loading condition, $\mathbf{S}_{h, \mathrm{iso,11}}$ and $\mathbf{S}_{h, \mathrm{iso,22}}$, with $\mathbf{S}_{h, \mathrm{iso,22}} = \mathbf{S}_{h, \mathrm{iso,33}}$. 26 data points were considered for training in this case. From Fig. \ref{fig:GPR_hyperelastic_1}(a), the model predictions are in excellent agreement with the training data, leading to relatively small errors as shown in Fig. \ref{fig:GPR_hyperelastic_1}(b). The mean percent relative error in this case is 4.75\%.

With a good fitting accuracy established from Fig. \ref{fig:GPR_hyperelastic_1}, the performance of our surrogate model is now analyzed in a wider testing regime. The testing regime in this case consists of uniaxial deformation mode (Eq. (\ref{eq:53})) with $\lambda \in [0.5, 1.5]$, and the simple shear deformation mode,

\begin{figure}[t]
    \centering
    \includegraphics[width=\textwidth]{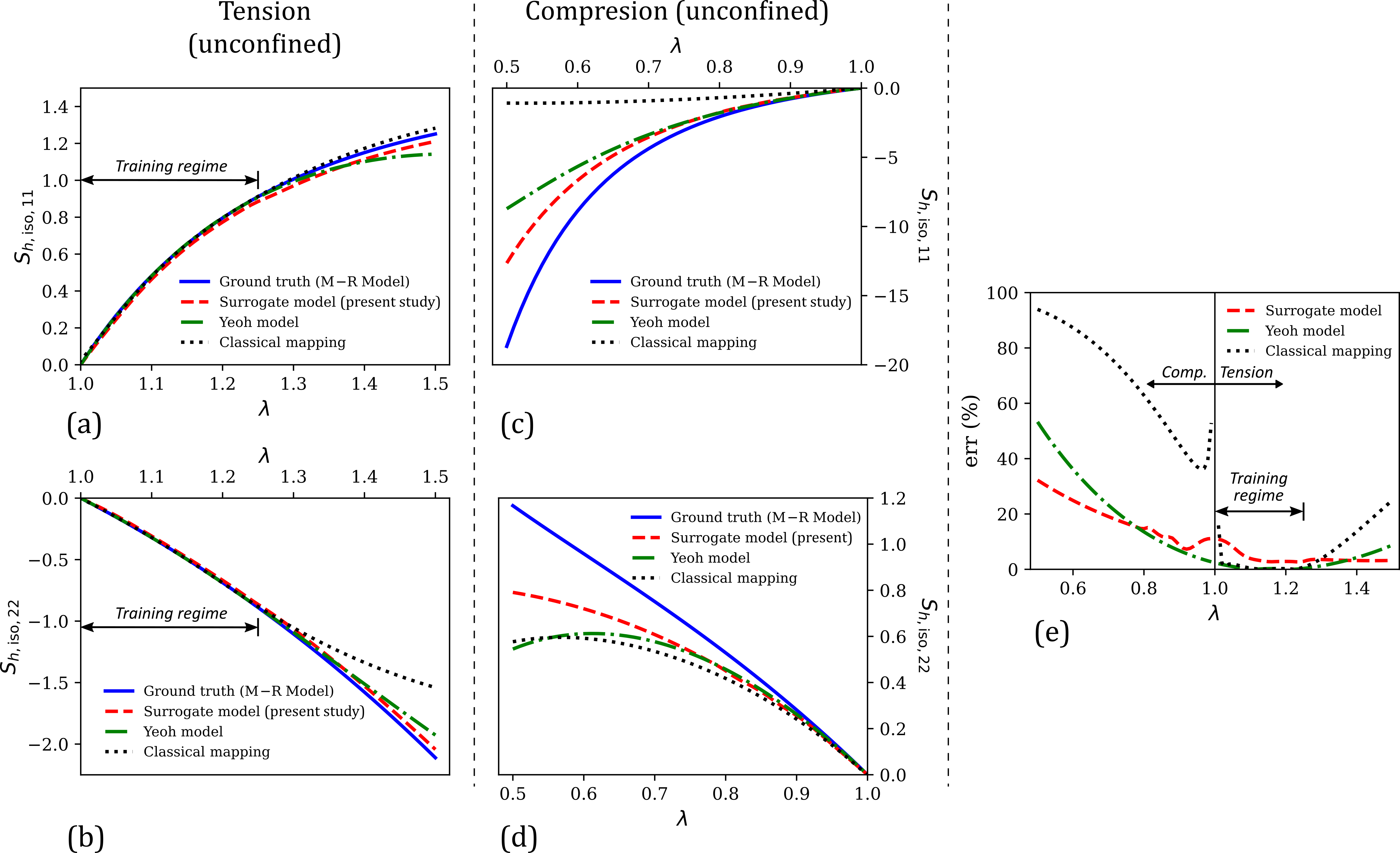}
    \caption{Comparison of the numerically generated isochoric hyperelastic stress--uniaxial stretch data from the Mooney--Rivlin model (i.e., ground truth) in the uniaxial testing regime ($\lambda \in [0.5, 1.5]$) with the corresponding predictions of our surrogate model, the Yeoh model, and the classical mapping model: (a) $\mathbf{S}_{h,\mathrm{iso,11}}$ versus $\lambda$ under uniaxial tension, (b) $\mathbf{S}_{h,\mathrm{iso,22}}$ versus $\lambda$ under uniaxial tension, (c) $\mathbf{S}_{h,\mathrm{iso,11}}$ versus $\lambda$ under uniaxial compression, and (d) $\mathbf{S}_{h,\mathrm{iso,22}}$ versus $\lambda$ under uniaxial compression. (e) Comparison of the percent relative error ($\mathrm{err}$) versus $\lambda$ responses of the present surrogate model, the Yeoh model, and the classical mapping model.}
    \label{fig:GPR_hyperelastic_2}
\end{figure}

\begin{equation} \label{eq:simple_shear}
    \mathbf{F}_{app} = \mathbf{I} + \gamma \mathbf{e}_1 \otimes \mathbf{E}_2 \quad \textrm{with} \hspace{2 pt} \gamma \in [0,0.5]
\end{equation}
where $\gamma$ is the shear strain. The surrogate model performance is compared with that of the classical mapping model (Eq. (\ref{eq:48})) and another polynomial hyperelastic model, the 2-parameter Yeoh model (henceforth, referred to simply as the Yeoh model). The Yeoh model (\cite{Yeoh:1993}) is given by
\begin{subequations} \label{eq:YH_model}
\begin{align}
        \Bar{W}_h^\mathrm{Y} = C_1 \left(\Bar{I}_1 - 3\right) + C_2 \left(\Bar{I}_1 - 3\right)^2
\end{align}
\begin{align}
        \mathbf{S}_{h,\mathrm{iso}}^\mathrm{Y} = 2 \frac{\partial \Bar{W}_h^\mathrm{Y}}{\partial \mathbf{C}} = J^{-2/3}\left[2 C_1 +  4 C_2 \left(\Bar{I}_1 - 3\right)\right] \mathrm{Dev}\mathbf{I}
\end{align}
\end{subequations}
The calibration of the Yeoh model using the uniaxial stress--stretch training data yielded $C_1 = 1.46$ and $C_2 = -0.21$.

Figure \ref{fig:GPR_hyperelastic_2} compares the stress versus uniaxial stretch predictions of the present surrogate model, the classical mapping model and the Yeoh model, with the ground truth (i.e., from the Mooney--Rivlin model that was utilized for generating training data). In the training regime of $\lambda \in [1, 1.25]$ (see Fig. \ref{fig:GPR_hyperelastic_2}(a-b)), all three models show an excellent agreement with the ground truth. In terms of the percent relative error that is shown in Fig. \ref{fig:GPR_hyperelastic_2}(e), the mean errors in the three model predictions in this regime are 4.75\% (surrogate model), 1.26\% (classical mapping model), and 0.50\% (Yeoh model).

In the overall testing regime, our surrogate model results in a markedly better performance than the other two models both in terms of the physical plausibility of predicted responses and the prediction accuracy. For example, the compressive stress--stretch responses predicted by the classical mapping model (Fig. \ref{fig:GPR_hyperelastic_2}(c-d)) exhibit unreasonable mechanical features: (i) the $\mathbf{S}_{h, \mathrm{iso,11}}-\lambda$ response shows a softening response that is uncharacteristic of hyperelastic soft materials, which typically exhibit compression--tension asymmetry in the loading direction (i.e., the material exhibits a highly nonlinear and stiffer response in compression than in tension) (\cite{Budday_etal:2017,Upadhyay_etal:2020a,Treloar:1944}), and (ii) the $\mathbf{S}_{h, \mathrm{iso,22}}-\lambda$ response starts to decrease following a maximum at approximately $\lambda = 0.57$, which violates the expected monotonicity in stress--stretch responses (prior to damage/failure). The latter behavior is also exhibited by the Yeoh model, for which the predicted $\mathbf{S}_{h, \mathrm{iso,22}}$ versus $\lambda$ response violates monotonicity at large compressive strains of $\lambda < 0.61$. Non-physical response predictions from conventional hyperelastic models have been observed in the literature for certain model parameter values and are attributed to the violation of the second law of thermodynamics by the model with those parameter values (\cite{Liu:2012,Upadhyay_etal:2019b,Fontenele_etal:2022}). Owing to its physics-informed formulation, our surrogate model prevents this issue and yields a physically-plausible mechanical response even at large deformations. Further, it results in a mean percent relative error of 10.98\% in the uniaxial testing regime, which is slightly lower compared to the $\overline{\mathrm{err}}$ of 11.08\% of the Yeoh model, and significantly lower than the $\overline{\mathrm{err}}$ of 49.94\% of the classical mapping model.

\begin{figure}[t]
    \centering
    \includegraphics[width=13 cm]{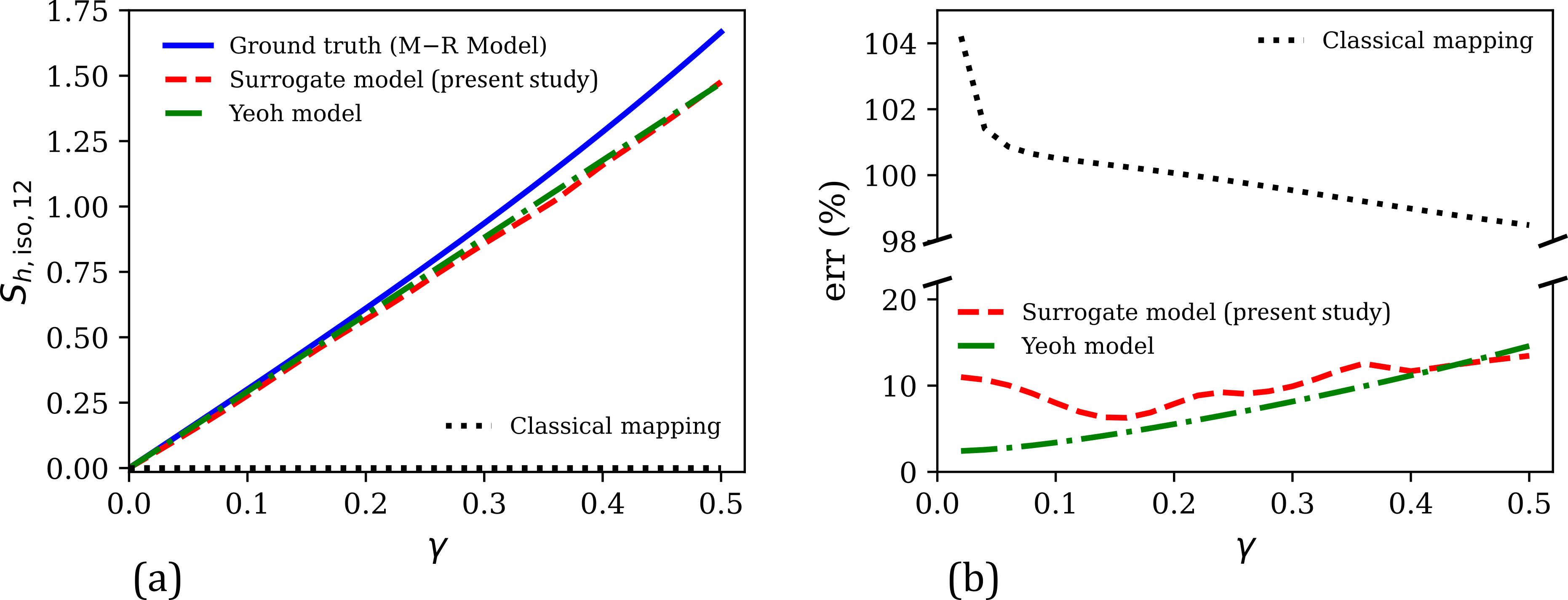}
    \caption{(a) Comparison of the numerically generated isochoric hyperelastic shear stress ($\mathbf{S}_{h,\mathrm{iso,12}}$)--shear strain ($\gamma$) data from the Mooney--Rivlin model (i.e., ground truth) in the shear testing regime ($\gamma \in [0, 0.5]$) with the corresponding predictions of our surrogate model, the Yeoh model, and the classical mapping model. (b) Comparison of the percent relative error ($\mathrm{err}$) versus $\gamma$ responses of the present surrogate model, the Yeoh model, and the classical mapping model.}
    \label{fig:GPR_hyperelastic_3}
\end{figure}
\begin{figure}[b]
    \centering
    \includegraphics[width=13 cm]{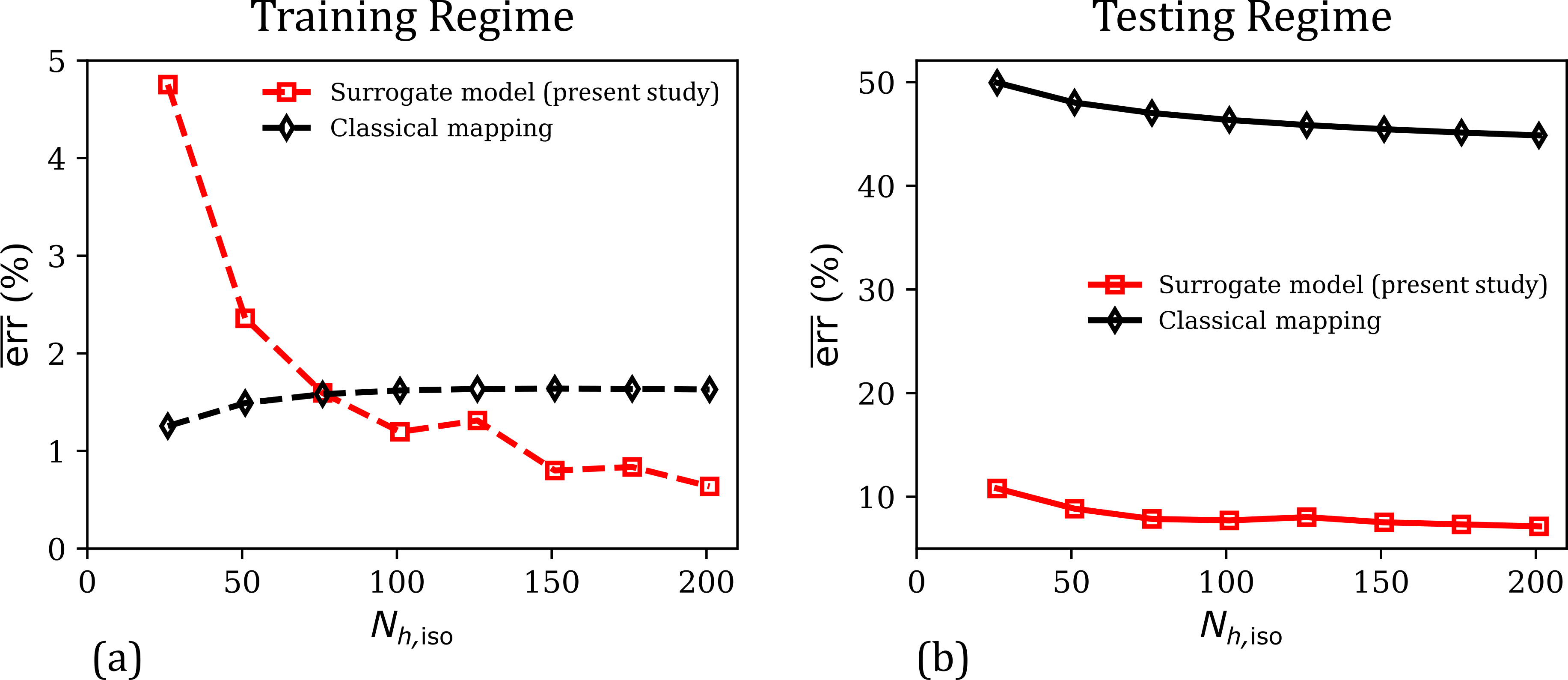}
    \caption{(a) Comparison of the evolution of mean percent relative error ($\overline{\mathrm{err}}$) in the predictions of the surrogate model and the classical mapping model in the training regime ($\lambda \in [1,1.25]$), as a function of the training dataset size ($N_{h,\mathrm{iso}}$). (b) The corresponding $\overline{\mathrm{err}}$ versus $N_{h,\mathrm{iso}}$ responses of the present surrogate model and the classical mapping model, for their predicted responses in the overall testing regime ($\lambda \in [0.5,1.5] \cup \gamma \in [0,0.5]$).}
    \label{fig:GPR_hyperelastic_4}
\end{figure}
Figure \ref{fig:GPR_hyperelastic_3}(a) shows the predicted shear stress versus shear strain responses from the three models and the ground truth. Here, the classical mapping model predicts zero stress regardless of the applied strain. This is attributed to the sole consideration of uniaxial deformation in the training dataset, in which non-diagonal components in the $\mathbf{C}$ and $\mathbf{S}_{h, \mathrm{iso}}$ tensors were always zero. As our surrogate model is based on the mapping of strain invariants with response coefficients, it does not suffer from this limitation and leads to accurate shear stress predictions as shown in Fig. \ref{fig:GPR_hyperelastic_3}(a). Note that even though the simple shear deformation mode generates multiple non-zero stress components (viz., $\mathbf{S}_{h, \mathrm{iso}, 11}$, $\mathbf{S}_{h, \mathrm{iso}, 12}$, $\mathbf{S}_{h, \mathrm{iso}, 22}$, and $\mathbf{S}_{h, \mathrm{iso}, 33}$), only the dominant 12-component is shown for brevity. In terms of the percent relative error metric that includes all non-zero stress components and is plotted in Fig. \ref{fig:GPR_hyperelastic_3}(b), our surrogate model leads to a maximum $\overline{\mathrm{err}}$ of 13.46\% across the shear deformation mode, with a mean error of 10.09\%. For comparison, the maximum $\overline{\mathrm{err}}$ for the Yeoh model is 14.59\%, while the mean $\overline{\mathrm{err}}$ is 7.54\%. Overall, it is seen that both the surrogate model of this study and the Yeoh model exhibit reasonable prediction accuracy across the entire testing regime (uniaxial + shear), but only our surrogate model results in stress--strain predictions that are physically plausible across all the investigated deformation modes.     

Figure \ref{fig:GPR_hyperelastic_4} shows the evolution of mean percent relative error $\overline{\mathrm{err}}$ in the training and overall testing regimes, with the training dataset size $N_{h,\mathrm{iso}}$. Similar to the hydrostatic loading case, our surrogate model results in a monotonically decreasing $\overline{\mathrm{err}}$ with $N_{h,\mathrm{iso}}$ response with asymptotic behavior in both cases. On the other hand, the classical mapping model shows an asymptotic decrease in $\overline{\mathrm{err}}$ only in the testing regime, while showing a slight increase in the training regime. Ultimately, while both the models show very good fitting accuracy for every investigated training dataset size, our surrogate model results in significantly more accurate predictions across multiple deformation modes. This ability to seamlessly transition between deformation modes without loss of accuracy is a distinctive feature of trustworthy constitutive models (\cite{Beda:2014}).
\subsection{The \texorpdfstring{$\widetilde{\mathcal{M}}_{v,\mathrm{iso}}$} {} Surrogate Model Under Dynamic Uniaxial and Shear Loading} \label{Section_5.3}
Thus far, the performance of the physics-informed data-driven constitutive model of this study has been evaluated only under quasi-static loading conditions, when the applied strain rate is so small that viscous dissipation remains negligible. Now, consider an isochoric dynamic uniaxial loading at a finite applied loading rate, equivalently following
\begin{subequations} \label{eq:57}
\begin{align}
        \mathbf{F}_{app} = \lambda \mathbf{e}_1 \otimes \mathbf{E}_1 + \frac{1}{\sqrt{\lambda}} \left(\mathbf{e}_2 \otimes \mathbf{E}_2 +  \mathbf{e}_3 \otimes \mathbf{E}_3\right) \quad \textrm{with} \hspace{2 pt} \lambda \in [1,1.5]
\end{align}

\begin{align}
        \dot{\mathbf{F}}_{app} = \dot{\lambda} \left[ \mathbf{e}_1 \otimes \mathbf{E}_1 - \frac{1}{2 \lambda^{3/2}} \left(\mathbf{e}_2 \otimes \mathbf{E}_2 +  \mathbf{e}_3 \otimes \mathbf{E}_3\right) \right] \quad \textrm{with} \hspace{2 pt} \dot{\lambda} \in [10,100]
\end{align}
\end{subequations}
where $\dot{\lambda}=d\lambda/dt$ is the applied uniaxial strain rate. At any given time $t$ during loading, $\lambda = \dot{\lambda}t$.

Dynamic loading of the form of Eq. (\ref{eq:57}) is applied during high strain rate tension tests on soft materials such as the tensile Kolsky bar experiment (e.g., see \cite{Upadhyay_etal:2021a,Yang_etal:2000}). The stress--strain (or stretch)--strain rate data from these tests combined with the stress--strain (or stretch) data from quasi-static uniaxial tests are then used to calibrate visco-hyperelastic constitutive models. As a number of visco-hyperelastic constitutive models (see \cite{Upadhyay_etal:2020} for a review) assume an additive decomposition of total stress into hyperelastic (rate-independent) and viscous overstress components (e.g., see Eq. (\ref{eq:14})), the quasi-static test data is exclusively used to calibrate the hyperelastic model ($\Bar{W}_h$) parameters, while the viscous overstress that is calculated by subtracting quasi-static stress--strain data from high strain rate stress--strain data is used to calibrate the model parameters of the viscoelastic component of the constitutive model (e.g., $\Bar{W}_v$ in Eq. (\ref{eq:14})). In this study, the 3-parameter Upadhyay--Subhash--Spearot (USS) viscous dissipation potential (\cite{Upadhyay_etal:2020,Upadhyay_etal:2021b}) is utilized to generate the viscous overstress training data for developing the $\widetilde{\mathcal{M}}_{v,\mathrm{iso}}$ surrogate model,
\begin{subequations} \label{eq:USS_model}
\begin{align}
        \Bar{W}_v^\mathrm{USS} = k_{11} \Bar{J}_2 \sqrt{\Bar{I}_1 - 3} + \frac{k_{21}}{c_{21}} \Bar{J}^{c_{21}}_5 \sqrt{\Bar{I}_2 - 3}
\end{align}
\begin{align}
        \mathbf{S}_{v,\mathrm{iso}}^\mathrm{USS} = 2 \frac{\partial \Bar{W}_v^\mathrm{USS}}{\partial \dot{\mathbf{C}}} =  J^{-2/3}\left[ \left( 4 k_{11} \sqrt{\Bar{I}_1 - 3} \right) \mathrm{Dev}(\dot{\Bar{\mathbf{C}}}) + \left(2 k_{21} \Bar{J}_5^{c_{21} - 1} \sqrt{\Bar{I}_2 - 3} \right)\mathrm{Dev} (\Bar{\mathbf{C}}\dot{\Bar{\mathbf{C}}} + \dot{\Bar{\mathbf{C}}}\Bar{\mathbf{C}})\right] 
\end{align}
\end{subequations}
where $k_{11}$ and $k_{21}$ are the linear and nonlinear rate sensitivity control parameters, respectively, and $c_{21}$ is the rate sensitivity index. We use $k_{11} = 1$, $k_{21} = 1$, and $c_{21} = 0.75$ for generating the training data. The expressions for invariants $\Bar{I}_1$, $\Bar{I}_2$, $\Bar{J}_2$, and $\Bar{J}_5$ are given in Eq. (\ref{eq:15}). Just like the quasi-static loading case, $J = J_{app} = \mathrm{det}(\mathbf{F}_{app})$, $\mathbf{C} = \mathbf{C}_{app} = \mathbf{F}_{app}^\mathrm{T} \mathbf{F}_{app}$, and $\Bar{\mathbf{C}} = \Bar{\mathbf{C}}_{app} = J^{-2/3} \mathbf{C}_{app}$. $\dot{\Bar{\mathbf{C}}} = \dot{\Bar{\mathbf{C}}}_{app}$ is the time rate of change of $\Bar{\mathbf{C}}_{app}$.

The training dataset $\mathcal{D}_{v,\mathrm{iso}} = \mathcal{D} = \{ \mathbf{C}_{app}, \dot{\mathbf{C}}_{app}, \mathbf{S}_{v,\mathrm{iso}}^{\mathrm{USS}}\}$ (see Eq. (\ref{eq:21})) generated from the USS model is used to train the $\widetilde{\mathcal{M}}_{v,\mathrm{iso}}$ surrogate model (Eq. (\ref{eq:28})). Similar to the previous quasi-static cases (involving $\mathbf{S}_\mathrm{vol}$ and $\mathbf{S}_{h,\mathrm{iso}}$), the normalization condition is implicitly enforced due to the consideration of the stress-free reference state in the training data. The second law of thermodynamics constraint, on the other hand, is enforced via C-GPR as the chosen regression method. Here, the inequality constraint on the hyperparameter optimization problem in Eq. (\ref{eq:42}) is enforced at all training input data points (i.e., the constraint points are the same as the training data points; $N_c = N_{v,\mathrm{iso}}$). The trained surrogate model can predict viscous overstress $\widetilde{\mathbf{S}}_{v,\mathrm{iso}}$ for any given applied tensors $\mathbf{C}$ and $\dot{\mathbf{C}}$.

\begin{figure}[t]
    \centering
    \includegraphics[width=\textwidth]{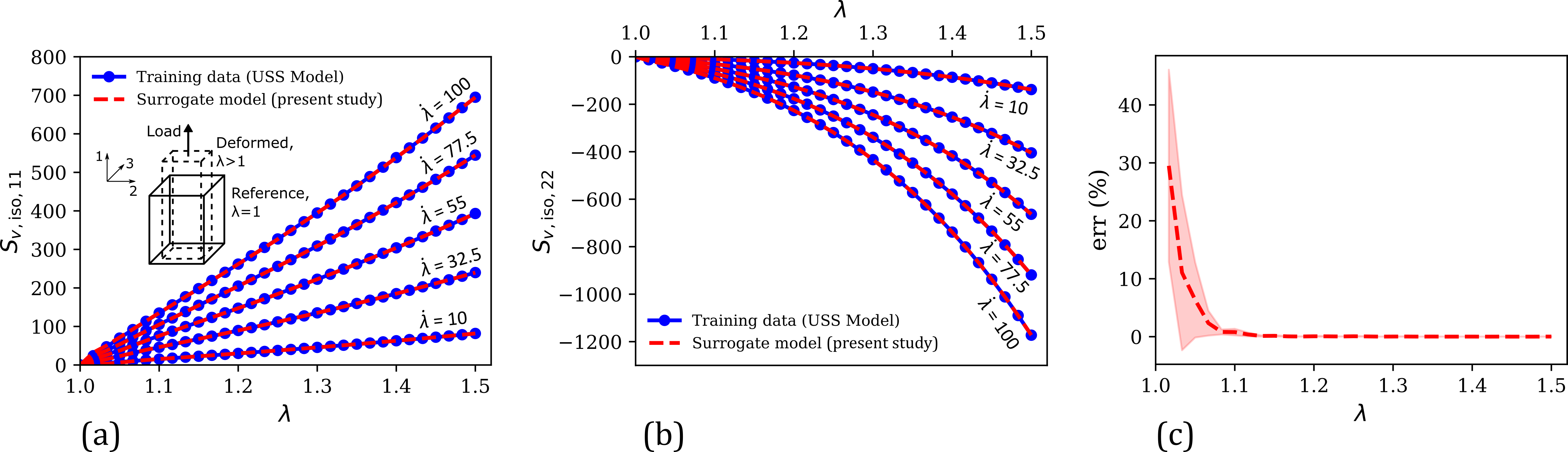}
    \caption{(a) Comparison of the numerically generated training data of isochoric viscous overstress component $\mathbf{S}_{v,\mathrm{iso,11}}$ versus uniaxial stretch ($\lambda$) at multiple stretch rates ($\dot{\lambda}$), with the corresponding surrogate model predictions. Inset shows a schematic illustration of the unconfined uniaxial tension deformation mode. (b) The corresponding $\mathbf{S}_{v,\mathrm{iso,11}}$ versus $\lambda$ responses at fixed $\dot{\lambda}$ values from the training data and the surrogate model. (c) Evolution of the percent relative error ($\mathrm{err}$) of surrogate model predictions versus $\lambda$ in the training regime (dashed line: mean $\mathrm{err}$ across all stretch rates; shaded region: mean $\pm$ standard deviation).}
    \label{fig:GPR_Visco-hyperelastic_1}
\end{figure}

Figure \ref{fig:GPR_Visco-hyperelastic_1} investigates the fitting response of the data-driven surrogate model in the training regime of uniaxial tension loading (Eq. (\ref{eq:53})) that is composed of 5 uniformly distributed stretch rates in the range of $\dot{\lambda} \in [10, 100]$ (viz., $\dot{\lambda} = 10, 32.5, 55, 77.5,$ and $100$), and 31 uniformly distributed stretch values in the range of $\lambda \in [1,1.5]$ for every stretch rate considered. The total number of training data points is thus $5 \times 31 = 155$. Both the non-zero stress-components under uniaxial tension, $\mathbf{S}_{v, \mathrm{iso,11}}$ and $\mathbf{S}_{v, \mathrm{iso,22}}$, are plotted in Figs. \ref{fig:GPR_Visco-hyperelastic_1}(a) and \ref{fig:GPR_Visco-hyperelastic_1}(b), respectively. An excellent agreement between model predictions and the stress--stretch--stretch rate training data is apparent from these plots. The corresponding scalar fitting error ($\mathrm{err}$) versus stretch response is shown in Fig. \ref{fig:GPR_Visco-hyperelastic_1}(c); here, the line represents the mean $\mathrm{err}$ values averaged across all the investigated stretch rates, and the shaded region represents the vertical range of mean $\pm$ standard deviation. The $\mathrm{err}$ versus $\lambda$ response starts at relatively high values at small stretches and then assumes consistently small values close to zero for $\lambda > 1.1$. The large $\mathrm{err}$ at small stretch values is attributed to the very small stresses in the small strain regime in the training data (i.e., the denominator in the formula for $\mathrm{err}$), which causes spuriously high $\mathrm{err}$ even when the absolute differences between true and predicted stresses are very small. Overall, the mean percent relative error $\overline{\mathrm{err}}$ across all the 155 training data points is 1.73\%, suggesting a very good model fitting accuracy. 

\begin{figure}[t]
    \centering
    \includegraphics[width=\textwidth]{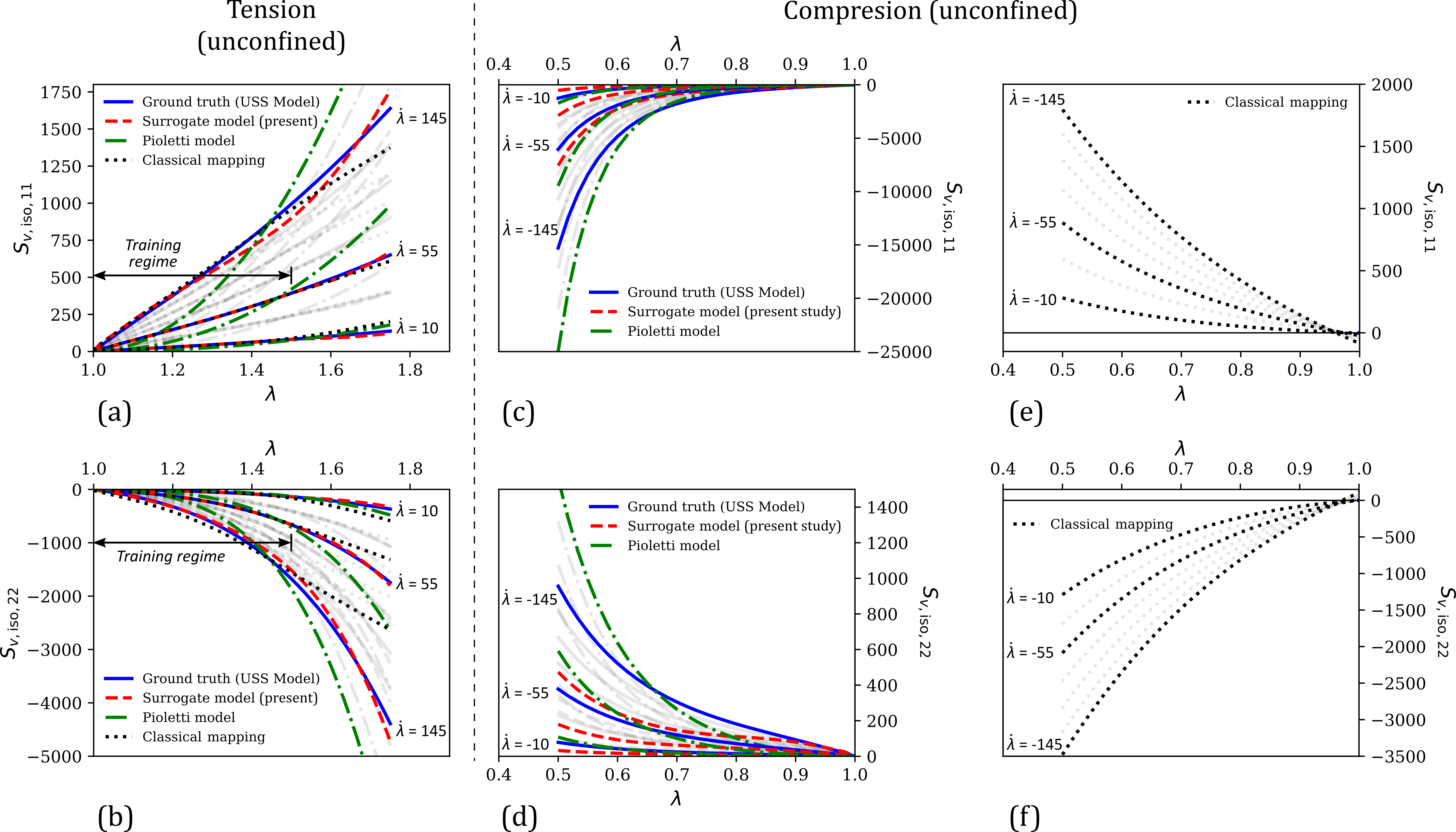}
    \caption{Comparison of the numerically generated isochoric viscous overstress--uniaxial stretch data at multiple stretch rates from the USS model (i.e., ground truth) in the uniaxial testing regime ($\lambda \in [0.5, 1.75]$, $\dot{\lambda} \in [-145,145]$) with the corresponding predictions of our surrogate model, the Pioletti model, and the classical mapping model: (a) $\mathbf{S}_{v,\mathrm{iso,11}}$ versus $\lambda$ under uniaxial tension, (b) $\mathbf{S}_{v,\mathrm{iso,22}}$ versus $\lambda$ under uniaxial tension, (c,e) $\mathbf{S}_{v,\mathrm{iso,11}}$ versus $\lambda$ under uniaxial compression, and (d,f) $\mathbf{S}_{v,\mathrm{iso,22}}$ versus $\lambda$ under uniaxial compression. Note: the light gray lines in the background are the data corresponding to the stretch rates $\dot{\lambda} = \pm32.5, \pm77.5, \pm100$ and $\pm122.5$, which were also considered in model testing but are not shown in this figure for clarity.}
    \label{fig:GPR_Visco-hyperelastic_2}
\end{figure}

The accuracy and physical plausibility of our data-driven model predictions are now investigated in a wider testing regime that consists of tension, compression, and simple shear deformation modes. Here, the two uniaxial deformation modes (i.e., tension and compression) comprise stretch rates in the range of $\dot{\lambda} \in [-145,145]$ and stretch values in the range of $\lambda \in [0.5,1.75]$ (cf. Eq. (\ref{eq:57})). Specific stretch rate values considered in model testing are: $\dot{\lambda} = \pm10, \pm32.5, \pm55, \pm77.5, \pm100, \pm122.5,$ and $\pm145$. Now, the dynamic simple shear deformation is given by
\begin{subequations} \label{eq:59}
\begin{align}
        \mathbf{F}_{app} = \mathbf{I} + \gamma \mathbf{e}_1 \otimes \mathbf{E}_2 \quad \textrm{with} \hspace{2 pt} \gamma \in [0,0.5]
\end{align}
\begin{align}
        \dot{\mathbf{F}}_{app} = \dot{\gamma} \mathbf{e}_1 \otimes \mathbf{E}_2 \quad \textrm{with} \hspace{2 pt} \dot{\gamma} \in [10,145]
\end{align}
\end{subequations}
where $\dot{\gamma}=d\gamma/dt$ is the applied shear strain rate. At any given time $t$ during loading, $\gamma = \dot{\gamma}t$. For model testing, we consider shear strain rates in the range of $\dot{\gamma} \in [10, 145]$ (specifically, $\dot{\gamma} = 10, 32.5, 55, 77.5, 100, 122.5,$ and $145$); for each strain rate, shear strains in the range of $\gamma \in [0, 0.5]$ are considered. The overall testing regime outstretches the training regime both in terms of the maximum stretch/strain and stretch-/strain-rate magnitudes and spans multiple deformation modes. Like the previous two quasi-static cases, the present model predictions are compared with those from the classical mapping model (Eq. (\ref{eq:28})) and an existing constitutive model from the literature. Here, we choose the Pioletti model (\cite{Pioletti_etal:1998,Pioletti_Rakotomanana:2000}),
\begin{subequations} \label{eq:PL_model}
\begin{align}
        \Bar{W}_v^\mathrm{PL} = \frac{\eta'}{4} \left(\Bar{I}_1 - 3\right) \Bar{J}_2
\end{align}
\begin{align}
        \mathbf{S}_{v,\mathrm{iso}}^\mathrm{PL} = 2 \frac{\partial \Bar{W}_v^\mathrm{PL}}{\partial \dot{\mathbf{C}}} = J^{-2/3} \eta' \left(\Bar{I}_1 - 3\right) \mathrm{Dev}(\dot{\Bar{\mathbf{C}}})
\end{align}
\end{subequations}
The Pioletti model was calibrated using the tensile stress--stretch--stretch rate training dataset, resulting in $\eta'$ = 6.94.
\begin{figure}[!b]
    \centering
    \includegraphics[width=6.5 cm]{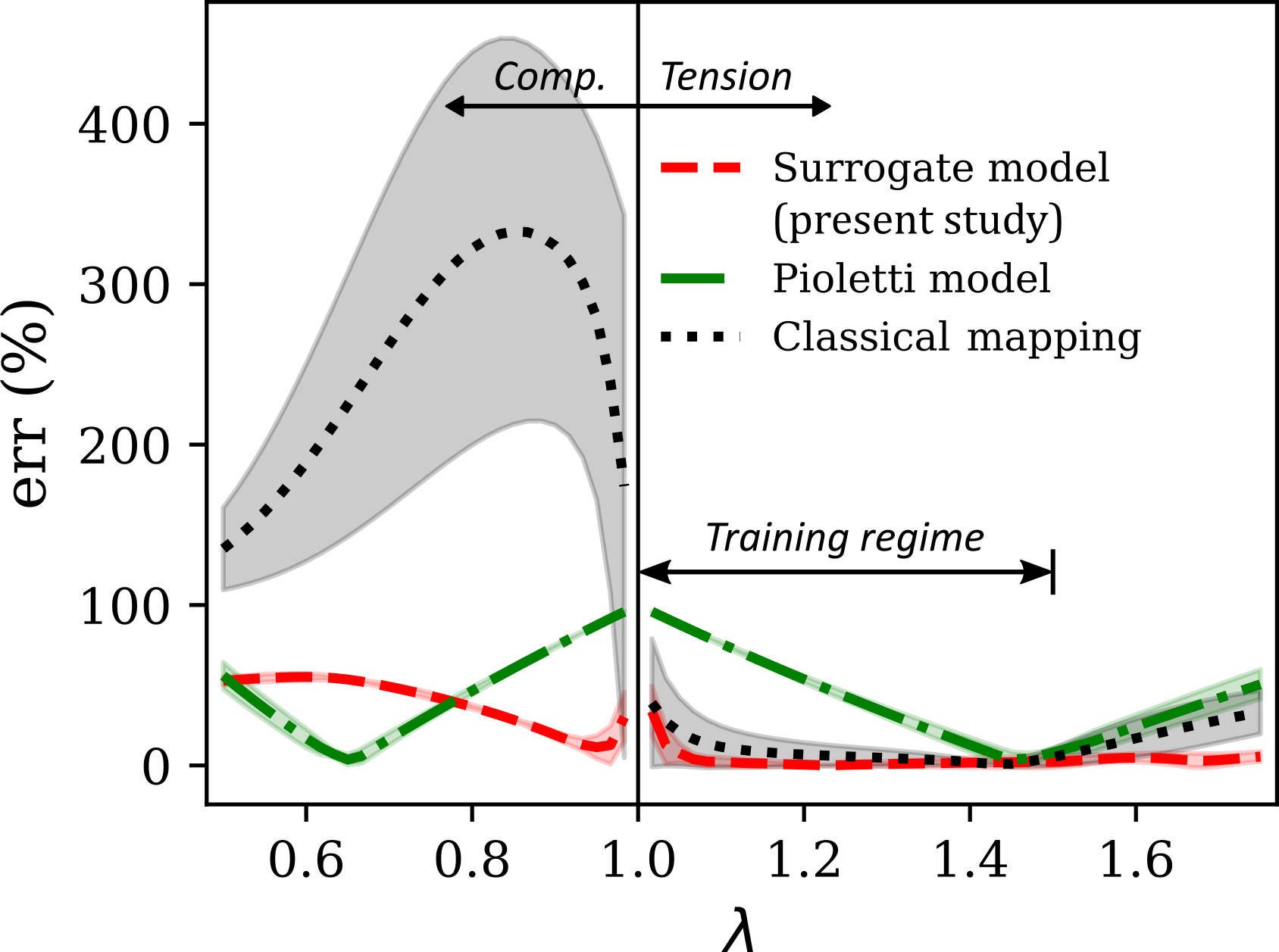}
    \caption{Comparison of the percent relative error ($\mathrm{err}$) versus $\lambda$ responses of the present surrogate model, the Pioletti model, and the classical mapping model. The lines represent mean $\mathrm{err}$ across all investigated strain rates, and the shaded regions represent mean $\pm$ standard deviation.}
    \label{fig:GPR_Visco-hyperelastic_3_A}
\end{figure}

Figures \ref{fig:GPR_Visco-hyperelastic_2}(a-b) compare the tensile stress versus stretch responses in the testing regime at three representative stretch rates (for clarity) as predicted by the present data-driven surrogate model, the classical mapping model, and the Pioletti model, with the ground truth (i.e., USS model predictions). Out of the three stretch rates shown in these figures, $\dot{\lambda}$ = 10 and $\dot{\lambda}$ = 55 belong to the training subset, while $\dot{\lambda}$ = 145 belongs to the testing dataset that was not considered during training. In the training regime (i.e., $\lambda \in [1, 1.5]$ and $\dot{\lambda} \in [10, 100]$), both the data-driven models exhibit excellent fitting performance as their predicted tensile stress--stretch responses nearly overlap the ground truth. The Pioletti model, on the other hand, results in a relatively poor fitting accuracy, which is attributed to its overly simple mathematical form with only one model parameter. In terms of the scalar percent error metric plotted in Fig. \ref{fig:GPR_Visco-hyperelastic_3_A} (averaged across all investigated strain rates), the average prediction error of our surrogate model in the training regime is the lowest at 1.73\%, followed by the classical mapping model (4.67\%) and the Pioletti model (40.59\%). Outside the training regime in the tension deformation mode when $\lambda > 1.5$ or $\dot{\lambda} > 100$ (see Figs. \ref{fig:GPR_Visco-hyperelastic_2}a-b), the Pioletti model remains the worst performing model among the three investigated models. This time, however, the stress predictions of the classical mapping model also differ considerably from the ground truth, especially at high stretch rates. Overall, in the tensile deformation mode, the mean percent relative error, $\overline{\mathrm{err}}$, of our surrogate model is just 3.40\%, which is considerably lower than that of the classical mapping model (12.30\%) and the Pioletti model (41.59\%).

In the compression deformation mode (Figs. \ref{fig:GPR_Visco-hyperelastic_2}(c-f)), the stress predictions of all three models show disagreement with the ground truth to varying degrees. Nevertheless, the stress--stretch plots of our data-driven surrogate model and the Pioletti model are physically reasonable and exhibit the expected compressive nature (i.e., negative $\mathbf{S}_{v, \mathrm{iso,11}}$ and $\mathbf{S}_{v, \mathrm{iso,22}}$), monotonicity at every investigated stretch rate, and compression--tension asymmetry in the 11-loading direction (i.e., stiffer response in compression than in tension). The classical mapping model, on the other hand, predicts tensile stresses under compression (see Figs. \ref{fig:GPR_Visco-hyperelastic_2} (e-f)), which is physically implausible. Owing to these unreasonable predictions, the mean percent relative error $\overline{\mathrm{err}}$ of this model, 253.14\%, is an order of magnitude higher than the corresponding mean errors of our surrogate model (39.69\%) and the Pioletti model (45.59\%).
\begin{figure}[t]
    \centering
    \includegraphics[width=6.5 cm]{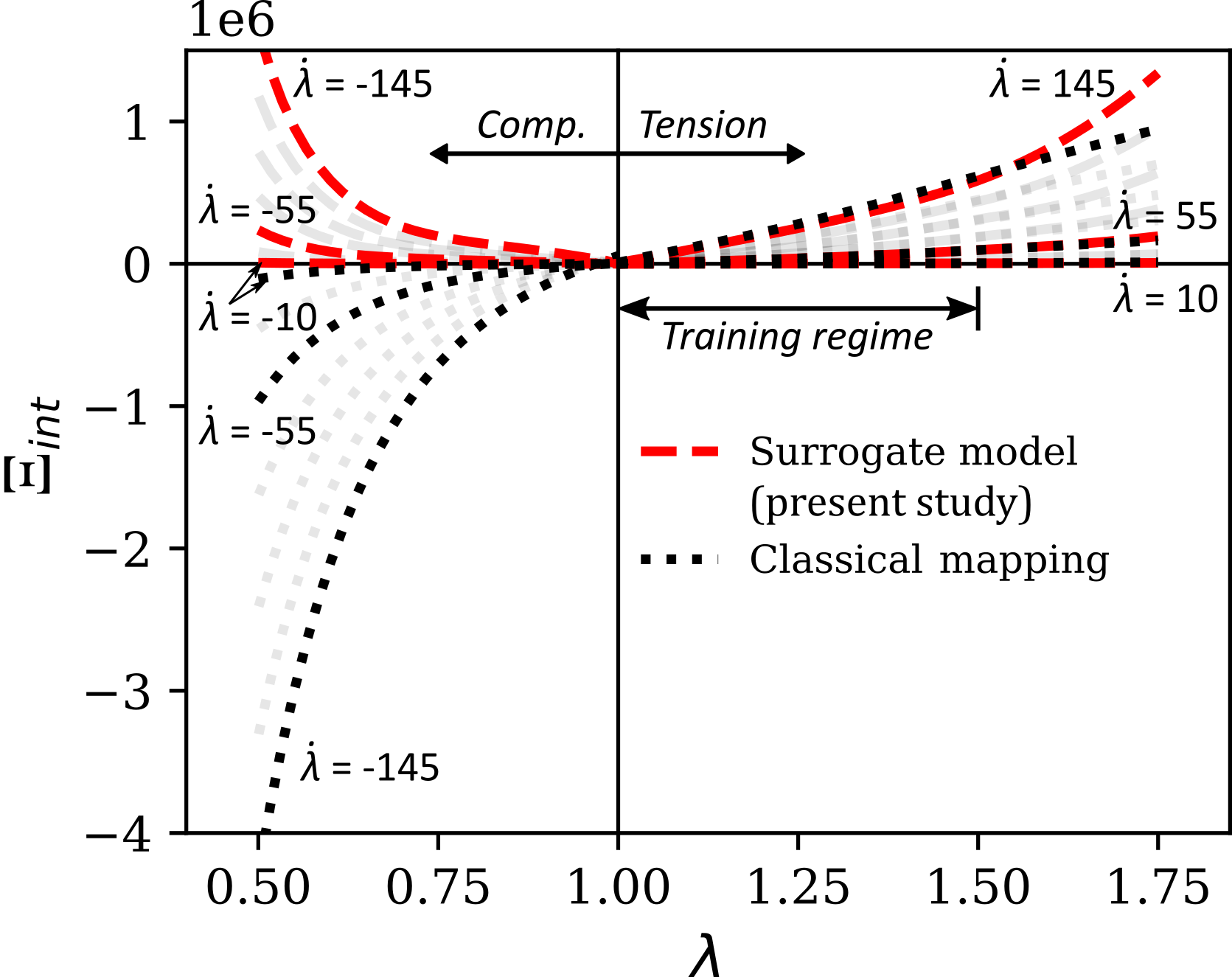}
    \caption{Comparison of the internal dissipation ($\Xi_{int}$) versus $\lambda$ responses of the present surrogate model and the classical ML-based mapping model. Note: the light gray lines in the background are the data corresponding to the stretch rates $\dot{\lambda} = \pm32.5, \pm77.5, \pm100$ and $\pm122.5$, which were also evaluated in model testing but are not shown in this figure for clarity.}
    \label{fig:GPR_Visco-hyperelastic_3_B}
\end{figure}
\begin{figure}[b]
    \centering
    \includegraphics[width=\textwidth]{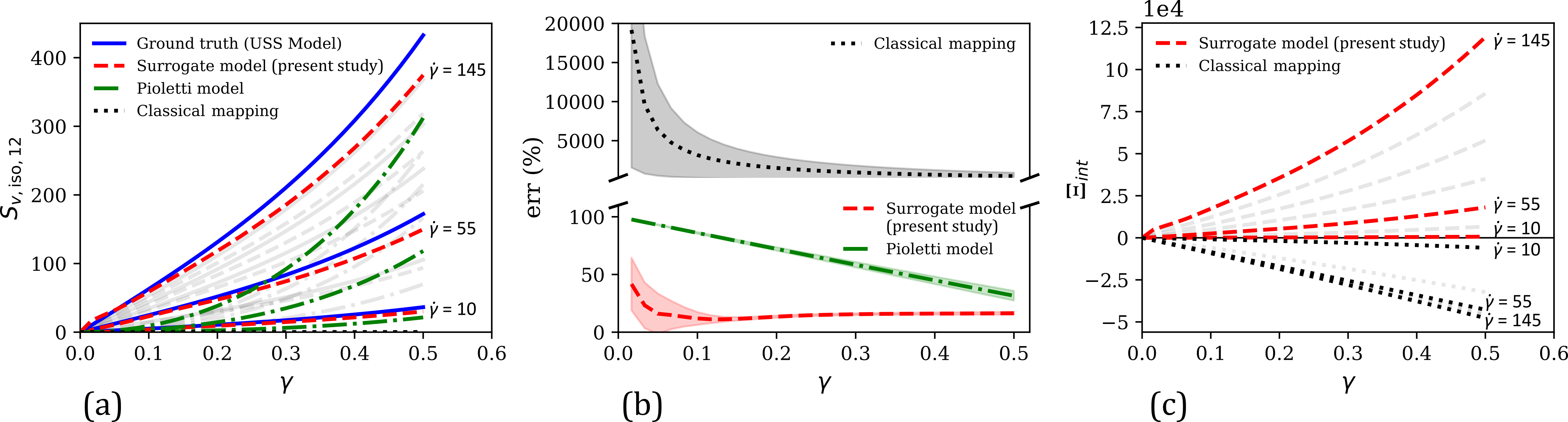}
    \caption{(a) Comparison of the numerically generated isochoric shear viscous overstress ($\mathbf{S}_{v,\mathrm{iso,12}}$)--shear strain ($\gamma$) data from the USS model (i.e., ground truth) in the shear testing regime ($\gamma \in [0, 0.5]$, $\dot{\gamma} \in [10,145]$) with the corresponding predictions of our surrogate model, the Pioletti model, and the classical mapping model. (b) The corresponding percent relative error ($\mathrm{err}$) versus $\gamma$ responses of these three models. The lines represent mean $\mathrm{err}$ across all investigated strain rates, and the shaded regions represent mean $\pm$ standard deviation. (c) Comparison of the internal dissipation ($\Xi_{int}$) versus $\gamma$ responses of the two data-driven models: our surrogate model and the classical mapping model. Note: in (a) and (c), the background light gray lines are the data corresponding to the strain rates $\dot{\gamma} = \pm32.5, \pm77.5, \pm100,$ and $\pm122.5$, which were also evaluated in model testing but are not shown for clarity.}
    \label{fig:GPR_Visco-hyperelastic_4}
\end{figure}

The predictions of the two data-driven models, the present surrogate model and the classical mapping model, are now analyzed in light of their conformity to the second law of thermodynamics in the testing regime. In this regard, Fig. \ref{fig:GPR_Visco-hyperelastic_3_B} plots the internal viscous dissipation $\Xi_{int}$ as a function of stretch at multiple stretch rates as predicted by these two models. In the tension deformation mode, both the models result in non-negative viscous dissipation, thus satisfying the second law of thermodynamics (see Eq. (\ref{eq:41}). However, in compression, the classical mapping model predictions show a negative internal dissipation at every investigated stretch rate, which violates the second law of thermodynamics. This causes the model to predict physically unreasonable stress--stretch--stretch rate predictions as were seen in Figs. \ref{fig:GPR_Visco-hyperelastic_2}(e-f). Owing to its physics-informed construction that enforces the non-negativity of viscous dissipation, our surrogate model prevents this issue and its predicted uniaxial mechanical response conforms to the second law of thermodynamics throughout the tensile testing regime.

\begin{figure}[t]
    \centering
    \includegraphics[width=13 cm]{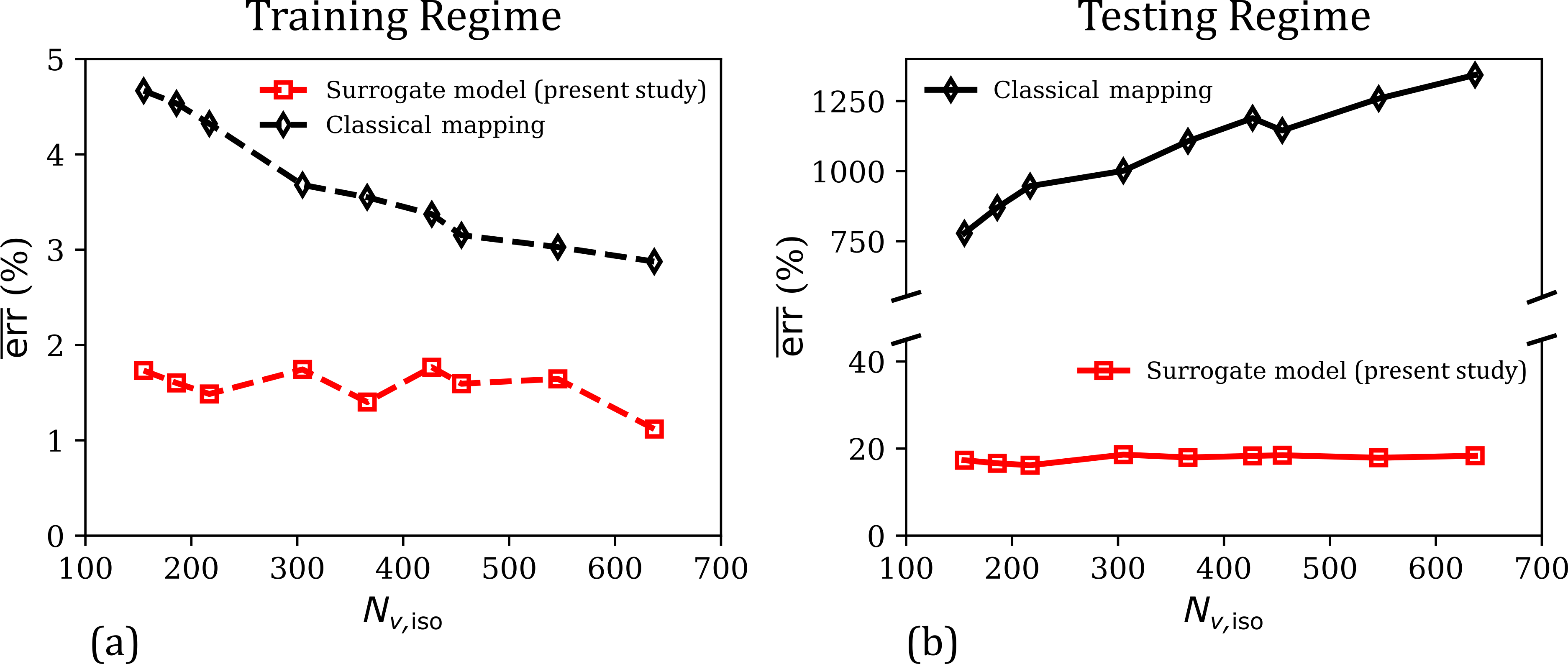}
    \caption{(a) Comparison of the evolution of mean percent relative error ($\overline{\mathrm{err}}$) in the predictions of the surrogate model and the classical mapping model in the training regime ($\lambda \in [1,1.5], \dot{\lambda} \in [10, 100]$), as a function of the training dataset size ($N_{v,\mathrm{iso}}$). (b) The corresponding $\overline{\mathrm{err}}$ versus $N_{v,\mathrm{iso}}$ responses of the present surrogate model and the classical mapping model, for their predicted responses in the overall testing regime ($[\lambda \in [0.5,1.75], \dot{\lambda} \in [-145, 145]] \cup [\gamma \in [0,0.5]], \dot{\gamma} \in [10,145]$).}
    \label{fig:GPR_Visco-hyperelastic_5}
\end{figure}
The performance of our surrogate model in the simple shear deformation mode is analyzed and compared with the Pioletti and classical mapping models in Fig. \ref{fig:GPR_Visco-hyperelastic_4}. Like the quasi-static hyperelastic case of Section \ref{Section_5.2}, the classical model predicts a zero shear stress component (i.e., $\mathbf{S}_{v, \mathrm{iso, 12}}$) regardless of the applied shear strain and strain rates (see Fig. \ref{fig:GPR_Visco-hyperelastic_4}(a)). This non-physical behavior is attributed to the absence of any non-zero non-diagonal components in the tensors $\mathbf{C}$, $\dot{\mathbf{C}}$, and $\mathbf{S}_{v, \mathrm{iso}}$ considered in the training dataset. Note, the diagonal stress components predicted by the classical mapping model in shear loading are non-zero. Among the other two models under comparison, both the present surrogate model and the Pioletti model lead to physically plausible predictions, but our surrogate model predictions are in better agreement with the ground truth compared to the Pioletti model predictions. The percent scalar error versus shear strain responses (averaged across investigated shear strain rates) of the three models are compared in Fig. \ref{fig:GPR_Visco-hyperelastic_4}(b). The mean percent error $\overline{\mathrm{err}}$ of our surrogate model across the shear deformation mode is 15.86\%, which is significantly lower than the corresponding mean error of the Pioletti model at 60.43\%, and more than two orders of magnitude lower compared to the $\overline{\mathrm{err}}$ of the classical mapping model, 2453.32\%. 

From Figs. \ref{fig:GPR_Visco-hyperelastic_4}(a-b), it is clear that among the two data-driven constitutive models under comparison, only the surrogate model of this work results in trustworthy predictions that maintain physical plausibility even under large strain and at high strain rates and more importantly at deformation modes completely unseen by the model, confirming its ability to generalize proficiently. Fig. \ref{fig:GPR_Visco-hyperelastic_4}(c) plots the internal viscous dissipation versus shear strain at multiple shear strain rates from these models. Unsurprisingly, the model predictions of our surrogate model show non-negative dissipation, thus obeying the second law of thermodynamics. Similar to the compression deformation mode, the classical mapping model in shear deformation exhibits negative dissipation that is inconsistent with the second law of thermodynamics. 

Finally, the effect of training dataset size $N_{v,\mathrm{iso}}$ on the mean percent relative error $\overline{\mathrm{err}}$ of the two data-driven models is studied in Fig. \ref{fig:GPR_Visco-hyperelastic_5}. Here, $N_{v,\mathrm{iso}}$ is the product of the number of distinct tensile stretch rates in the range of $\dot{\lambda} \in [10, 100]$ considered during training and the number of stretch values in the range of $\lambda \in [1,1.5]$ for every stretch rate case (e.g., in Fig. \ref{fig:GPR_Visco-hyperelastic_1}, $N_{v,\mathrm{iso}}$ was equal to $5 \times 31 = 155$). We consider nine $N_{v,\mathrm{iso}}$ values: 155 (5 (stretch rates) $\times$ 31 (stretch values)), 186 ($6 \times 31$), 217 ($7 \times 31$), 305 ($5 \times 61$), 366 ($6 \times 61$), 427 ($7 \times 61$), 455 ($5 \times 91$), 546 ($6 \times 91$), and 637 ($7 \times 91$). The effect of $N_{v,\mathrm{iso}}$ on $\overline{\mathrm{err}}$ values computed for the training regime (i.e., the fitting error) and for the overall testing regime (tension $+$ compression $+$ shear) are shown in Fig. \ref{fig:GPR_Visco-hyperelastic_5}(a) and Fig. \ref{fig:GPR_Visco-hyperelastic_5}(b), respectively. From these plots, the mean percent relative errors of both the data-driven models in the training regime are very small (maximum $\overline{\mathrm{err}}$ value of less than 5\%), demonstrating a consistently high fitting accuracy. Further, both the $\overline{\mathrm{err}}$ versus $N_{v,\mathrm{iso}}$ responses exhibit a generally decreasing trend. In the testing regime, the responses of the two data-driven models differ significantly. On one hand, the classical mapping model leads to extremely high mean percent relative errors that generally increase with the training dataset size. On the other hand, our data-driven model consistently results in reasonably small mean percent relative errors (< 20\%) for all investigated training dataset sizes, revealing a good prediction performance even when data availability is limited.

%
\section{Summary and Discussion} \label{Section_6}
This work presents a framework for the development of physics-informed data-driven constitutive models to describe the short-time, strain-rate-dependent mechanical response of soft materials. A major motivation of this work is the limitations of the traditional continuum thermodynamics-based and the machine learning-based constitutive models: while the former have a limited fitting and prediction accuracy owing to their fixed mathematical form and require expert model selection based on experimental observations (e.g., choice of the hyperelastic model), the latter require exorbitant amounts of training data (experiments that would be required to uniformly sample strain space are not physically plausible) and generally result in poor out-of-sample predictions (poor generalization performance). Our proposed framework takes a significant step toward eliminating these limitations by combining the physics-informed nature of continuum thermodynamics with the highly accurate and flexible regression capability of supervised ML. The result is a fully data-driven constitutive model that can capture complex material response features with high accuracy without any expert intervention, yields physically reasonable and accurate out-of-sample predictions, and can be trained with a small amount of training data that is achievable from simple contemporary experiments. As some of these points where previously exhibited for hyperelasticity and elastoplasticity this work outlines the extension of this general framework toward capturing the rate-dependent response of viscoelastic materials at finite deformations.

The formulation of our data-driven constitutive model is based on the generalized stress--strain--strain rate equations of the continuum thermodynamics-based framework of external state-variable driven viscous dissipation-based visco-hyperelasticity. In these equations, the total stress is additively decomposed into volumetric, isochoric hyperelastic, and isochoric viscous overstress components. Each of these stress components is written as linear combinations of the components of an integrity basis. This type of linear relationship allowed us to propose three data-driven surrogate model mappings---$\widetilde{\mathcal{M}}_{\mathrm{vol}}$, $\widetilde{\mathcal{M}}_{h,\mathrm{iso}}$, and $\widetilde{\mathcal{M}}_{v,\mathrm{iso}}$---to capture each of the three stress components, respectively. These surrogate models map strain / strain rate invariants to the coefficients of the integrity basis that make up their corresponding stress components. It is shown that this type of model construction ensures key physics-based constraints on the predicted response: principles of local action, determinism, material frame-indifference, the balance of angular momentum, isotropic material symmetry, and limited memory. Further, owing to the exact inference property of the GPR supervised learning method (both standard and C-GPR) and the special inequality constraint capability of C-GPR, the proposed surrogate models also respect the normalization condition and the second law of thermodynamics.

The performance of each of the three surrogate models that form our constitutive model was evaluated by fitting them to a small numerically-generated training dataset, each obtained from one deformation mode---corresponding to common experimental protocols---and then applying the trained model to predict material responses in a significantly wider testing regime comprising multiple distinct deformation modes. In every case, our model's performance was compared with those of a traditional continuum thermodynamics-based constitutive model and a classical ML-based mapping model. The results showed that our models provide critical improvements in describing material responses compared to the other modeling frameworks. For example, from the results of the $\widetilde{\mathcal{M}}_{\mathrm{vol}}$ model under hydrostatic loading, it was seen that our surrogate model predictions are in excellent agreement with the ground truth both in the training regime of confined compression ($J \in [0.75,1]$) and in the overall testing regime comprising confined compression and tension ($J \in [0.5,1.5]$). In comparison, the traditional volumetric neo-Hookean model resulted in large prediction errors in the testing regime, and the classical mapping model resulted in large errors as well as a physically unreasonable softening response in confined tension.

The pathology of predicting unreasonable / thermodynamically unstable physical responses in deformation modes that are not considered in training was consistently noted from the classical mapping approach throughout this study. For example, in the isochoric quasi-static loading case that was considered to evaluate the $\widetilde{\mathcal{M}}_{h,\mathrm{iso}}$ surrogate model, the classical mapping model trained under uniaxial tension violated stress-stretch monotonicity and compression--tension asymmetry under uniaxial compression, and resulted in a zero shear stress regardless of the applied deformation under simple shear loading. In the isochoric dynamic loading case considered for analyzing the $\widetilde{\mathcal{M}}_{v,\mathrm{iso}}$ surrogate model, the classical mapping model trained under rate-dependent uniaxial tension predicted tensile stresses under compression as well as no shear stresses under simple shear. Notably, most efforts in ML-enabled constitutive models in the literature follow the so-called classical approach of mapping strain components to stress components, which is evidently insufficient for performing tasks that traditional phenomenological and micromechanical modeling excelled at. 

Note that in the isochoric quasi-static loading case, even the traditional Yeoh model resulted in a physically unreasonable softening response in compression. Physically unreasonable model predictions are known to be caused by thermodynamic consistency issues (\cite{Upadhyay_etal:2019b}). Unsurprisingly, in the isochoric dynamic loading case, it was seen that the classical mapping model predictions violated the second law of thermodynamics through negative internal dissipation in both the compression and shear deformation modes (the two deformation modes not considered in training). Unlike these alternative model types, our physics-informed data-driven surrogate models resulted in physically-reasonable mechanical responses across all the investigated deformation modes (training and testing) and their prediction errors were consistently at the lowest level (among the three model types).

The physics-informed nature of our constitutive model not only promotes accurate and physically reasonable predictions but also eliminates the requirement of large training datasets by restricting the solution space of possible surrogate model parameters based on physical laws (\cite{Karniadakis_etal:2021}). Notably, our model works in both the low-data regime and in the limited-data regime (due to experimental limitations) allowing it to extrapolate beyond stress states \emph{seen} by simple experiments. For each of the three surrogate models of this work, a reasonably good prediction accuracy over the full testing regime was obtained with relatively small-sized training datasets like the ones commonly seen in the experimental literature. In addition, the prediction accuracy generally improved when the training dataset size was increased (notably, still in the limited-data regime, utilizing synthetic data in deformation modes consistent with easily accessible experiments). On the other hand, classical ML-based mapping models resulted in very high prediction errors at low training dataset sizes, and these errors did not necessarily decrease with increasing data volume. In fact, in the hydrostatic and the isochoric dynamic loading cases, the prediction errors of these black-box models increased when large datasets were employed for training them, which coincided with highly non-physical predictions in the testing regime. Overall, even the lowest mean percent relative error of the classical mapping models, in any case, was several times the peak mean percent relative error of our surrogate models.

Recall that the proposed data-driven constitutive model is only applicable to short-time responses that are dominant under high strain rate deformation. Under sufficiently slow loading rates, long-time effects such as relaxation and creep become important. In addition, our current model enforces material isotropy as one of the physics-based constraints. While many soft materials (e.g., hydrogels, tissues, and elastomers) are considered isotropic, some are anisotropic (e.g., the white matter of the brain (\cite{Eskandari_etal:2021})). In future work, efforts to add long-time effects and anisotropy modeling capabilities in the current framework will be pursued, as well as the utilization of direct experimental data.

\section*{Acknowledgements} \label{Acknowledgements}
This research was supported by the National Institute of Neurological Disorders and Stroke of the National Institutes of Health under Award Number U01NS112120. NB gratefully acknowledges support by the Air Force Office of Scientific Research under award number FA9550-22-1-0075.
\appendix
\renewcommand{\thesection}{Appendix A.}
\setcounter{table}{0}
\renewcommand{\thetable}{A.\arabic{table}}

\section{Derivation of the generalized stress equations of the visco-hyperelastic constitutive model} \label{Appendix_A}
Starting with Eq. (\ref{eq:14}), the total stress in a visco-hyperelastic soft material is written as
\begin{equation} \label{eq:A1}
    \mathbf{S} = \mathbf{S}_\mathrm{vol} + \mathbf{S}_{h,\mathrm{iso}} + \mathbf{S}_{v,\mathrm{iso}} = 2\frac{\partial U(J)}{\partial \mathbf{C}} + 2\frac{\partial \Bar{W}_h(\Bar{I}_1,\Bar{I}_2)}{\partial\mathbf{C}} + 2\frac{\partial \Bar{W}_v(\Bar{I}_1,\Bar{I}_2,\Bar{J}_1,\Bar{J}_2,\Bar{J}_3,\Bar{J}_4,\Bar{J}_5,\Bar{J}_6,\Bar{J}_7)}{\partial\dot{\mathbf{C}}}
\end{equation}
Expanding individual stress components once using the chain rule,
\begin{equation} \label{eq:A2}
    \begin{split}
        \mathbf{S} &= \mathbf{S}_\mathrm{vol} + \mathbf{S}_{h,\mathrm{iso}} + \mathbf{S}_{v,\mathrm{iso}} \\ &= 2 \frac{dU(J)}{dJ}\frac{\partial J}{\partial \textbf{C}} + 2 \sum_{j=1}^2\frac{\partial \Bar{W}_h(\Bar{I}_1,\Bar{I}_2)}{\partial \Bar{I}_j}\frac{\partial \Bar{I}_j}{\partial \mathbf{C}} + 2 \sum_{k=1}^7\frac{\partial \Bar{W}_v(\Bar{I}_1,\Bar{I}_2,\Bar{J}_1,\Bar{J}_2,\Bar{J}_3,\Bar{J}_4,\Bar{J}_5,\Bar{J}_6,\Bar{J}_7)}{\partial \Bar{J}_k}\frac{\partial \Bar{J}_k}{\partial \dot{\mathbf{C}}}  
    \end{split}
\end{equation}
Further expanding the derivatives of invariants (i.e., $\partial \Bar{I}_j/\partial \mathbf{C}$ and $\partial \Bar{J}_k/\partial \dot{\mathbf{C}}$) in the above equation via the chain rule (see expressions of these invariants in \ref{eq:15}),
\begin{equation} \label{eq:A3}
    \begin{split}
        \mathbf{S} &= \mathbf{S}_{\textrm{vol}} + \mathbf{S}_{h,\mathrm{iso}} + \mathbf{S}_{v,\mathrm{iso}} \\ &= 2 \frac{dU(J)}{dJ}\frac{\partial J}{\partial \mathbf{C}} + 2 \sum_{j=1}^2 \frac{\partial \Bar{W}_h}{\partial \Bar{I}_j}\left(\left(\frac{\partial \Bar{\mathbf{C}}}{\partial \mathbf{C}}\right)^{\!\!\mathrm{T}}:\frac{\partial \Bar{I}_j}{\partial \Bar{\mathbf{C}}}\right) + 2 \sum_{k=1}^7\frac{\partial \Bar{W}_v}{\partial \Bar{J}_k}\left(\left(\frac{\partial \dot{\Bar{\mathbf{C}}}}{\partial \dot{ \mathbf{C}}}\right)^{\!\!\mathrm{T}}:\frac{\partial \Bar{J}_k}{\partial \dot{\Bar{\mathbf{C}}}}\right)
    \end{split}
\end{equation}
The individual stress components in Eq. (\ref{eq:A3}) can be simplified using the following results\footnote{The following standard results are used to derive the second part of Eq. (\ref{eq:A4}): \\ Given a scalar $\phi$ and second order tensors \textbf{Y} and \textbf{Z}, we have $\frac{\partial (\alpha\mathbf{Y})}{\partial\mathbf{Z}} = \alpha \frac{\partial\mathbf{Y}}{\partial\mathbf{Z}} + \mathbf{Y} \otimes \frac{\partial \alpha}{\partial\mathbf{Z}}$. \\$\dot{\Bar{\mathbf{C}}} = \frac{d(J^{-2/3}\mathbf{C})}{dt} = J^{-2/3} \dot{\mathbf{C}} - \frac{2}{3}J^{-5/3}\dot{J}\mathbf{C}$, where $\dot{J} = \frac{J}{2} \mathrm{tr}(\mathbf{F}^{-\mathrm{T}}\dot{\mathbf{C}}\mathbf{F}^{-1})$\\ $\frac{\partial \dot{J}}{\partial \dot{\mathbf{C}}} = \frac{J}{2}\mathbf{C}^{-1}$.},
\begin{equation} \label{eq:A4}
    \frac{\partial J}{\partial \mathbf{C}} = \frac{\partial (\mathrm{det}\mathbf{C})^{1/2}}{\partial \mathbf{C}} = \frac{1}{2}J\mathbf{C}^{-1}, \quad \frac{\partial \Bar{\mathbf{C}}}{\partial \mathbf{C}} = \frac{\partial \dot{\Bar{\mathbf{C}}}}{\partial \dot{\mathbf{C}}} =J^{-2/3} \left(\mathbbm{I} - \frac{1}{3}\mathbf{C}\otimes\mathbf{C}^{-1}\right) = J^{-2/3}\mathbbm{P}^\mathrm{T}
\end{equation}
$\mathbbm{P}$ is called the referential fourth order projection tensor (\cite{Holzapfel:2000}),
\begin{equation} \label{eq:A5}
    \mathbbm{P} \coloneqq \mathbbm{I} - \frac{1}{3}\mathbf{C}^{-1}\otimes\mathbf{C}
\end{equation}
, $\mathbbm{I}$ being the fourth order unit tensor (in index notation, $\mathbbm{I}_{ijkl} = \delta_{ik}\delta_{jl}$, where $\delta$ is the Kronecker delta symbol). The tensor $\mathbbm{P}$ projects any second order tensor to its referential deviatoric component, i.e., $\mathbbm{P}:\textbf{Z} = \textrm{Dev}\textbf{Z}$, where \textbf{Z} is an arbritrary second order tensor. Remember, $\mathrm{Dev}\mathbf{Z} = \mathbf{Z} - \frac{1}{3}(\mathbf{Z}:\mathbf{C})\mathbf{C}^{-1}$.

By substituting Eq. (\ref{eq:A4}) in Eq. (\ref{eq:A3}) and invoking the projection property of $\mathbbm{P}$, we have
\begin{equation} \label{eq:A6}
    \begin{split}
        \mathbf{S} &= \mathbf{S}_{\mathrm{vol}} + \mathbf{S}_{h,\mathrm{iso}} + \mathbf{S}_{v,\mathrm{iso}} \\ &= J\frac{dU}{dJ}\mathbf{C}^{-1} + J^{-2/3} \left[ 2 \sum_{j=1}^2 \frac{\partial \Bar{W}_h}{\partial \Bar{I}_j} \mathrm{Dev}\left(\frac{\partial \Bar{I}_j}{\partial \Bar{\mathbf{C}}}\right)\right] + J^{-2/3} \left[ 2 \sum_{k=1}^7 \frac{\partial \Bar{W}_v}{\partial \Bar{J}_k} \mathrm{Dev} \left(\frac{\partial \Bar{J}_k}{\partial \dot{\Bar{\mathbf{C}}}}\right)\right]
    \end{split}
\end{equation}
The derivatives of invariants $\Bar{I}_j$ with respect to $\Bar{\mathbf{C}}$ and those of invariants $\Bar{J}_k$ with respect to $\Bar{\mathbf{C}}$ are given as
\begin{subequations} \label{eq:A7}
\begin{align}
    \frac{\partial\Bar{I}_1}{\partial \Bar{\mathbf{C}}} = \mathbf{I},\quad \frac{\partial\Bar{I}_2}{\partial \Bar{\mathbf{C}}} = \Bar{I}_1\mathbf{I} - \Bar{\mathbf{C}}
\end{align}
\begin{align}
    \frac{\partial\Bar{J}_1}{\partial \dot{\Bar{\mathbf{C}}}} = \mathbf{I}, \quad \frac{\partial\Bar{J}_2}{\partial \dot{\Bar{\mathbf{C}}}} = 2\dot{\Bar{\mathbf{C}}}, \quad \frac{\partial\Bar{J}_3}{\partial \dot{\Bar{\mathbf{C}}}} = \Bar{J}_3\dot{\Bar{\mathbf{C}}}^{-1}
\end{align}
\begin{align}
    \frac{\partial\Bar{J}_4}{\partial \dot{\Bar{\mathbf{C}}}} = \Bar{\mathbf{C}}, \quad \frac{\partial\Bar{J}_5}{\partial \dot{\Bar{\mathbf{C}}}} = (\Bar{\mathbf{C}}\dot{\Bar{\mathbf{C}}} + \dot{\Bar{\mathbf{C}}}\Bar{\mathbf{C}}), \quad \frac{\partial\Bar{J}_6}{\partial \dot{\Bar{\mathbf{C}}}} = \Bar{\mathbf{C}}^2 = \Bar{I}_1\Bar{\mathbf{C}} - \Bar{I}_2\mathbf{I} + \Bar{\mathbf{C}}^{-1}, \quad \frac{\partial\Bar{J}_7}{\partial \dot{\Bar{\mathbf{C}}}} = (\Bar{\mathbf{C}}^{2}\dot{\Bar{\mathbf{C}}} + \dot{\Bar{\mathbf{C}}}\Bar{\mathbf{C}}^{2})
\end{align}
\end{subequations}
Note, the result $\Bar{\mathbf{C}}^2 = \Bar{I}_1\Bar{\mathbf{C}} - \Bar{I}_2\mathbf{I} + \Bar{\mathbf{C}}^{-1}$ is due to the Cayley--Hamilton theorem (\cite{Holzapfel:2000}). By substituting Eq. (\ref{eq:A7}) in Eq. (\ref{eq:A6}), we get
\begin{equation} \label{eq:A8}
    \begin{split}
        \mathbf{S} &= \mathbf{S}_{\mathrm{vol}} + \mathbf{S}_{h,\mathrm{iso}} + \mathbf{S}_{v,\mathrm{iso}} \\ &= J\frac{dU}{dJ}\mathbf{C}^{-1} + J^{-2/3} \left[ 2 \left(\frac{\partial \Bar{W}_h}{\partial \Bar{I}_1} + \Bar{I}_1 \frac{\partial \Bar{W}_h}{\partial \Bar{I}_2} \right)\mathrm{Dev}(\mathbf{I}) - 2 \frac{\partial \Bar{W}_h}{\partial \Bar{I}_2} \mathrm{Dev}(\Bar{\mathbf{C}})\right] \\& \quad + J^{-2/3} \Bigg[2 \frac{\partial \Bar{W}_v}{\partial \Bar{J}_1} \mathrm{Dev}(\mathbf{I}) + 4 \frac{\partial \Bar{W}_v}{\partial \Bar{J}_2} \mathrm{Dev}(\dot{\Bar{\mathbf{C}}}) + 2 \frac{\partial \Bar{W}_v}{\partial \Bar{J}_3} \mathrm{Dev}(\dot{\Bar{\mathbf{C}}}^{-1}) + 2 \frac{\partial \Bar{W}_v}{\partial \Bar{J}_4} \mathrm{Dev}(\Bar{\mathbf{C}}) \\& \quad + 2 \frac{\partial \Bar{W}_v}{\partial \Bar{J}_5} \mathrm{Dev}(\Bar{\mathbf{C}}\dot{\Bar{\mathbf{C}}} + \dot{\Bar{\mathbf{C}}}\Bar{\mathbf{C}}) + 2 \frac{\partial \Bar{W}_v}{\partial \Bar{J}_6} (\Bar{I}_1\mathrm{Dev}(\Bar{\mathbf{C}}) + \mathrm{Dev}(\Bar{\mathbf{C}}^{-1}) - \Bar{I}_2\mathrm{Dev}(\mathbf{I})) \\& \quad + 2 \frac{\partial \Bar{W}_v}{\partial \Bar{J}_7} \mathrm{Dev}(\Bar{\mathbf{C}}^2 \dot{\Bar{\mathbf{C}}} + \dot{\Bar{\mathbf{C}}}\Bar{\mathbf{C}}^2)\Bigg]
    \end{split}
\end{equation}
From Eq. (\ref{eq:A8}), the three stress-components that constitute the total stress in a visco-hyperelastic soft material can be expressed as
\begin{equation} \label{eq:A9}
    \mathbf{S}_\mathrm{vol} = \zeta_1(J)\mathbf{C}^{-1}
\end{equation}
\begin{equation} \label{eq:A10}
    \mathbf{S}_{h,\mathrm{iso}} = J^{-2/3} \left[ \Gamma_1(\Bar{I}_1,\Bar{I}_2) \mathrm{Dev}(\mathbf{I}) + \Gamma_2(\Bar{I}_1,\Bar{I}_2) \mathrm{Dev}(\Bar{\mathbf{C}}) \right]
\end{equation}
\vspace{-12 pt}
\begin{equation} \label{eq:A11}
\begin{split}
    \mathbf{S}_{v,\mathrm{iso}} = {}& J^{-2/3} \bigl[ \Phi_1(\Bar{I}_1,\Bar{I}_2, \Bar{J}_1, \dots, \Bar{J}_7) \mathrm{Dev}(\mathbf{I}) + \Phi_2(\Bar{I}_1,\Bar{I}_2, \Bar{J}_1, \dots, \Bar{J}_7) \mathrm{Dev}(\Bar{\mathbf{C}}) + {}\\& \Phi_3(\Bar{I}_1,\Bar{I}_2, \Bar{J}_1, \dots, \Bar{J}_7) \mathrm{Dev}({\Bar{\mathbf{C}}}^{-1}) + \Phi_4(\Bar{I}_1,\Bar{I}_2, \Bar{J}_1, \dots, \Bar{J}_7) \mathrm{Dev} (\dot{\Bar{\mathbf{C}}}) + {} \\&\Phi_5(\Bar{I}_1,\Bar{I}_2, \Bar{J}_1, \dots, \Bar{J}_7) \mathrm{Dev} ({\dot{\Bar{\mathbf{C}}}}^{-1}) +\Phi_6(\Bar{I}_1,\Bar{I}_2, \Bar{J}_1, \dots, \Bar{J}_7) \mathrm{Dev} (\Bar{\mathbf{C}}\dot{\Bar{\mathbf{C}}} + \dot{\Bar{\mathbf{C}}}\Bar{\mathbf{C}}) + {} \\&\Phi_7(\Bar{I}_1,\Bar{I}_2, \Bar{J}_1, \dots, \Bar{J}_7) \mathrm{Dev} (\Bar{\mathbf{C}}^{2}\dot{\Bar{\mathbf{C}}} + \dot{\Bar{\mathbf{C}}}\Bar{\mathbf{C}}^{2})\bigr]
\end{split}
\end{equation}
where
\begin{subequations} \label{eq:A12}
    \begin{align}
        \zeta_1(J) = J\frac{dU}{dJ}
    \end{align}
    \begin{align}
        \Gamma_1(\Bar{I}_1,\Bar{I}_2) = 2 \left(\frac{\partial \Bar{W}_h}{\partial \Bar{I}_1} + \Bar{I}_1 \frac{\partial \Bar{W}_h}{\partial \Bar{I}_2} \right), \quad \Gamma_2(\Bar{I}_1,\Bar{I}_2) = -2 \frac{\partial \Bar{W}_h}{\partial \Bar{I}_2} 
    \end{align}
    \begin{gather}
        \Phi_1(\Bar{I}_1,\Bar{I}_2, \Bar{J}_1, \dots, \Bar{J}_7) = 2 \left(\frac{\partial \Bar{W}_v}{\partial \Bar{J}_1} - \Bar{I}_2 \frac{\partial \Bar{W}_v}{\partial \Bar{J}_6} \right), \quad \Phi_2(\Bar{I}_1,\Bar{I}_2, \Bar{J}_1, \dots, \Bar{J}_7) = 2 \left(\frac{\partial \Bar{W}_v}{\partial \Bar{J}_4} + \Bar{I}_1 \frac{\partial \Bar{W}_v}{\partial \Bar{J}_6} \right), \\
        \Phi_3(\Bar{I}_1,\Bar{I}_2, \Bar{J}_1, \dots, \Bar{J}_7) = 2 \frac{\partial \Bar{W}_v}{\partial \Bar{J}_6}, \quad \Phi_4(\Bar{I}_1,\Bar{I}_2, \Bar{J}_1, \dots, \Bar{J}_7) = 4 \frac{\partial \Bar{W}_v}{\partial \Bar{J}_2}, \quad \Phi_5(\Bar{I}_1,\Bar{I}_2, \Bar{J}_1, \dots, \Bar{J}_7) = 2 \Bar{J}_3 \frac{\partial \Bar{W}_v}{\partial \Bar{J}_3}, \nonumber\\ \Phi_6(\Bar{I}_1,\Bar{I}_2, \Bar{J}_1, \dots, \Bar{J}_7) = 2 \frac{\partial \Bar{W}_v}{\partial \Bar{J}_5}, \quad \Phi_7(\Bar{I}_1,\Bar{I}_2, \Bar{J}_1, \dots, \Bar{J}_7) = 2 \frac{\partial \Bar{W}_v}{\partial \Bar{J}_7}\nonumber
    \end{gather}
\end{subequations}
Equations (\ref{eq:A9}--\ref{eq:A11}) constitute the alternative form of the generalized functional constitutive equation in Eq. (\ref{eq:14}) and is utilized in the physics-informed mapping approach of this study. Of course, traditional constitutive models have explicit expressions for coefficients $\zeta_1$, $\Gamma_1$, $\Gamma_2$, $\Phi_1$, $\dots$, $\Phi_7$ that are functions of their model parameters. Table A1 lists out these coefficients for several commonly used visco-hyperelastic constitutive models. Our physics-informed data-driven model is not limited by the fitting ability of such explicit mathematical expressions and can learn the mapping between invariants and the coefficients directly from the experimental data.
\begin{table}[ht]
    \caption{Coefficients of the integrity bases in commonly employed volumetric energy density functions ($U$): the Simo--Miehe model (\cite{Simo_Miehe:1992}), the volumetric neo-Hookean model (\cite{DeRooij_Kuhl:2016}), and the volumetric Ogden model (\cite{Ogden:1997})}
    \label{table:A1}
    \begin{tabular}{lcc}
    \toprule
    \scriptsize{\textbf{Model}}               & \scriptsize{$\boldsymbol{U}$}                                                           & \scriptsize{$\boldsymbol{\zeta_1}$}                                      \\
    \midrule
    \scriptsize{Simo--Miehe model}             & \scriptsize{$\frac{\kappa}{2} \left( \frac{J^2 - 1}{2} - \ln{J} \right)$}               & \scriptsize{$\frac{\kappa}{2} \left( J^2 - 1 \right)$}                   \\
    \scriptsize{Volumetric neo-Hookean model} & \scriptsize{$\frac{\kappa}{2} \left(J - 1\right)^2$}                                    & \scriptsize{$\kappa J \left( J - 1 \right)$}                             \\
    \scriptsize{Volumetric Ogden model}        & \scriptsize{$\frac{\kappa}{\beta^2} \left(\frac{1}{J^\beta} - 1 + \beta \ln{J}\right)$} & \scriptsize{$\frac{\kappa}{\beta} \left( 1 - \frac{1}{J^\beta} \right)$} \\
    \bottomrule
    \end{tabular}
\end{table}
\begin{table}[ht]
    \caption{Coefficients of the integrity bases in commonly employed hyperelastic strain energy density functions ($\Bar{W}_h$): the neo-Hookean model (\cite{Rivlin:1948b,Treloar:1943}), the Mooney--Rivlin model and the generalized Rivlin models (\cite{Mooney:1940,Rivlin:1948}), the Yoeh model (\cite{Yeoh:1993}), the Gent model (\cite{Gent:1996}), and the Gent--Gent model (\cite{Pucci_Saccomandi:2002})}
    \label{table:A2}
    \resizebox{\textwidth}{!}{%
    \begin{tabular}{lccc}
    \toprule
    \textbf{Model}           & $\boldsymbol{\Bar{W}_h}$                                                                                                           & $\boldsymbol{\Gamma_1}$                                                             & $\boldsymbol{\Gamma_2}$                          \\
    \midrule
    Neo-Hookean model       & $A_{10}\left(\Bar{I}_1 - 3\right)$                                                                                                 & $2A_{10}$                                                                           & $0$                                              \\
    Mooney--Rivlin model     & $A_{10}\left(\Bar{I}_1 - 3\right) + A_{01}\left(\Bar{I}_2 - 3\right)$                                                              & $2 \left( A_{10} + \Bar{I}_1 A_{01}\right)$                                         & $- 2 A_{01}$                                     \\
    Generalized Rivlin model & $A_{10}\left(\Bar{I}_1 - 3\right) + A_{01}\left(\Bar{I}_2 - 3\right) + A_{11}\left(\Bar{I}_1 - 3\right)\left(\Bar{I}_2 - 3\right)$ & $2 ( A_{10} + \Bar{I}_1 A_{01} + A_{11} (\Bar{I}_1^2 - 3\Bar{I}_1 + \Bar{I}_2 -3))$ & $-2 \left(A_{01} + A_{11}(\Bar{I}_1 - 3)\right)$ \\
    Yeoh model               & $C_1 \left(\Bar{I}_1 - 3\right) + C_2 \left(\Bar{I}_1 - 3\right)^2$                                                                & $2 C_1 +  4 C_2 \left(\Bar{I}_1 - 3\right)$                                         & $0$                                              \\
    Gent model               & $-\frac{\mu J_m}{2}\ln{\left(1 - \frac{\Bar{I}_1 - 3}{J_m}\right)}$                                                                & $\frac{\mu}{1 - \frac{\Bar{I}_1 - 3}{J_m}}$                                         & $0$                                              \\
    Gent--Gent model         & $-\frac{\mu J_m}{2}\ln{\left(1 - \frac{\Bar{I}_1 - 3}{J_m}\right)} + \frac{3C_2}{2}\ln{\left(\frac{\Bar{I}_2}{3}\right)}$          & $\frac{\mu}{1 - \frac{\Bar{I}_1 - 3}{J_m}} + 3 C_2 \frac{\Bar{I}_1}{\Bar{I}_2}$     & $-3\frac{C_2}{\Bar{I}_2}$ \\
    \bottomrule
    \end{tabular}%
    }
\end{table}
\begin{table}[ht]
    \caption{Coefficients of the integrity bases in commonly employed viscous dissipation potentials ($\Bar{W}_v$): the Pioletti model (\cite{Pioletti_etal:1998}), the generalized Pioletti model (\cite{Kulkarni_etal:2016}), and the Upadhyay--Subhash--Spearot (USS) model (\cite{Upadhyay_etal:2020,Upadhyay_etal:2021b})}
    \label{table:A3}
    \resizebox{\textwidth}{!}{%
    \begin{tabular}{lcccccccc}
    \toprule
    \textbf{Model}                   & $\boldsymbol{\Bar{W}_v}$                                                                                & $\boldsymbol{\Phi_1}$ & $\boldsymbol{\Phi_2}$ & $\boldsymbol{\Phi_3}$ & $\boldsymbol{\Phi_4}$                   & $\boldsymbol{\Phi_5}$ & $\boldsymbol{\Phi_6}$                                  & $\boldsymbol{\Phi_7}$ \\
    \midrule
    Pioletti Model                   & $\frac{\eta'}{4} \left(\Bar{I}_1 - 3\right) \Bar{J}_2$                                                  & $0$                   & $0$                   & $0$                   & $\eta'\left(\Bar{I}_1 - 3\right)$       & $0$                   & $0$                                                    & $0$                   \\
    Generalized Pioletti Model       & $\eta \left(\Bar{I}_1 - 3\right)^\beta \Bar{J}_2$                                                       & $0$                   & $0$                   & $0$                   & $4\eta\beta(\Bar{I}_1 - 3)^{\beta - 1}$ & $0$                   & $0$                                                    & $0$                   \\
    Upadhyay--Subhash--Spearot model & $k_{11} \Bar{J}_2 \sqrt{\Bar{I}_1 - 3} + \frac{k_{21}}{c_{21}} \Bar{J}^{c_{21}}_5 \sqrt{\Bar{I}_2 - 3}$ & $0$                   & $0$                   & $0$                   & $4 k_{11} \sqrt{\Bar{I}_1 - 3}$         & $0$                   & $2 k_{21} \Bar{J}_5^{c_{21} - 1} \sqrt{\Bar{I}_2 - 3}$ & $0$ \\
    \bottomrule
    \end{tabular}%
    }
\end{table}
\appendix
\renewcommand{\thesection}{Appendix B.}
\section{Invariants in the undeformed state of a material} \label{Appendix_B}
In the undeformed state of a material when $\mathbf{F} = \mathbf{C} = \mathbf{I}$, the invariant $J$ is unity and thus, the tensors $\mathbf{C}$ and $\Bar{\mathbf{C}}$ become identical. The deformation rate tensors $\dot{\mathbf{C}}$ and $\dot{\Bar{\mathbf{C}}}$, however, are not equal. $\dot{\Bar{\mathbf{C}}}$ in general is given by
\begin{equation} \label{eq:A13}
    \dot{\Bar{\mathbf{C}}} = J^{-2/3}\dot{\mathbf{C}} - \frac{2}{3}J^{-5/3}\dot{J}\mathbf{C}, \quad \textrm{where} \hspace{2 pt} \dot{J} = \frac{J}{2}\mathrm{tr}\left( \mathbf{F}^{-\mathrm{T}} \dot{\mathbf{C}} \mathbf{F}^{-1}\right)
\end{equation}

By substituting the undeformed state values of $\mathbf{F}$, $\mathbf{C}$, and $J$ in Eq. (\ref{eq:A13}), the relationship between the tensors $\dot{\mathbf{C}}$ and $\dot{\Bar{\mathbf{C}}}$ in the undeformed state becomes
\begin{equation} \label{eq:A14}
    \textrm{Undeformed state:} \; \dot{\Bar{\mathbf{C}}} = \dot{\mathbf{C}} - \frac{1}{3}\mathrm{tr}(\dot{\mathbf{C}})\mathbf{I}
\end{equation}

Using the result $\Bar{\mathbf{C}} = \mathbf{C} = \mathbf{I}$ and the expression for $\Bar{\mathbf{C}}$ from Eq. (\ref{eq:A14}), the invariants (see Eq. (\ref{eq:15})) in the undeformed state are obtained as
\begin{subequations} \label{eq:A15}
\begin{align}
    \Bar{I}_1 = 3,\quad \Bar{I}_2 = 3
\end{align}
\begin{align}
    \Bar{J}_1 = 0, \quad \Bar{J}_2 = \frac{2}{3}\mathrm{tr}\left( \dot{\mathbf{C}}^2\right), \quad \Bar{J}_3 = \mathrm{det} \left(\dot{\mathbf{C}} - \frac{1}{3}\mathrm{tr}(\dot{\mathbf{C}})\mathbf{I}\right)
\end{align}
\begin{align}
    \Bar{J}_4 = 0, \quad \Bar{J}_5 = \frac{2}{3}\mathrm{tr}\left( \dot{\mathbf{C}}^2\right), \quad \Bar{J}_6 = 0, \quad \Bar{J}_7 = \frac{2}{3}\mathrm{tr}\left( \dot{\mathbf{C}}^2\right)
\end{align}
\end{subequations}

Notice that among the above invariants, only $\Bar{J}_1$, $\Bar{J}_4$, and $\Bar{J}_6$ vanish in the undeformed state. The strain invariants $\Bar{I}_1$ and $\Bar{I}_2$ assume a non-zero but constant value of 3. Lastly, the invariants $\Bar{J}_2$, $\Bar{J}_3$, $\Bar{J}_5$, and $\Bar{J}_7$ remain functions of the applied loading rate (also note: $\Bar{J}_2=\Bar{J}_5=\Bar{J}_7$ under this condition).

\bibliography{bib.bib}

\end{document}


\maketitle
\section{Data-driven Model Predictions in the Confined Tension Regime} \label{Section_SM1}
In the Section 5.1 of the main manuscript on "Hydrostatic (Confined Compression Loading)", two data-driven constitutive models, the present physics-informed surrogate model and a classical mapping model, were trained using stress--strain data under the confined compression deformation mode. A small dataset comprising of 26 data-points was considered in the training process. Fig. SM1 shown below plots the stress predictions of these data-driven models in the uniaxial testing regime (i.e., compression $+$ tension) when they are trained using a much larger training dataset of 201 data-points. 
\begin{figure}[t!]
    \centering
    \includegraphics[width=13 cm]{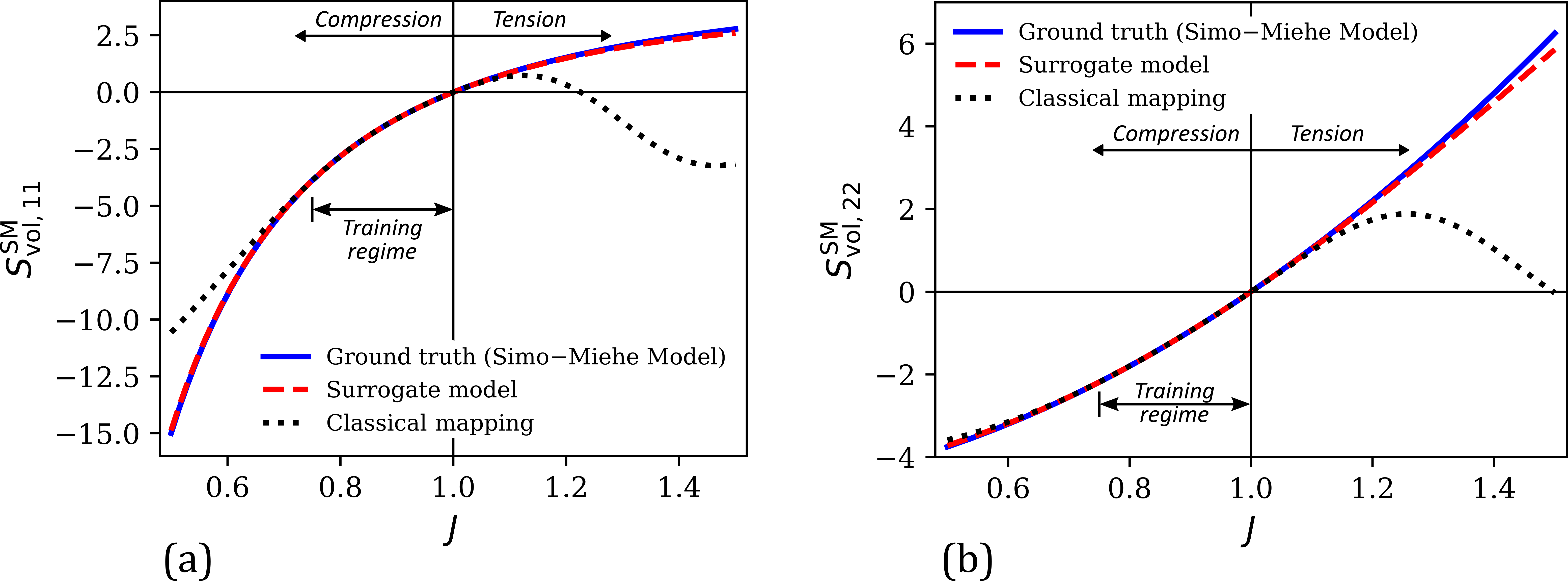}
    \caption{Stress predictions of data-driven models in the uniaxial testing regime (i.e., compression $+$ tension) when they are trained using a training dataset of 201 data-points.}
    \label{fig:GPR_hydrostatic_SM}
\end{figure}